\documentclass[12pt]{article}
\usepackage{amsfonts,graphicx,amsmath,amsthm,amssymb,epsfig}
\usepackage{float}
\allowdisplaybreaks
\begin{document}

\title{\bf Estimating the Role of Bag Constant and Modified Theory on
Anisotropic Stellar Models}
\author{Tayyab Naseer \thanks{tayyabnaseer48@yahoo.com;
tayyab.naseer@math.uol.edu.pk}~ and M. Sharif \thanks{msharif.math@pu.edu.pk} \\
Department of Mathematics and Statistics, The University of Lahore,\\
1-KM Defence Road Lahore-54000, Pakistan.}

\date{}
\maketitle

\begin{abstract}
In this article, we are devoted to discuss different compact stars
admitting anisotropic interiors in a particular modified theory of
gravity. For this purpose, a spherically symmetric metric is adopted
to formulate the field equations corresponding to two different
$f(\mathcal{R},\mathcal{T},\mathcal{Q})$ models, where
$\mathcal{Q}=\mathcal{R}_{\alpha\gamma}\mathcal{T}^{\alpha\gamma}$.
Since the field equations contain extra degrees of freedom, we
choose Finch-Skea metric and MIT bag model equation of state to make
them solvable. We also use matching conditions to calculate a
constant triplet in the chosen ansatz. The resulting solutions are
then graphically analyzed for particular values of the bag constant
and model parameter in the interior of 4U 1820-30 compact star. The
viability and stability of the modified models are also checked
through certain tests. Further, we calculate the values of model
parameter through the vanishing radial pressure constraint that
correspond to the observed data (radii and masses) of eight
different star candidates. Finally, we conclude that our models I
and II are in well-agreement with the conditions needed for
physically relevant interiors to exist.
\end{abstract}

\section{Introduction}

General Relativity (GR) has long been considered the prevailing
gravitational theory within the scientific community, addressing a
multitude of challenges. However, it falls short in providing a
comprehensive understanding of the rapid cosmic expansion. To
grapple with these enigmatic cosmic phenomena, such as the
accelerated expansion and the presence of dark matter, scientists
have recently explored various extensions to GR. Recent observations
have suggested the existence of a repulsive force responsible for
this cosmic acceleration, commonly referred to the dark energy,
driving galaxies apart from each other. The $f(\mathcal{R})$ theory
stands as the first ever extension to GR, achieved by a modification
of the Einstein-Hilbert action, wherein the Ricci scalar
$\mathcal{R}$ is replaced by its generic function \cite{1}. In the
quest to understand celestial structures, researchers have employed
various techniques within this modified framework to examine the
stability and viability \cite{2}-\cite{2f}. Beyond its applications
to celestial bodies, $f(\mathcal{R})$ gravity models have played a
pivotal role in addressing different cosmological concerns,
including the late-time evolution of the universe, the inflationary
era, etc. \cite{3}-\cite{5a}.

In a study conducted by Bertolami and his colleagues \cite{10}, a
search to unreveal the cosmos' intriguing facets prompted an
exploration of a novel concept of the matter-geometry coupling. For
this, they incorporated the impact of geometric properties into the
matter Lagrangian within the context of $f(\mathcal{R})$ theory.
This pioneering approach piqued the curiosity of various
astronomers, who redirected their focus towards understanding the
accelerated expansion of the universe. Building on this foundation,
recent developments at the action level have led to the emergence of
modified theories of gravity, capturing the keen interest of
astrophysicists. Harko et al. \cite{20} introduced the first theory
based on this concept, known as $f(\mathcal{R},\mathcal{T})$
gravity, with $\mathcal{T}$ representing the trace of the
energy-momentum tensor (EMT). The inclusion of $\mathcal{T}$ gives
rise to non-conservation phenomena and has sparked comprehensive
investigations into the physical feasibility of self-gravitating
structures, resulting in a plethora of noteworthy findings
\cite{21}-\cite{21h}. Haghani et al. \cite{22} introduced a more
intricate functional that depends on $\mathcal{R}$, $\mathcal{T}$
and $\mathcal{R}_{\alpha\gamma}\mathcal{T}^{\alpha\gamma}$. Within
this framework, they delved into the study of a cosmic era marked by
exponential expansion and explored three different models to assess
their physical viability. To further enrich their analysis, they
employed the Lagrange multiplier technique to compute conserved EMT
within this theory.

The development of this modified proposal hinges on the
incorporation of the factor $\mathcal{Q}$ to enforce a robust
non-minimal matter-geometry coupling in self-gravitating systems.
Altering the Einstein-Hilbert action offers a potential avenue for
elucidating the roles of dark energy and dark matter without relying
on exotic fluid distributions. While alternative extensions of GR,
such as $f(\mathcal{R},\L_m)$ and $f(\mathcal{R},\mathcal{T})$
gravitational theories, also include matter Lagrangian with
arbitrary interactions, their functionals may not be considered the
most generalized forms providing a comprehensive understanding of
the impact of the coupling on self-gravitating objects in certain
scenarios. Notably, in the case where the internal composition of a
star exhibits $\mathcal{T}=0$, the preservation of coupling effects
within $f(\mathcal{R},\mathcal{T})$ theory fades away. However, such
phenomenon does not occur in the context of
$f(\mathcal{R},\mathcal{T},\mathcal{Q})$ gravity. The non-conserved
energy-momentum tensor in this theory results in an additional
force, leading to an end of of test particles' motion along geodesic
paths. This force also contributes to elucidating galactic rotation
curves.

Sharif and Zubair \cite{22a} adopted a different approach,
investigating two mathematical models $\mathcal{R}+\beta\mathcal{Q}$
and $\mathcal{R}(1+\beta\mathcal{Q})$ with distinct choices for the
fluid Lagrangian, specifically $\L_m=\rho$ and $-P$. In doing so,
they examined various properties of black holes and their associated
thermodynamic laws, subsequently establishing constraints to assess
the viability of these models. Odintsov and Saez-Gomez \cite{23}
explored the impact of non-constant matter distributions, revealing
that such variations can lead to a pure de Sitter universe in the
context of $f(\mathcal{R},\mathcal{T},\mathcal{Q})$ theory. Ayuso et
al. \cite{24} introduced scalar and vector fields to investigate the
modified field equations. Their analysis unveiled the intricate
non-linearity stemming from conformal and arbitrary non-minimal
matter-geometry couplings. The numerical solutions and stability
criteria for different models, along with the use of a perturbation
function, have been discussed in detail \cite{25}. Notably, the
choice of the matter Lagrangian was found to yield diverse results,
particularly depending on the radial/tangential components of
pressure \cite{25a}. Furthermore, the curvature tensor was
orthogonally decomposed in terms of
$f(\mathcal{R},\mathcal{T},\mathcal{Q})$ EMT and the Weyl tensor,
resulting in some scalar functions for both charged and uncharged
fluids, which bear significance in the study of celestial systems
\cite{26}-\cite{26e}. Extracting the solutions of modified field
equations through various techniques led to model several
anisotropic systems, which were found to be consistent and
physically valid \cite{27a}-\cite{27aaa}.

Within the vast and enigmatic expanse of the cosmos, stars stand as
fundamental constituents of our galaxy. Astrophysicists have long
been captivated by the composition of these celestial bodies,
focusing on the exploration of their evolution. Among these
appealing entities, neutron stars emerged as particularly
intriguing, stemming from the gravitational collapse of massive
stars. They possess a mass ranging from one to three times that of
our solar system and hold a dense core comprised of newly formed
neutrons. These neutrons exert a pressure capable of counteracting
the attractive force of gravity, preventing further collapse.
Although the concept of a neutron star was initially postulated in
1934 \cite{28}, observational confirmation took several years, as
these structures often elude direct detection, emitting insufficient
radiations. In addition to neutron stars, the theoretical realm
includes another enigmatic entity, the quark star, situated between
the extremes of a black hole and a neutron star. These hypothetical
objects are believed to contain strange, down and up quarks within
their cores \cite{29a}-\cite{29h}.

The matter determinants characterizing self-gravitating interiors,
whether isotropic or anisotropic, are typically described in terms
of the energy density and pressure. These attributes are
interrelated in some specific way, with one of the well-established
constraint being the MIT bag model equation of state (EoS)
\cite{30}. Notably, the bag model effectively captures the
characteristics of quark star-like structures such as 4U 1820-30,
RXJ 185635-3754, PSR 0943+10, Her X-1, SAX J 1808.4-3658, 4U
1728-34, etc. On the other hand, such a description is not
attainable with a neutron star EoS \cite{33a}. Physicists have
extensively employed the MIT bag model to unreveal the inner
composition of strange systems \cite{33b}-\cite{34aa}. Demorest et
al. \cite{34b} conducted a detailed analysis of the compact
structure PSR J1614-2230, concluding that only this model supports
such incredibly dense objects. Rahaman et al. \cite{35} further
investigated specific stellar compositions using this model in
conjunction with an interpolating function, while Sharif and his
collaborators \cite{38,38c} extended this work across various
modified gravity scenarios from which they yield physically stable
models.

Numerous methodologies have been presented in scientific literature
to address the challenges posed by the Einstein and modified field
equations. These approaches encompass the utilization of a specific
EoS, well-defined metric potentials, or a vanishing complexity
condition, etc. In the realm of radial/time metric components,
Finch-Skea ansatz \cite{38ca} has garnered considerable attention
from researchers as a valuable tool for the investigation of
anisotropic compact structures. Banerjee et al. \cite{38m} presented
a family of interior solutions analogous to BTZ exterior by making
use of Finch-Skea metric and found physically viable results even in
lower dimensions. Hansraj and Maharaj \cite{38n} produced three
different classes of analytic solutions to the Einstein-Maxwell
field equations and confirmed the fulfillment of stability criteria.
The physical analysis indicated that this analogue can be used to
model physically realistic charged compact stars. Researchers, both
within the realms of GR and in modified theories, have delved into
this particular metric solution through different approaches like
the gravitational decoupling and the consideration of complexity
factor, etc., to discuss self-gravitating charged/uncharged systems
\cite{38o}-\cite{38r}.

This study is dedicated to evaluate the physical validity of the
Finch-Skea anisotropic solutions within the framework of
$f(\mathcal{R},\mathcal{T},\mathcal{R}_{\alpha\gamma}\mathcal{T}^{\alpha\gamma})$
theory. The structure of this paper is as follows. Section
\textbf{2} provides some basics of this modified theory and explores
the corresponding field equations for two distinct models.
Additionally, we adopt the MIT bag model EoS to characterize the
internal composition of the considered 4U 1820-30 star. In section
\textbf{3}, we consider the Finch-Skea spacetime and employ junction
conditions to calculate the values of an unknown triplet
$(d_1,d_2,d_3)$, within this ansatz. Section \textbf{4} undertakes a
graphical examination of the physical characteristics of the
resulting solutions by fixing the range of the model parameter. In
section \textbf{5}, we determine the values of the model parameter
which are compatible with the estimated data of different compact
stars. Lastly, section \textbf{6} provides a summary of our
findings.

\section{Modified $f(\mathcal{R},\mathcal{T},\mathcal{R}_{\alpha\gamma}\mathcal{T}^{\alpha\gamma})$
Gravity and MIT Bag Model}

The modification of the Einstein-Hilbert action for
$f(\mathcal{R},\mathcal{T},\mathcal{R}_{\alpha\gamma}\mathcal{T}^{\alpha\gamma})$
theory (with $\kappa=8\pi$) is given by \cite{23}
\begin{equation}\label{g1}
\mathbb{S}=\int\sqrt{-g}
\left\{\frac{f(\mathcal{R},\mathcal{T},\mathcal{R}_{\alpha\gamma}\mathcal{T}^{\alpha\gamma})}{16\pi}
+\L_{m}\right\}d^{4}x,
\end{equation}
where the Ricci scalar is replaced by the modified functional and
$\L_{m}$ represents the fluid Lagrangian density. Giving variation
in the action \eqref{g1} with respect to the metric tensor yields
the field equations in this framework as
\begin{equation}\label{g2}
\mathcal{G}_{\alpha\gamma}=8\pi\mathcal{T}_{\alpha\gamma}^{(\mathrm{eff})}=\frac{8\pi\mathcal{T}_{\alpha\gamma}}
{f_{\mathcal{R}}-\L_{m}f_{\mathcal{Q}}}+\mathcal{T}_{\alpha\gamma}^{(\mathrm{cor})},
\end{equation}
connecting the geometry of the spacetime structure with the fluid
content in its interior. The geometry and modified fluid
distribution can be expressed by the Einstein tensor
$\mathcal{G}_{\alpha\gamma}$ and EMT
$\mathcal{T}_{\alpha\gamma}^{(\mathrm{eff})}$, respectively. The
functional in the action \eqref{g1} modifies the fluid sector,
containing the EMT of GR along with some correction terms,
represented by $\mathcal{T}_{\alpha\gamma}^{(\mathrm{cor})}$ as
\begin{eqnarray}\nonumber
\mathcal{T}_{\alpha\gamma}^{(\mathrm{cor})}&=&-\frac{1}{\bigg(\L_{m}f_{\mathcal{Q}}-f_{\mathcal{R}}\bigg)}
\left[\left(f_{\mathcal{T}}+\frac{1}{2}\mathcal{R}f_{\mathcal{Q}}\right)\mathcal{T}_{\alpha\gamma}
-\left\{\L_{m}f_{\mathcal{T}}-\frac{\mathcal{R}}{2}\bigg(\frac{f}{\mathcal{R}}-f_{\mathcal{R}}\bigg)\right.\right.\\\nonumber
&+&\left.\frac{1}{2}\nabla_{\varrho}\nabla_{\sigma}(f_{\mathcal{Q}}\mathcal{T}^{\varrho\sigma})\right\}g_{\alpha\gamma}
-\frac{1}{2}\Box(f_{\mathcal{Q}}\mathcal{T}_{\alpha\gamma})-(g_{\alpha\gamma}\Box-
\nabla_{\alpha}\nabla_{\gamma})f_{\mathcal{R}}\\\label{g4}
&-&2f_{\mathcal{Q}}\mathcal{R}_{\varrho(\alpha}\mathcal{T}_{\gamma)}^{\varrho}
+\nabla_{\varrho}\nabla_{(\alpha}[\mathcal{T}_{\gamma)}^{\varrho}
f_{\mathcal{Q}}]+2(f_{\mathcal{Q}}\mathcal{R}^{\varrho\sigma}+\left.f_{\mathcal{T}}g^{\varrho\sigma})\frac{\partial^2\L_{m}}
{\partial g^{\alpha\gamma}\partial g^{\varrho\sigma}}\right],
\end{eqnarray}
where $f_{\mathcal{R}},~f_{\mathcal{T}}$ and $f_{\mathcal{Q}}$ are
the partial derivatives of the functional
$f(\mathcal{R},\mathcal{T},\mathcal{Q})$ with reference to its
arguments. Also, we represent the covariant derivative and
D'Alambert operator by $\nabla_\varrho$ and $\Box\equiv
\frac{1}{\sqrt{-g}}\partial_\alpha\big(\sqrt{-g}g^{\alpha\gamma}\partial_{\gamma}\big)$,
respectively. Moreover, the matter Lagrangian for such fluid
distributions can be taken as $\L_{m}=-\rho$, where $\rho$ is the
energy density. This leads to the vanishing of the last term of
Eq.\eqref{g4} \cite{22}. The equivalence principle is broken in this
theory due to the non-conservation of the fluid EMT (i.e.,
$\nabla_\alpha \mathcal{T}^{\alpha\gamma}\neq 0$) until we involve
an additional force. This force affects the motion of particles in
the self-gravitating systems, changing their path from geodesic to
non-geodesic. Consequently, we obtain
\begin{align}\nonumber
\nabla^\alpha
\mathcal{T}_{\alpha\gamma}&=\frac{2}{2f_\mathcal{T}+\mathcal{R}f_\mathcal{Q}+16\pi}\bigg[\nabla_\alpha
\big(f_\mathcal{Q}\mathcal{R}^{\varrho\alpha}\mathcal{T}_{\varrho\gamma}\big)-\mathcal{G}_{\alpha\gamma}\nabla^\alpha
\big(f_\mathcal{Q}\L_m\big)+\nabla_\gamma
\big(\L_mf_\mathcal{T}\big)\\\label{g4a}
&-\frac{1}{2}\nabla_\gamma\mathcal{T}^{\varrho\sigma}\big(f_\mathcal{T}g_{\varrho\sigma}+f_\mathcal{Q}\mathcal{R}_{\varrho\sigma}\big)
-\frac{1}{2}\big\{\nabla^{\alpha}(\mathcal{R}f_{\mathcal{Q}})+2\nabla^{\alpha}f_{\mathcal{T}}\big\}\mathcal{T}_{\alpha\gamma}\bigg].
\end{align}

The EMT serves as a fundamental tool in exploring the internal
characteristics and composition of self-gravitating systems, making
it particularly indispensable in the field of astrophysics. A myriad
of celestial bodies, spanning a wide spectrum of massive structures
in the universe, is believed to exhibit pressure anisotropy.
Consequently, the EMT assumes paramount importance for
astrophysicists engaged in deciphering the intricate details of
stellar model evolution. Its application extends beyond the confines
of conventional isotropic systems, allowing researchers to gain
valuable insights into the gravitational interactions and behaviors
of anisotropic matter distributions. The anisotropic EMT is
mathematically defined as follows
\begin{equation}\label{g5}
\mathcal{T}_{\alpha\gamma}=(\rho+P_t) \mathrm{U}_{\alpha}
\mathrm{U}_{\gamma}+P_t
g_{\alpha\gamma}+\left(P_r-P_t\right)\mathrm{V}_\alpha\mathrm{V}_\gamma,
\end{equation}
where $P_t$ and $P_r$ are the tangential and radial pressures,
respectively. Furthermore, $\mathrm{V}_{\alpha}$ symbolizes the
four-vector and $\mathrm{U}_\alpha$ being the four-velocity. The
field equations for $f(\mathcal{R},\mathcal{T},\mathcal{Q})$ theory
derive the trace as
\begin{align}\nonumber
&3\nabla^{\varrho}\nabla_{\varrho}
f_\mathcal{R}-\mathcal{T}(8\pi+f_\mathcal{T})-\mathcal{R}\left(\frac{\mathcal{T}}{2}f_\mathcal{Q}-f_\mathcal{R}\right)+\frac{1}{2}
\nabla^{\varrho}\nabla_{\varrho}(f_\mathcal{Q}\mathcal{T})\\\nonumber
&+\nabla_\alpha\nabla_\varrho(f_\mathcal{Q}\mathcal{T}^{\alpha\varrho})
-2f+(\mathcal{R}f_\mathcal{Q}+4f_\mathcal{T})\L_m+2\mathcal{R}_{\alpha\varrho}\mathcal{T}^{\alpha\varrho}f_\mathcal{Q}\\\nonumber
&-2g^{\gamma\sigma} \frac{\partial^2\L_m}{\partial
g^{\gamma\sigma}\partial
g^{\alpha\varrho}}\left(f_\mathcal{T}g^{\alpha\varrho}+f_\mathcal{Q}R^{\alpha\varrho}\right)=0.
\end{align}
The $f(\mathcal{R},\mathcal{T})$ gravity is attained from the above
equation when setting $f_\mathcal{Q}$ to zero. In addition,
$f_\mathcal{T}=0$ results in the $f(\mathcal{R})$ theory.

The line element allows us to explore the gravitational field and
the way it warps spacetime within massive objects. It plays a
pivotal role in describing the conditions prevailing in the interior
regions of stars, which are essential for comprehending their
structure, evolution, and ultimately, their fate. In this regard, we
consider the interior metric representing static spherical spacetime
as follows
\begin{equation}\label{g6}
ds^2=-e^{A_1} dt^2+e^{{A_2}} dr^2+r^2d\theta^2+r^2\sin^2\theta
d\phi^2,
\end{equation}
where $A_1=A_1(r)$ and ${{A_2}}={{A_2}}(r)$. The four quantities now
take the form
\begin{equation}\label{g7}
\mathrm{V}^\alpha=\delta^\alpha_1 e^{-\frac{{{A_2}}}{2}}, \quad
\mathrm{U}^\alpha=\delta^\alpha_0 e^{-\frac{A_1}{2}},
\end{equation}
satisfying $\mathrm{U}^\alpha \mathrm{U}_{\alpha}=-1$ and
$\mathrm{V}^\alpha \mathrm{U}_{\alpha}=0$.

Our universe is presently undergoing a rapid expansion phase and
consists of a vast number of stars existing in a non-linear realm.
However, studying their behavior through linear analysis can provide
valuable insights for astronomers seeking to comprehend the
formation of these massive structures more effectively. Moreover,
the non-minimally matter-geometry coupled term
$\mathcal{R}_{\alpha\gamma}\mathcal{T}^{\alpha\gamma}$ makes this
theory significantly complicated in comparison with
$f(\mathcal{R},\L_m)$ and $f(\mathcal{R},\mathcal{T})$ scenarios.
The motivation behind exploring such coupling in the realm of
theoretical physics stems from a desire to refine our understanding
of the fundamental interactions shaping the universe. This coupling
involves the intricate interplay between matter fields and the
geometry of spacetime, offering a nuanced perspective beyond the
conventional minimal coupling scenarios. Researchers are driven by
the quest to figure out the mysteries of cosmic evolution through
such interactions. Non-minimal coupling models may provide a more
comprehensive framework to describe phenomena like dark matter and
dark energy, offering potential insights into the accelerated
expansion of the universe. To address this, we shall examine two
specific functional forms as follows \cite{22}
\begin{itemize}
\item \textbf{Model I:}\quad $f(\mathcal{R},\mathcal{T},\mathcal{R}_{\alpha\gamma}\mathcal{T}^{\alpha\gamma})=\mathcal{R}+\beta
\mathcal{R}_{\alpha\gamma}\mathcal{T}^{\alpha\gamma}$,
\item \textbf{Model II:}\quad $f(\mathcal{R},\mathcal{T},\mathcal{R}_{\alpha\gamma}\mathcal{T}^{\alpha\gamma})=\mathcal{R}(1+\beta
\mathcal{R}_{\alpha\gamma}\mathcal{T}^{\alpha\gamma})$,
\end{itemize}
where $\beta$ denotes a real-valued coupling parameter. It is worth
noting that various values of this parameter, well within the
observed range, ensure the physical viability of compact star
models. Haghani et al. \cite{22} extensively examined these models
and explored the evolution of the scale factor as well as
deceleration parameter. They deduced that when $\beta>0$, the
solution exhibits an oscillatory behavior characterized by
alternating phases of expansion and contraction. This dynamic
pattern reflects the influence of positive energy density on the
evolution of the system. Conversely, when $\beta<0$, the evolution
of the universe takes on a different character, with the scale
factor demonstrating a dependence of a hyperbolic cosine function.
By focusing on these models, Sharif and Zubair \cite{22a} delved
into isotropic distributions, deriving scientifically acceptable
values for the coupling parameter. The value of $\mathcal{Q}$ in
terms of geometric components is given by
\begin{eqnarray}\nonumber
\mathcal{Q}&=&e^{-{A_2}}\bigg[\frac{\rho}{4}\left(A_1'^2-A_1'{A_2}'+2A_1''+\frac{4A_1'}{r}\right)
+P_t\left(\frac{{A_2}'}{r}-\frac{A_1'}{r}+\frac{2e^{A_2}}{r^2}-\frac{2}{r^2}\right)\\\nonumber
&-&\frac{P_r}{4}\left(A_1'^2-A_1'{A_2}'+2A_1''+\frac{4{A_2}'}{r}\right)\bigg],
\end{eqnarray}
where prime shows differentiation with respect to $r$.

The modified model I yields the independent field equations
\eqref{g2} for the anisotropic fluid \eqref{g5} given by
\begin{align}\nonumber
8\pi\rho&=e^{-{A_2}}\bigg[\frac{{A_2}'}{r}+\frac{e^{A_2}}{r^2}-\frac{1}{r^2}+\beta\bigg\{\rho\bigg(\frac{3A_1'{A_2}'}{8}-\frac{A_1'^2}{8}
+\frac{{A_2}'}{r}+\frac{e^{A_2}}{r^2}-\frac{3A_1''}{4}-\frac{3A_1'}{2r}\\\nonumber
&-\frac{1}{r^2}\bigg)-\rho'\bigg(\frac{{A_2}'}{4}-\frac{1}{r}-A_1'\bigg)+\frac{\rho''}{2}+P_r\bigg(\frac{A_1'{A_2}'}{8}
-\frac{A_1'^2}{8}-\frac{A_1''}{4}+\frac{{A_2}'}{2r}+\frac{{A_2}''}{2}\\\label{g8}
&-\frac{3{A_2}'^2}{4}\bigg)+\frac{5{A_2}'P'_r}{4}-\frac{P''_r}{2}+P_t\bigg(\frac{{A_2}'}{2r}-\frac{A_1'}{2r}+\frac{3e^{A_2}}{r^2}
-\frac{1}{r^2}\bigg)-\frac{P'_t}{r}\bigg\}\bigg],\\\nonumber 8\pi
P_r&=e^{-{A_2}}\bigg[\frac{A_1'}{r}-\frac{e^{A_2}}{r^2}+\frac{1}{r^2}+\beta\bigg\{\rho\bigg(\frac{A_1'{A_2}'}{8}+\frac{A_1'^2}{8}
-\frac{A_1''}{4}-\frac{A_1'}{2r}\bigg)-\frac{A_1'\rho'}{4}-P_r\\\nonumber
&\times\bigg(\frac{5A_1'^2}{8}-\frac{7A_1'{A_2}'}{8}+\frac{5A_1''}{4}-\frac{7{A_2}'}{2r}+\frac{A_1'}{r}-{A_2}'^2
-\frac{e^{A_2}}{r^2}+\frac{1}{r^2}\bigg)+P'_r\bigg(\frac{A_1'}{4}+\frac{1}{r}\bigg)\\\label{g8a}
&-P_t\bigg(\frac{{A_2}'}{2r}-\frac{A_1'}{2r}+\frac{3e^{A_2}}{r^2}-\frac{1}{r^2}\bigg)+\frac{P'_t}{r}\bigg\}\bigg],\\\nonumber
8\pi
P_t&=e^{-{A_2}}\bigg[\frac{A_1'^2}{4}-\frac{A_1'{A_2}'}{4}+\frac{A_1''}{2}-\frac{{A_2}'}{2r}+\frac{A_1'}{2r}
+\beta\bigg\{\rho\bigg(\frac{A_1'^2}{8}+\frac{A_1'{A_2}'}{8}-\frac{A_1''}{4}-\frac{A_1'}{2r}\bigg)\\\nonumber
&-\frac{A_1'\rho'}{4}+P_r\bigg(\frac{A_1'^2}{8}-\frac{A_1'{A_2}'}{8}+\frac{A_1''}{4}-\frac{{A_2}'}{2r}-\frac{{A_2}''}{2}
+\frac{3{A_2}'^2}{4}\bigg)-\frac{5{A_2}'P'_r}{4}+\frac{P''_r}{2}\\\label{g8b}
&-P_t\bigg(\frac{A_1'^2}{4}-\frac{A_1'{A_2}'}{4}+\frac{A_1''}{2}-\frac{{A_2}'}{r}+\frac{A_1'}{r}\bigg)
-P'_t\bigg(\frac{{A_2}'}{4}-\frac{A_1'}{4}-\frac{3}{r}\bigg)+\frac{P''_t}{2}\bigg\}\bigg].
\end{align}
On the other hand, they become for model II as
\begin{align}\nonumber
8\pi\rho&=e^{-{A_2}}\bigg[\frac{{A_2}'}{r}+\frac{e^{A_2}}{r^2}-\frac{1}{r^2}+\beta\bigg\{\rho\bigg(\bigg(\frac{{A_2}'}{r}+\frac{e^{A_2}}{r^2}
-\frac{1}{r^2}\bigg)\tau_1-\mathcal{R}\bigg(\frac{1}{r^2}-\frac{{A_2}'}{r}-\frac{e^{A_2}}{r^2}\\\nonumber
&+\frac{\mathcal{R}e^{A_2}}{2}-\frac{3A_1'^2}{8}-\frac{3A_1'}{2r}+\frac{5A_1'{A_2}'}{8}-\frac{3A_1''}{4}\bigg)
+\mathcal{R}'\bigg(\frac{{A_2}'}{2}-\frac{1}{r}\bigg)-\frac{\mathcal{R}''}{2}-\tau_4\bigg(\frac{2}{r}\\\nonumber
&-\frac{{A_2}'}{2}\bigg)-\tau_7\bigg)+\rho'\bigg(\tau_1\bigg(\frac{{A_2}'}{2}-\frac{2}{r}\bigg)-\mathcal{R}\bigg(\frac{1}{r}-\frac{{A_2}'}{4}\bigg)
-\mathcal{R}'-2\tau_4\bigg)-\rho''\bigg(\tau_1\\\nonumber
&+\frac{\mathcal{R}}{2}\bigg)-P_r\bigg(\bigg(\frac{1}{r^2}-\frac{{A_2}'}{r}-\frac{e^{A_2}}{r^2}\bigg)\tau_2+\mathcal{R}\bigg(\frac{A_1'^2}{8}
-\frac{1}{r^2}-\frac{A_1'{A_2}'}{8}+\frac{{A_2}'}{2r}+\frac{A_1''}{4}\bigg)\\\nonumber
&-\mathcal{R}'\bigg(\frac{2}{r}-\frac{{A_2}'}{2}\bigg)-\frac{\mathcal{R}''}{2}+\tau_5\bigg(\frac{2}{r}-\frac{{A_2}'}{2}\bigg)+\tau_8\bigg)
-P'_r\bigg(\tau_2\bigg(\frac{2}{r}-\frac{{A_2}'}{2}\bigg)+\mathcal{R}\\\nonumber
&\times\bigg(\frac{{A_2}'}{4}-\frac{2}{r}\bigg)-\mathcal{R}'+2\tau_5\bigg)-P''_r\bigg(\tau_2-\frac{\mathcal{R}}{2}\bigg)
-P_t\bigg(\bigg(\frac{1}{r^2}-\frac{{A_2}'}{r}-\frac{e^{A_2}}{r^2}\bigg)\tau_3\\\nonumber
&+\mathcal{R}\bigg(\frac{A_1'}{2r}+\frac{1}{r^2}-\frac{{A_2}'}{2r}\bigg)+\frac{\mathcal{R}'}{r}+\tau_6\bigg(\frac{2}{r}-\frac{{A_2}'}{2}\bigg)
+\tau_9\bigg)-P'_t\bigg(\tau_3\bigg(\frac{2}{r}-\frac{{A_2}'}{2}\bigg)\\\label{g8c}
&+\frac{\mathcal{R}}{2}+2\tau_6\bigg)-P''_t\tau_3\bigg\}\bigg],\\\nonumber
8\pi
P_r&=e^{-{A_2}}\bigg[\frac{A_1'}{r}-\frac{e^{A_2}}{r^2}+\frac{1}{r^2}+\beta\bigg\{\rho\bigg(\bigg(\frac{A_1'}{r}-\frac{e^{A_2}}{r^2}
+\frac{1}{r^2}\bigg)\tau_1-\mathcal{R}\bigg(\frac{e^{A_2}}{r^2}-\frac{1}{r^2}-\frac{A_1'}{r}\\\nonumber
&+\frac{A_1'^2}{8}+\frac{A_1'}{2r}-\frac{A_1'{A_2}'}{8}+\frac{A_1''}{4}\bigg)
+\frac{\mathcal{R}'A_1'}{4}+\tau_4\bigg(\frac{2}{r}+\frac{A_1'}{2}\bigg)\bigg)+\rho'\bigg(\tau_1\bigg(\frac{2}{r}
+\frac{A_1'}{2}\bigg)\\\nonumber
&+\frac{\mathcal{R}A_1'}{4}\bigg)+P_r\bigg(\bigg(\frac{1}{r^2}+\frac{A_1'}{r}-\frac{e^{A_2}}{r^2}\bigg)\tau_2
-\mathcal{R}\bigg(\frac{\mathcal{R}e^{A_2}}{2}-\frac{3A_1'^2}{8}+\frac{1}{r^2}+\frac{A_1'}{r}+\frac{3{A_2}'}{2r}\\\nonumber
&+\frac{3A_1'{A_2}'}{8}-\frac{3A_1''}{4}\bigg)-\mathcal{R}'\bigg(\frac{1}{r}+\frac{A_1'}{4}\bigg)
+\tau_5\bigg(\frac{2}{r}+\frac{A_1'}{2}\bigg)\bigg)+P'_r\bigg(\tau_2\bigg(\frac{2}{r}+\frac{A_1'}{2}\bigg)\\\nonumber
&-\mathcal{R}\bigg(\frac{A_1'}{4}+\frac{1}{r}\bigg)\bigg)-P_t\bigg(\bigg(\frac{A_1'}{r}+\frac{e^{A_2}}{r^2}-\frac{1}{r^2}\bigg)\tau_3
+\mathcal{R}\bigg(\frac{{A_2}'}{2r}-\frac{1}{r^2}-\frac{A_1'}{2r}\bigg)+\frac{\mathcal{R}'}{r}\\\label{g8d}
&-\tau_6\bigg(\frac{2}{r}+\frac{A_1'}{2}\bigg)\bigg)+P'_t\bigg(\tau_3\bigg(\frac{2}{r}
+\frac{A_1'}{2}\bigg)-\frac{\mathcal{R}}{2}\bigg)\bigg\}\bigg],\\\nonumber
8\pi
P_t&=e^{-{A_2}}\bigg[\frac{A_1''}{2}-\frac{A_1'{A_2}'}{4}+\frac{A_1'^2}{4}+\frac{A_1'}{2r}-\frac{{A_2}}{2r}
+\beta\bigg\{\rho\bigg(\tau_1\bigg(\frac{A_1''}{2}-\frac{A_1'{A_2}'}{4}+\frac{A_1'^2}{4}\\\nonumber
&+\frac{A_1'}{2r}-\frac{{A_2}}{2r}\bigg)-\mathcal{R}\bigg(\frac{A_1'{A_2}'}{8}-\frac{A_1''}{4}-\frac{A_1'^2}{8}+\frac{{A_2}'}{2r}\bigg)
-\bigg(\frac{{A_2}'}{2}-\frac{1}{r}-\frac{A_1'}{2}\bigg)\tau_4+\tau_7\\\nonumber
&-\frac{\mathcal{R}'A_1'}{4}\bigg)-\rho'\bigg(\tau_1\bigg(\frac{{A_2}'}{2}-\frac{1}{r}-\frac{A_1'}{2}\bigg)+\frac{\mathcal{R}A_1'}{4}-2\tau_4\bigg)
+\rho''\tau_1-P_r\bigg(\bigg(\frac{{A_2}'}{2r}\\\nonumber
&-\frac{A_1''}{2}-\frac{A_1'^2}{4}-\frac{A_1'}{2r}+\frac{A_1'{A_2}'}{4}\bigg)\tau_2+\mathcal{R}\bigg(\frac{A_1'^2}{8}+\frac{A_1'}{2r}-\frac{A_1'{A_2}'}{8}
+\frac{A_1''}{4}\bigg)+\mathcal{R}'\\\nonumber
&\times\bigg(\frac{A_1'}{2}+\frac{1}{r}-\frac{{A_2}'}{4}\bigg)+\frac{\mathcal{R}''}{2}+\tau_5\bigg(\frac{{A_2}'}{2}-\frac{1}{r}-\frac{A_1'}{2}\bigg)
-\tau_8\bigg)-P'_r\bigg(\tau_2\bigg(\frac{{A_2}'}{2}\\\nonumber
&-\frac{1}{r}-\frac{A_1'}{2}\bigg)+\mathcal{R}\bigg(\frac{A_1'}{2}+\frac{1}{r}-\frac{{A_2}'}{4}\bigg)+\mathcal{R}'-2\tau_5\bigg)+P''_r\bigg(\tau_2-\frac{\mathcal{R}}{2}\bigg)
+P_t\bigg(\tau_3\\\nonumber
&\times\bigg(\frac{A_1''}{2}-\frac{A_1'{A_2}'}{4}+\frac{A_1'^2}{4}+\frac{A_1'}{2r}-\frac{{A_2}}{2r}\bigg)-\mathcal{R}\bigg(\frac{\mathcal{R}e^{A_2}}{2}-\frac{2}{r^2}+\frac{{A_2}'}{r}-\frac{A_1'}{r}
+\frac{2e^{A_2}}{r^2}\bigg)\\\nonumber
&-\mathcal{R}'\bigg(\frac{A_1'}{4}-\frac{{A_2}'}{4}\bigg)-\frac{\mathcal{R}''}{2}-\tau_6\bigg(\frac{{A_2}'}{2}-\frac{1}{r}-\frac{A_1'}{2}\bigg)+\tau_9\bigg)-P'_t\bigg(\tau_3\bigg(\frac{{A_2}'}{2}-\frac{1}{r}\\\label{g8e}
&-\frac{A_1'}{2}\bigg)+\mathcal{R}\bigg(\frac{A_1'}{4}-\frac{{A_2}'}{4}+\frac{2}{r}\bigg)+\mathcal{R}'-2\tau_6\bigg)+P''_t\bigg(\tau_3-\frac{\mathcal{R}}{2}\bigg)\bigg\}\bigg],
\end{align}
where $\tau_i^{'s},~i=1,2,3,...,9$ are provided in Appendix
$\mathbf{A}$. Misner and Sharp \cite{41b} proposed a formula that
helps to calculate the mass of self-gravitating spherical objects.
This is given by
\begin{equation}\nonumber
m(r)=\frac{r}{2}\big(1-g^{\alpha\gamma}r_{,\alpha}r_{,\gamma}\big),
\end{equation}
giving rise to
\begin{equation}\label{g12a}
m(r)=\frac{r}{2}\big(1-e^{-{A_2}}\big).
\end{equation}

Within any geometric configuration, physical parameters such as
pressure and energy density can be intricately linked through some
specific relationships known as EoSs. The end of a massive star's
life cycle may result in the formation of a white dwarf, neutron
star, or black hole, with neutron stars standing out as particularly
captivating celestial formations within our universe. Moreover,
under varying density conditions in their cores, these structures
can undergo transformations into quark stars or black holes
\cite{33b,41c,41d}. Notably, these celestial objects boast
incredibly dense cores, generating strong gravitational fields
around them, despite their relatively small sizes. Since the sets of
equations \eqref{g8}-\eqref{g8b} and \eqref{g8c}-\eqref{g8e} possess
matter variables and its derivatives coupled with geometric terms,
obtaining solutions for these equations can be a non-trivial task.
Consequently, the imposition of certain constraints becomes
imperative to derive the requisite solutions. Here, we adopt the MIT
bag model EoS to delve into the physical characteristics of diverse
quark candidates \cite{30}. We define the quark pressure as
\begin{equation}\label{g13}
P_r=\sum_{\iota=s,u,d}P^\iota-\mathrm{B_c},
\end{equation}
with $\mathrm{B_c}$ being the bag constant. Further, the subscripts
$s,~u$ and $d$ correspond to strange, up and down flavors of quarks,
respectively. The quark energy density is linked with its pressure
by the amount $\rho^\iota=3P^\iota$. Hence, the total energy density
can mathematically be expressed as
\begin{equation}\label{g14}
\rho=\sum_{\iota=s,u,d}\rho^\iota+\mathrm{B_c}.
\end{equation}
Joining the above two equations results in the following
\begin{equation}\label{g14a}
P_r=\frac{1}{3}\left(\rho-4\mathrm{B_c}\right).
\end{equation}

Researchers investigated several quark interiors with the help of
the above equation and deduced the values of the bag constant which
are compatible with the observed data \cite{41f,41h}. Equations
\eqref{g8} and \eqref{g8a} now become after combining with EoS
\eqref{g14a} as
\begin{align}\nonumber
\rho&=\bigg[\beta\bigg(\frac{9A_1''}{8}-\frac{e^{{A_2}}}{r^2}+\frac{1}{r^2}-\frac{{A_2}''}{8}-\frac{5A_1'{A_2}'}{8}-\frac{{A_2}'^2}{16}
-\frac{7{A_2}'}{2r}+\frac{3A_1'^2}{16}+\frac{7A_1'}{4r}\bigg)\\\nonumber
&+8\pi
e^{{A_2}}\bigg]^{-1}\bigg[\frac{3}{4}\bigg(\frac{{A_2}'}{r}+\frac{A_1'}{r}\bigg)+\mathrm{B_c}\bigg\{8\pi
e^{A_2}-\beta\bigg(\frac{4{A_2}'}{r}-\frac{3A_1'^2}{4}-\frac{3A_1''}{2}+\frac{{A_2}''}{2}\\\label{g14b}
&+\frac{{A_2}'^2}{4}+A_1'{A_2}'-\frac{A_1'}{r}+\frac{e^{A_2}}{r^2}-\frac{1}{r^2}\bigg)\bigg\}\bigg],\\\nonumber
P_r&=\bigg[\beta\bigg(\frac{9A_1''}{8}-\frac{e^{{A_2}}}{r^2}+\frac{1}{r^2}-\frac{{A_2}''}{8}-\frac{5A_1'{A_2}'}{8}-\frac{{A_2}'^2}{16}
-\frac{7{A_2}'}{2r}+\frac{3A_1'^2}{16}+\frac{7A_1'}{4r}\bigg)\\\nonumber
&+8\pi
e^{{A_2}}\bigg]^{-1}\bigg[\frac{1}{4}\bigg(\frac{{A_2}'}{r}+\frac{A_1'}{r}\bigg)-\mathrm{B_c}\bigg\{8\pi
e^{A_2}-\beta\bigg(\frac{A_1'{A_2}'}{2}
+\frac{{A_2}'}{r}-\frac{2A_1'}{r}\\\label{g14c}
&+\frac{e^{A_2}}{r^2}-A_1''-\frac{1}{r^2}\bigg)\bigg\}\bigg],
\end{align}
whilst Eqs.\eqref{g8c} and \eqref{g8d} take the form
\begin{align}\nonumber
\rho&=\bigg[\beta\bigg\{\frac{3}{4}\bigg(\frac{{A_2}'}{r}+\frac{A_1'}{r}\bigg)\bigg(\tau_1+\frac{\tau_2}{3}\bigg)
+\frac{3}{8}\big({A_2}'+A_1'\big)\bigg(\tau_4+\frac{\tau_5}{3}\bigg)
+\mathcal{R}\bigg(\frac{A_1''}{2}+\frac{{A_2}'}{4r}\\\nonumber
&-\frac{7A_1'{A_2}'}{16}+\frac{A_1'^2}{16}+\frac{5A_1'}{4r}-\frac{\mathcal{R}e^{A_2}}{2}\bigg)+\mathcal{R}'\bigg(\frac{A_1'}{8}-\frac{1}{2r}
+\frac{{A_2}'}{8}\bigg)-\frac{\mathcal{R''}}{4}-\frac{\tau_8}{4}\bigg\}\\\nonumber
&-8\pi
e^{{A_2}}\bigg]^{-1}\bigg[-\frac{3}{4}\bigg(\frac{{A_2}'}{r}+\frac{A_1'}{r}\bigg)-8\pi\mathrm{B_c}e^{A_2}-\beta\mathrm{B_c}\bigg\{\bigg(\frac{{A_2}'}{r}
+\frac{A_1'}{r}\bigg)\tau_2-\mathcal{R}\bigg(\frac{A_1'^2}{2}\\\nonumber
&+\frac{\mathcal{R}e^{A_2}}{2}+\frac{A_1'}{r}+\frac{A_1'{A_2}'}{4}+\frac{2{A_2}'}{r}-\frac{A_1''}{2}\bigg)-\mathcal{R}'\bigg(\frac{A_1'}{4}-\frac{1}{r}
+\frac{{A_2}'}{4}\bigg)+\frac{\mathcal{R''}}{2}-\tau_8\\\label{g14e}
&+\tau_5\bigg(\frac{{A_2}'}{2}+\frac{A_1'}{2}\bigg)\bigg\}\bigg],\\\nonumber
P_r&=\bigg[\beta\bigg\{\frac{3}{4}\bigg(\frac{{A_2}'}{r}+\frac{A_1'}{r}\bigg)\bigg(\tau_1+\frac{\tau_2}{3}\bigg)+\frac{3}{8}\big({A_2}'
+A_1'\big)\bigg(\tau_4+\frac{\tau_5}{3}\bigg)+\mathcal{R}\bigg(\frac{A_1''}{2}+\frac{{A_2}'}{4r}\\\nonumber
&-\frac{7A_1'{A_2}'}{16}+\frac{A_1'^2}{16}+\frac{5A_1'}{4r}-\frac{\mathcal{R}e^{A_2}}{2}\bigg)+\mathcal{R}'\bigg(\frac{A_1'}{8}-\frac{1}{2r}
+\frac{{A_2}'}{8}\bigg)-\frac{\mathcal{R''}}{4}-\frac{\tau_8}{4}\bigg\}\\\nonumber
&-8\pi
e^{{A_2}}\bigg]^{-1}\bigg[-\frac{1}{4}\bigg(\frac{{A_2}'}{r}+\frac{A_1'}{r}\bigg)+8\pi\mathrm{B_c}e^{A_2}-\beta\mathrm{B_c}\bigg\{\bigg(\frac{{A_2}'}{r}
+\frac{A_1'}{r}\bigg)\tau_1+\mathcal{R}\bigg(\frac{A_1'^2}{4}\\\nonumber
&-\frac{\mathcal{R}e^{A_2}}{2}+\frac{{A_2}'}{r}-\frac{A_1'{A_2}'}{2}+\frac{A_1''}{2}+\frac{2A_1'}{r}\bigg)+\mathcal{R}'\bigg(\frac{A_1'}{4}-\frac{1}{r}
+\frac{{A_2}'}{4}\bigg)-\frac{\mathcal{R''}}{2}-\tau_7\\\label{g14f}
&+\tau_4\bigg(\frac{{A_2}'}{2}+\frac{A_1'}{2}\bigg)\bigg\}\bigg].
\end{align}
One can determine the tangential pressure for model I by
incorporating the energy density \eqref{g14b} and the radial
pressure \eqref{g14c} into Eq.\eqref{g8b}. Similarly, for model II,
this can be obtained by utilizing together Eqs.\eqref{g8e},
\eqref{g14e} and \eqref{g14f}. In this study, we present solutions
to the field equations for both models using MIT bag model, with a
particular range of the coupling parameter. This analysis provides
valuable insights into the behavior of strange quark celestial
bodies.

\section{Finch-Skea Spacetime and Matching Conditions}

Since the field equations still involve extra unknowns, we turn our
attention to the Finch-Skea spacetime, an ansatz that has garnered
considerable attention in the realm of astrophysics. The metric
coefficients of this line element are given by \cite{38ca}
\begin{equation}\label{g15}
e^{A_1}=\frac{1}{4}\big(2d_1+d_2\sqrt{d_3}r^2\big)^2, \quad\quad
e^{{A_2}}=d_3r^2+1,
\end{equation}
comprising a triplet ($d_1,d_2,d_3$), need to be determined. We
shall use the matching criteria in the following to determine their
values in terms of radius and mass of a compact star. Since multiple
metrics have been proposed in the literature, one must check whether
the ansatz under consideration is physically acceptable or not. For
this, a criteria has been proposed \cite{41j} and we take
derivatives of both time/radial components in order to verify it as
follows
\begin{align}\nonumber
A_1'(r)&=\frac{4 d_2 \sqrt{d_3} r}{d_2 \sqrt{d_3} r^2+2 d_1}, \quad
A_1''(r)=\frac{8 d_1 d_2 \sqrt{d_3}-4 d_2^2 d_3 r^2}{\left(d_2
\sqrt{d_3} r^2+2 d_1\right)^2},\\\nonumber {A_2}'(r)&=\frac{2 d_3
r}{d_3 r^2+1},\quad {A_2}''(r)=\frac{2 d_3}{d_3 r^2+1}-\frac{4 d_3^2
r^2}{\left(d_3 r^2+1\right)^2}.
\end{align}
We observe from the above expressions that $A_1'(0)=0={A_2}'(0)$ and
$A_1''(0)$, ${A_2}''(0)$$>$$0$ everywhere ($r=0$ being star's
center). Hence, the acceptance of both potentials defined in
Eq.\eqref{g15} is validated.

The matching of the interior and exterior regions at the
hypersurface offers a set of conditions that prove to be an
immensely valuable tool for understanding the overall structure of
celestial bodies. Ensuring a correspondence between the
characteristics of both these spacetimes is indeed necessary. This
entails considerations such as the charge/uncharge distribution,
whether the regions are static or in a state of dynamical evolution,
and other relevant parameters. In the light of these factors, the
Schwarzschild metric with $\mathrm{M}$ representing the total mass
emerges as the most suitable choice for describing the exterior
geometry. We take it as
\begin{equation}\label{g20}
ds^2=-\left(1-\frac{2\mathrm{M}}{r}\right)dt^2+\frac{dr^2}{\left(1-\frac{2\mathrm{M}}{r}\right)}
+r^2d\theta^2+r^2\sin^2\theta d\phi^2.
\end{equation}
The matching conditions are mainly based on two fundamental forms.
The first of them assures the continuity of metric coefficients and
their derivatives of both regions across the spherical boundary.
This gives
\begin{eqnarray}\label{g21}
g_{tt}&{_{=}^{\Sigma}}&e^{A_1(\emph{R})}=\frac{1}{4}\big(2d_1+d_2\sqrt{d_3}\emph{R}^2\big)^2=1-\frac{2\mathrm{M}}{\emph{R}},\\\label{g21a}
g_{rr}&{_{=}^{\Sigma}}&e^{{A_2}(\emph{R})}=d_3\emph{R}^2+1
=\bigg(1-\frac{2\mathrm{M}}{\emph{R}}\bigg)^{-1},\\\label{g22}
\frac{\partial g_{tt}}{\partial
r}&{_{=}^{\Sigma}}&A_1'(\emph{R})=\frac{4d_2\sqrt{d_3}\emph{R}}{2d_1+d_2\sqrt{d_3}\emph{R}^2}=\frac{2\mathrm{M}}{\emph{R}\big(\emph{R}-2\mathrm{M}\big)}.
\end{eqnarray}
Solving the above equations simultaneously, the triplet is
calculated as
\begin{align}\label{g23}
d_1=\frac{2\emph{R}-5\mathrm{M}}{2\emph{R}\sqrt{\emph{R}-2\mathrm{M}}},\quad
d_2=\frac{1}{\emph{R}^{\frac{3}{2}}}\sqrt{\frac{\mathrm{M}}{2}},\quad
d_3=\frac{2\mathrm{M}}{\emph{R}^2\big(\emph{R}-2\mathrm{M}\big)}.
\end{align}
An alternative method for determining these constants is to take the
initial two constraints described in Eqs.\eqref{g21} and
\eqref{g21a}, in addition to the condition of vanishing radial
pressure at the spherical boundary, i.e., $P_r~{_=^\Sigma}~0$. By
employing these three equations, distinct set of values for the
triplet can be derived. However, we shall use the later condition
for estimating the values of the model parameter.

\section{Graphical Interpretation of Constructed Solutions}

This section is devoted to graphically analyze our both resulting
solutions for a particular 4U 1820-30 compact body in this modified
theory. The preliminary data of this star is provided as the radius
$\emph{R}=9.1 \pm 0.4~km$ and mass $\mathrm{M}=1.58 \pm
0.06~\mathrm{M}_{\bigodot}$ \cite{41k}. Further, we conduct a
comprehensive analysis of the considered model, encompassing an
array of critical properties to ascertain the physical viability of
the developed solutions. These properties include the metric
potentials, anisotropy, energy bounds, mass and redshift, etc. The
investigation also extends to the stability of the obtained models,
taking into account the specific range of the model parameter
$\beta$ that can be positive or negative.

There are some of our works in which the positive values of $\beta$
do not provide physically acceptable results as the behavior of the
fluid triplet appears to be the negative \cite{27da,27d}. Thus we
required to adopt its negative values to obtain physically relevant
compact interiors. However, this is not the case in this article.
Both positive as well as negative values of this parameter can be
chosen that shall later be shown through graphical analysis.
Moreover, we plot these physical properties for different values of
the bag constant and find good results for
$\mathrm{B_c}=92~MeV/fm^3$. Hence, all the plots in this paper
correspond to this value. It is important to emphasize that a valid
solution should exhibit metric functions that are not only
monotonically increasing but also devoid of singularities. This
consistency is clearly evident in the graphical representations of
the temporal and radial metric coefficients \eqref{g15}, as
illustrated in Figure \textbf{1}.
\begin{figure}[h!]\center
\epsfig{file=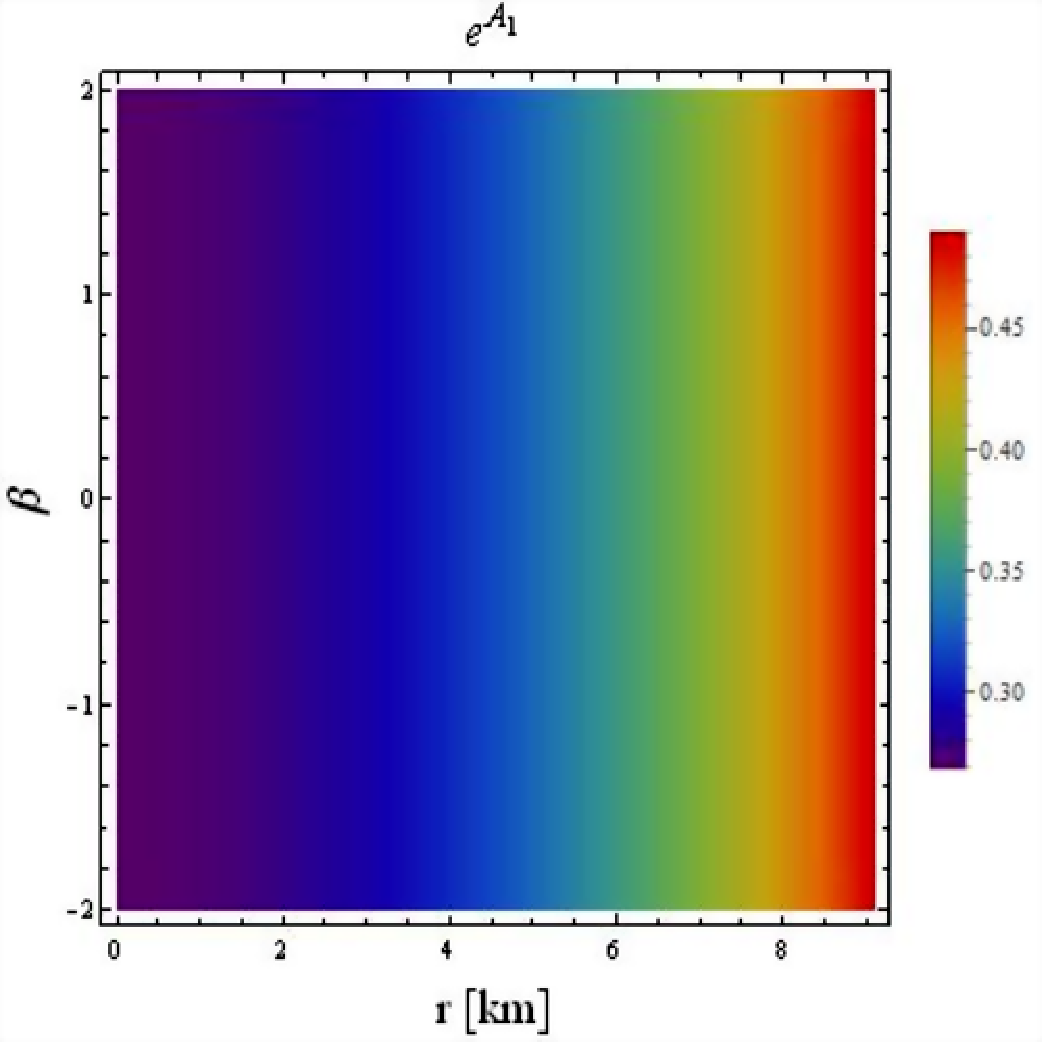,width=.4\linewidth}\epsfig{file=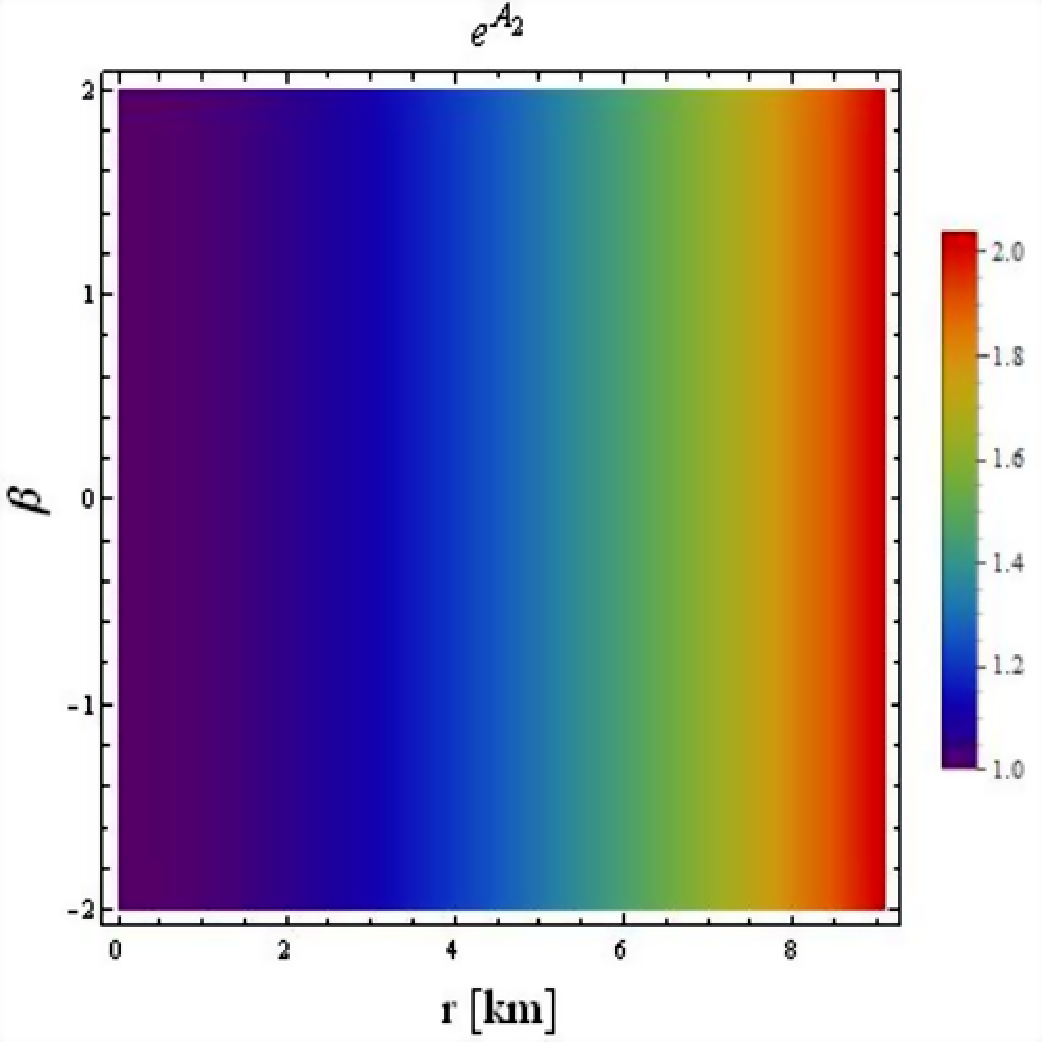,width=.4\linewidth}
\caption{Metric potentials \eqref{g15} versus $\beta$ and $r$.}
\end{figure}

\subsection{Matter Determinants}

The concentration of matter within the core of a geometric structure
is a fundamental need in order to validate the corresponding
solution. This implies that the density and pressures must be
maximum and finitely positive in the center and decrease outwards.
In this case of anisotropic fluid, the energy density and the
radial/tangential pressures are important factors to assess. To this
end, we perform a comparative analysis for models I and II in Figure
\textbf{2}. Notably, the analysis reveals that the matter variables
exhibit behavior consistent with physically acceptable models. This
observation lends support to the presence of highly dense cores
within the framework of modified gravity. Examining the first row of
Figure \textbf{2}, it becomes apparent that model I generates
slightly dense structures. Moreover, the remaining graphs in the
same Figure demonstrate that model II acquires higher values of both
pressure components within the considered stellar object.
Importantly, the regular conditions for both models are verified and
the corresponding plots are included in Figures \textbf{3} and
\textbf{4}.
\begin{figure}[h!]\center
\epsfig{file=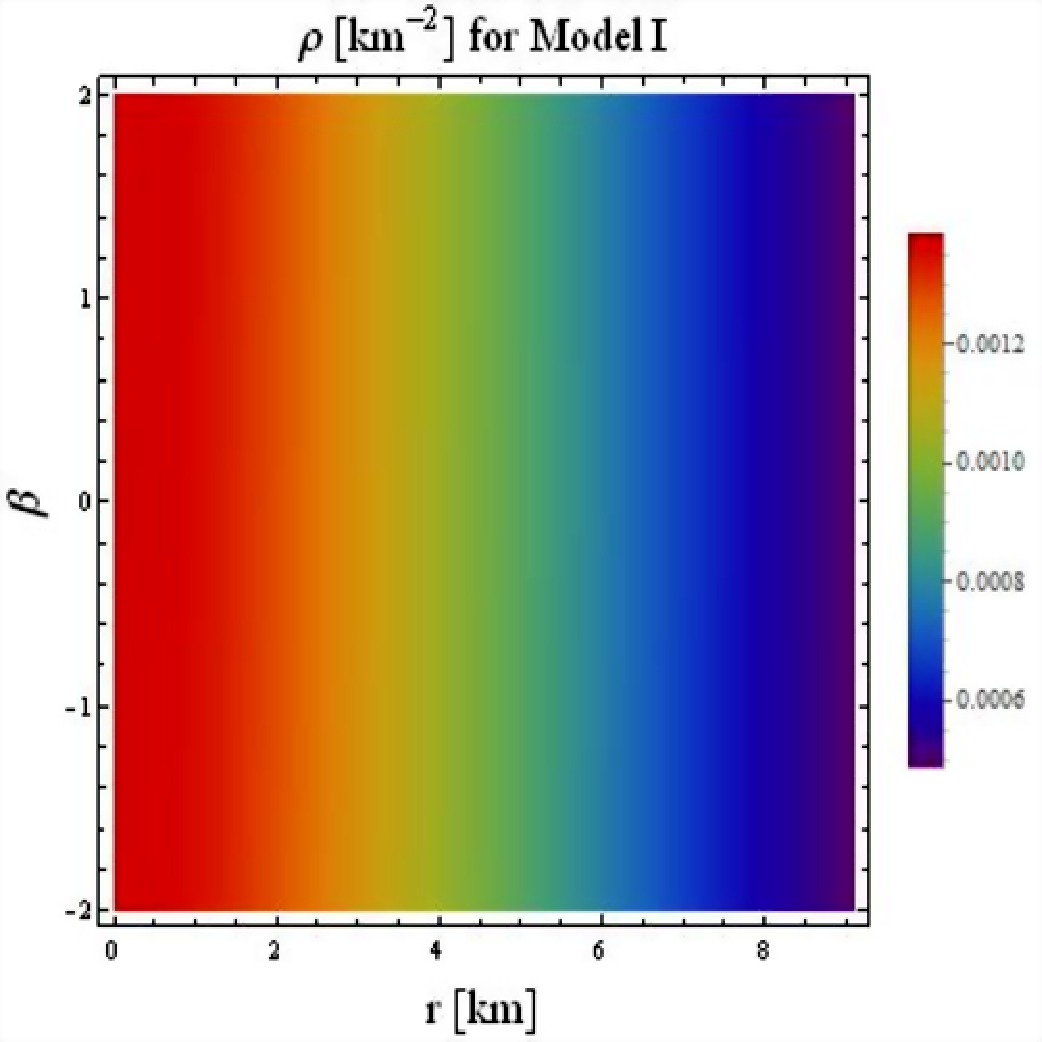,width=.4\linewidth}\epsfig{file=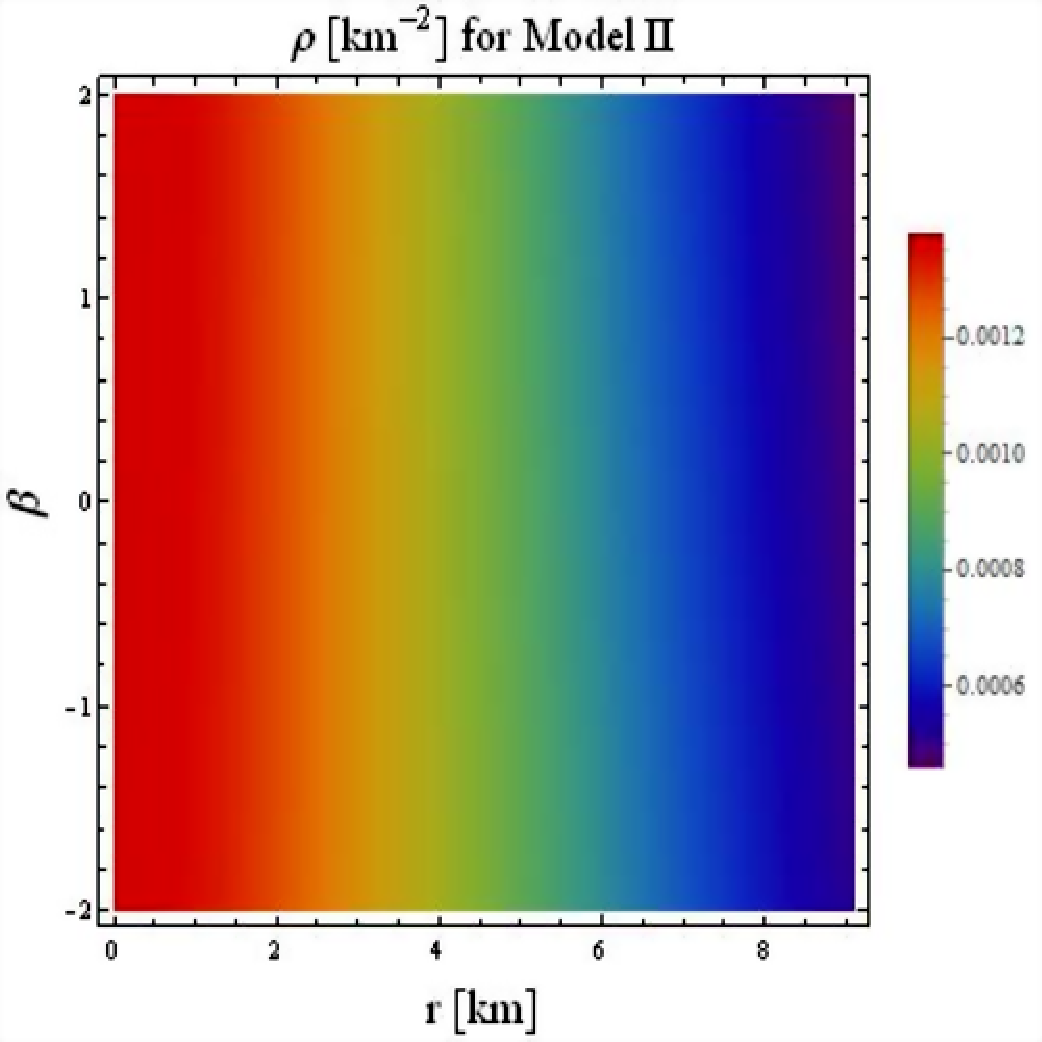,width=.4\linewidth}
\epsfig{file=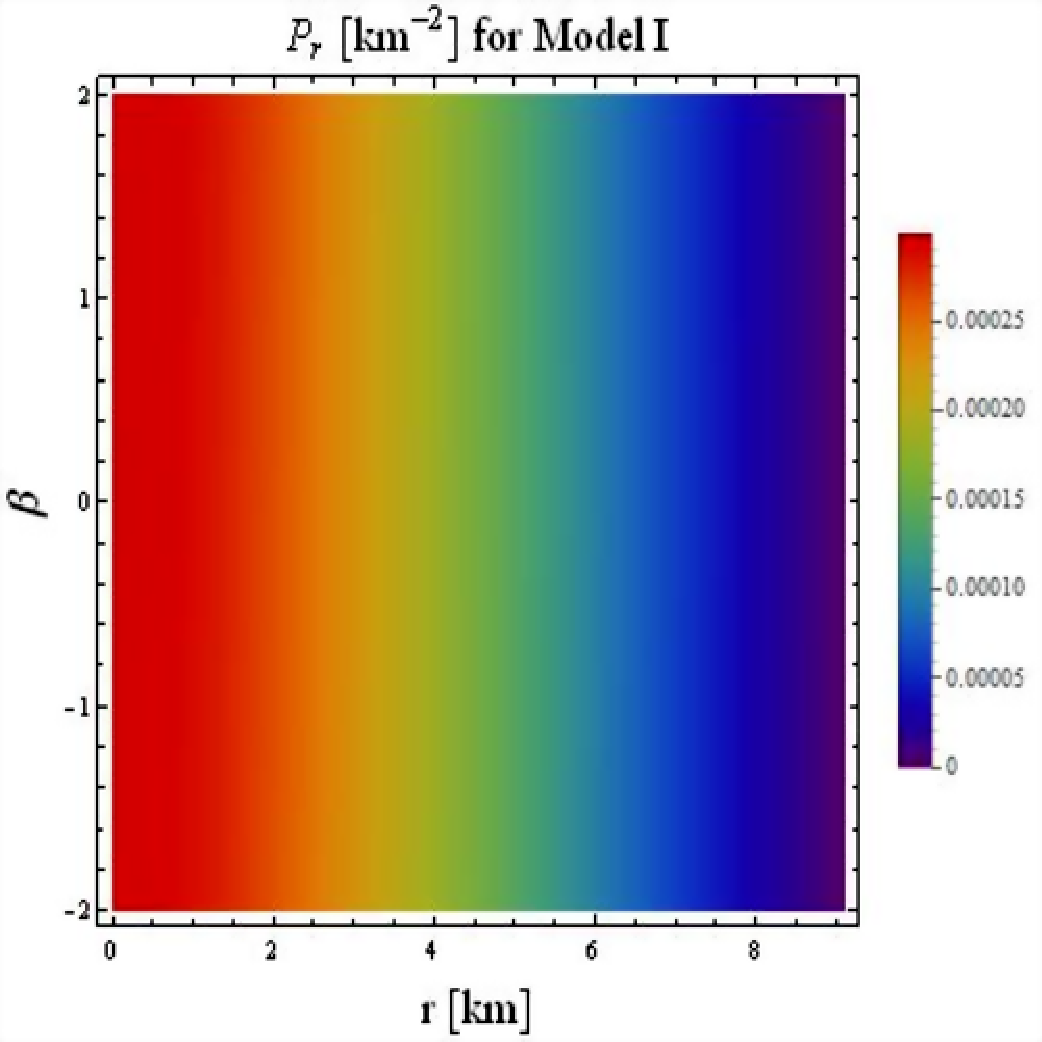,width=.4\linewidth}\epsfig{file=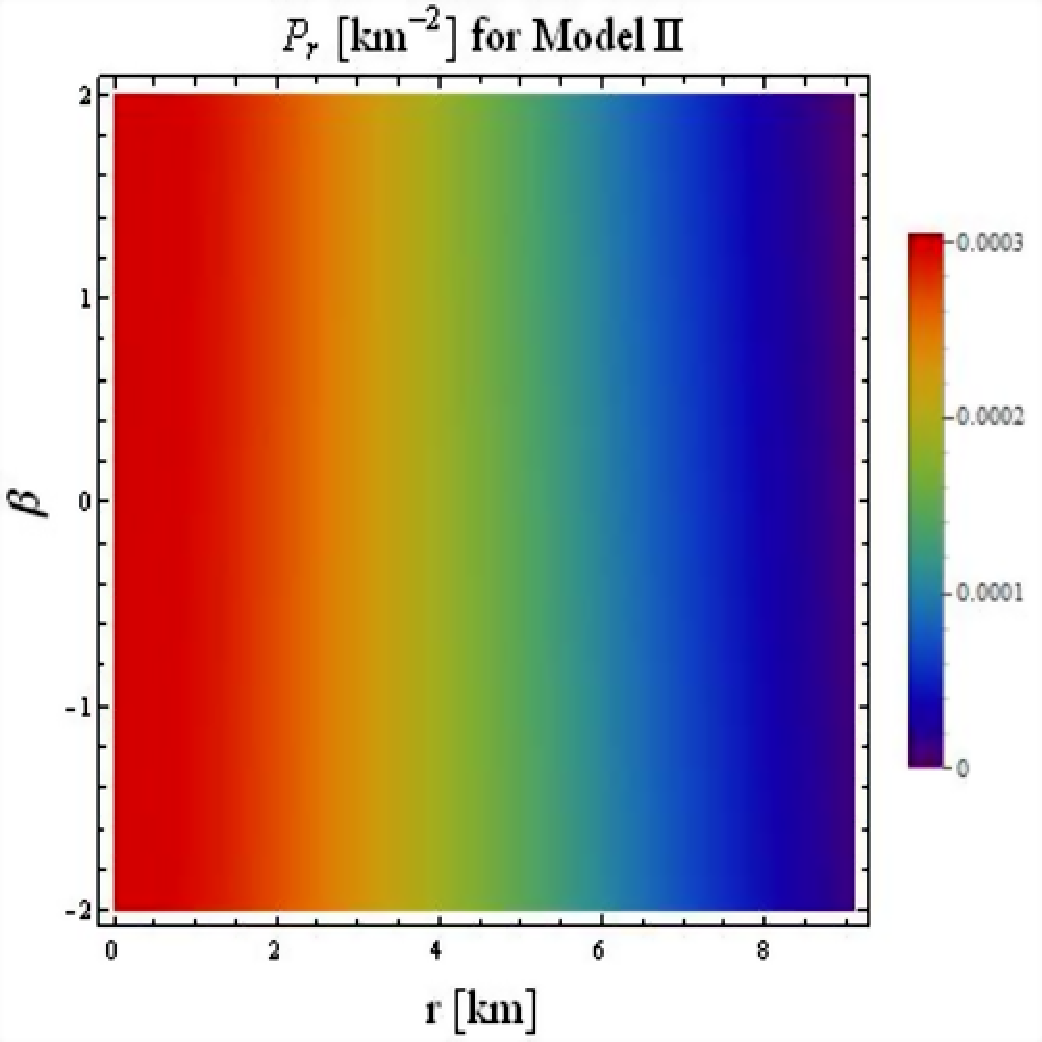,width=.4\linewidth}
\epsfig{file=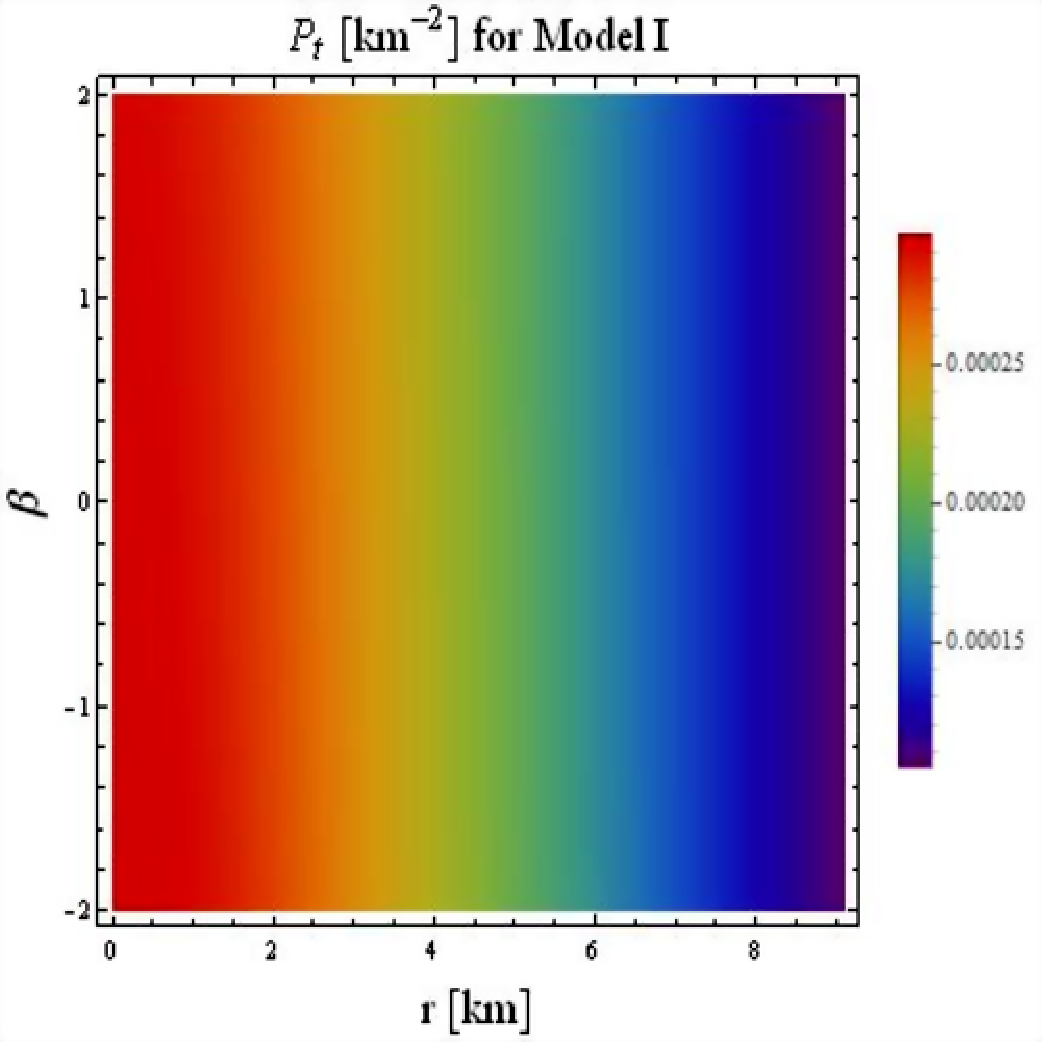,width=.4\linewidth}\epsfig{file=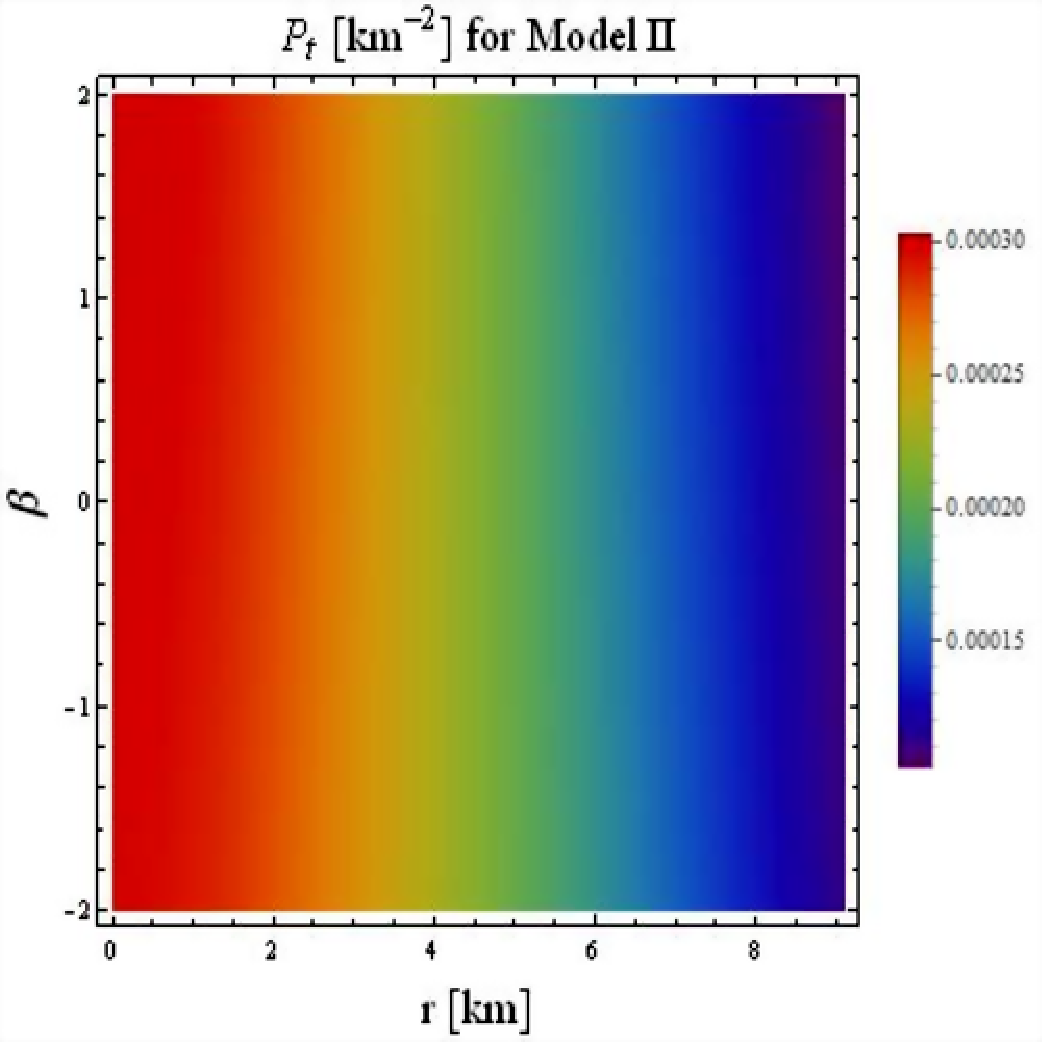,width=.4\linewidth}
\caption{Matter variables versus $\beta$ and $r$.}
\end{figure}
\begin{figure}[h!]\center
\epsfig{file=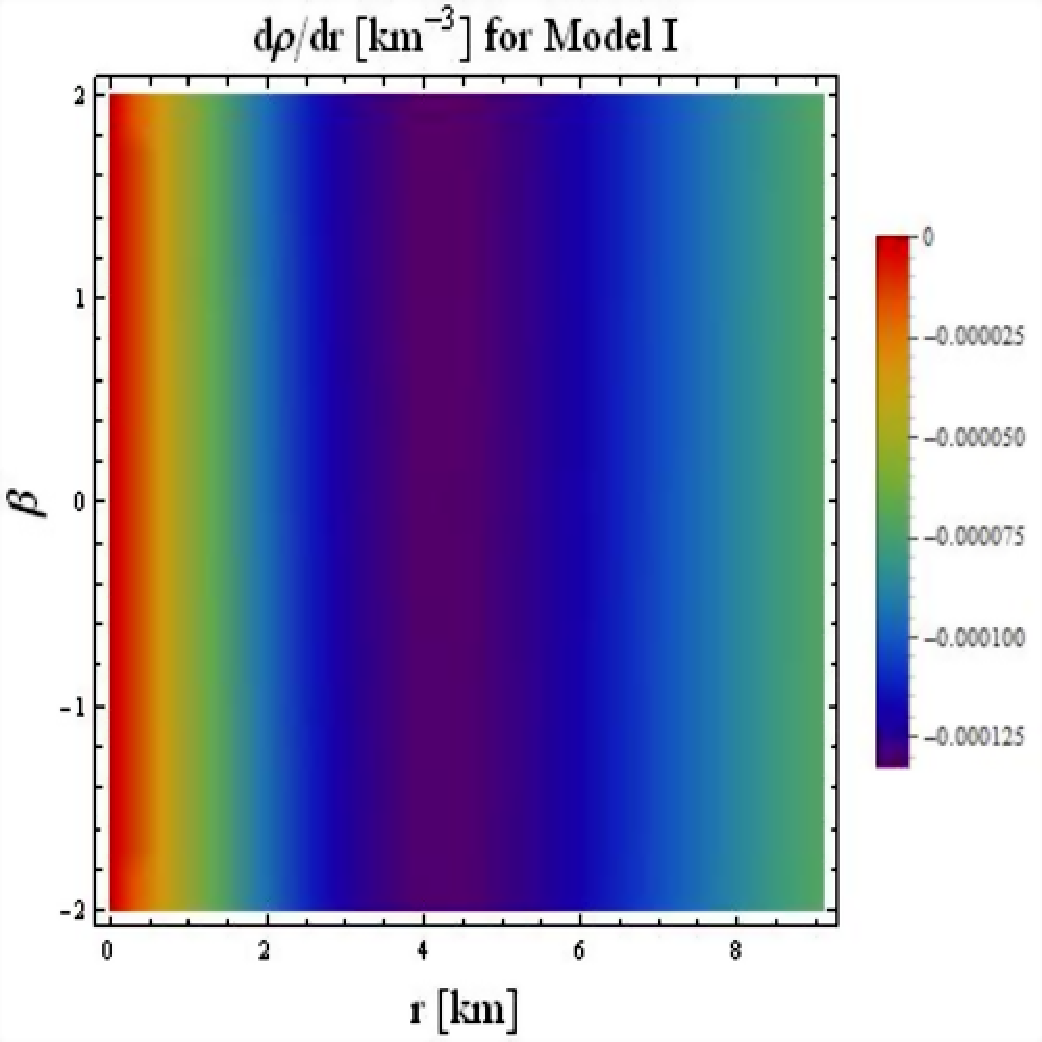,width=.4\linewidth}\epsfig{file=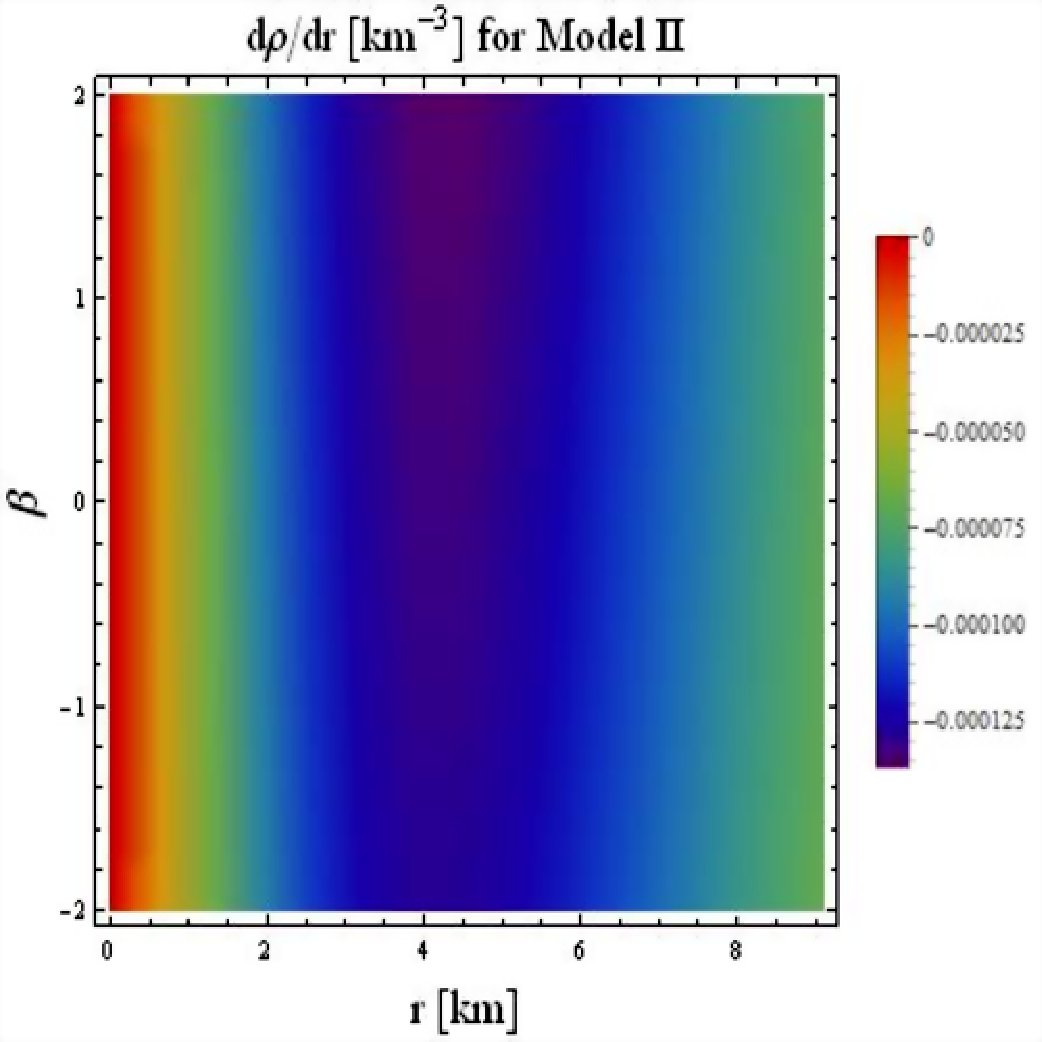,width=.4\linewidth}
\epsfig{file=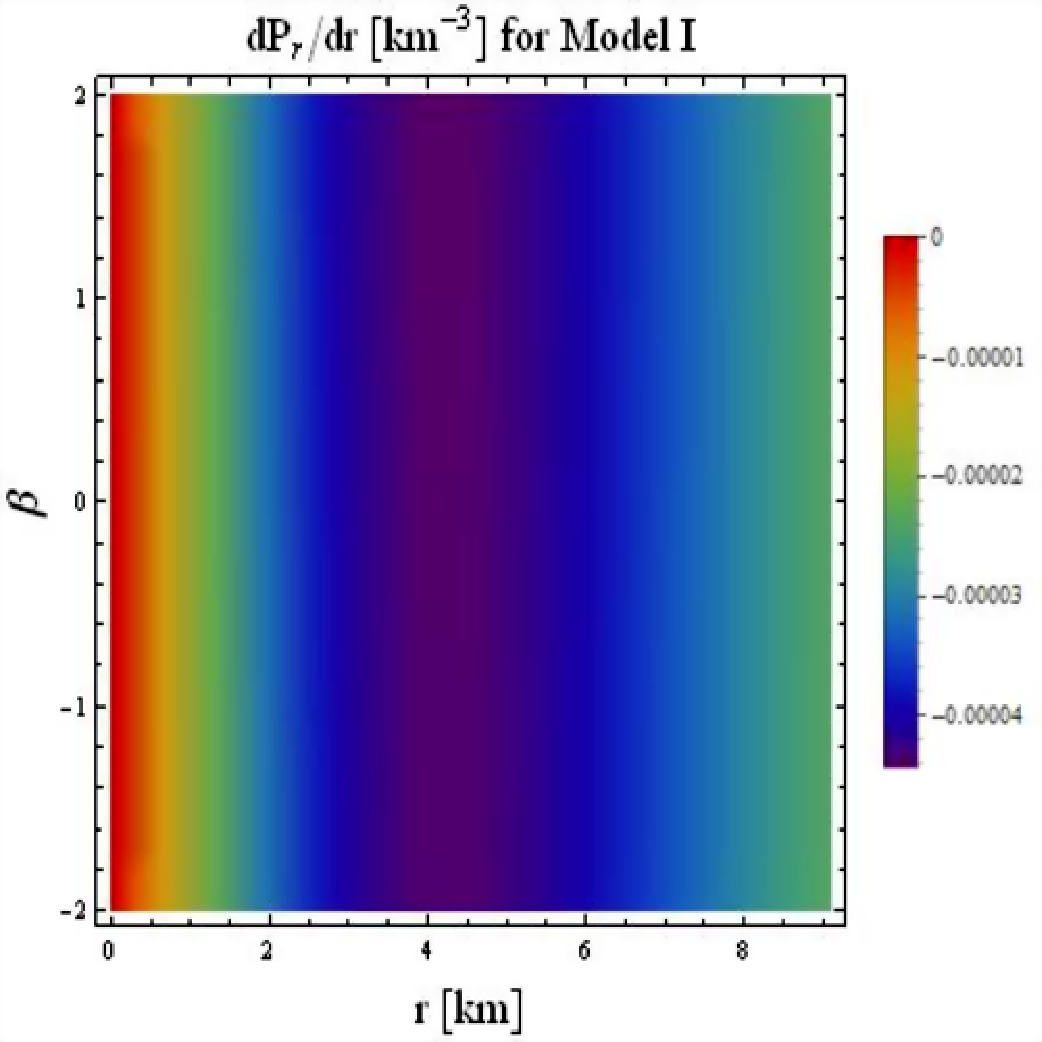,width=.4\linewidth}\epsfig{file=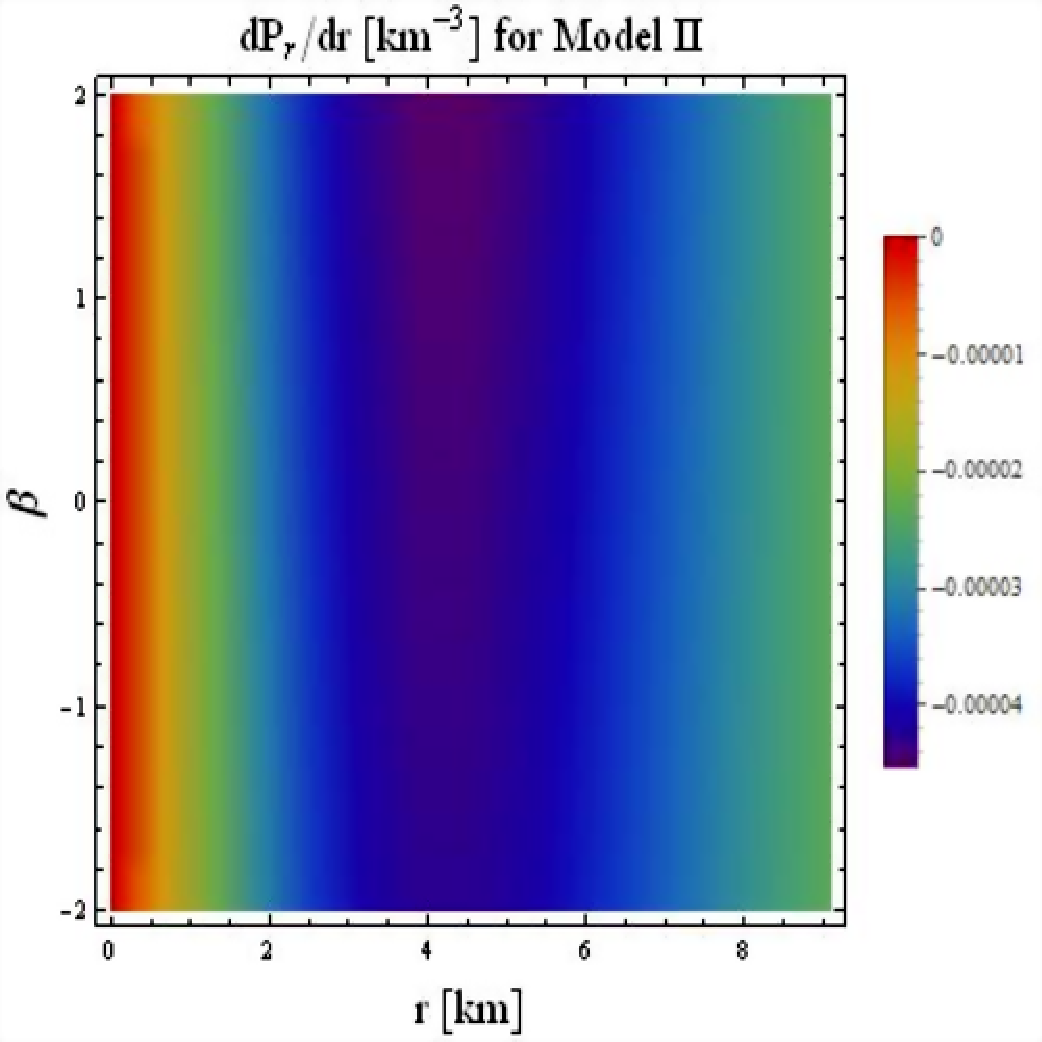,width=.4\linewidth}
\epsfig{file=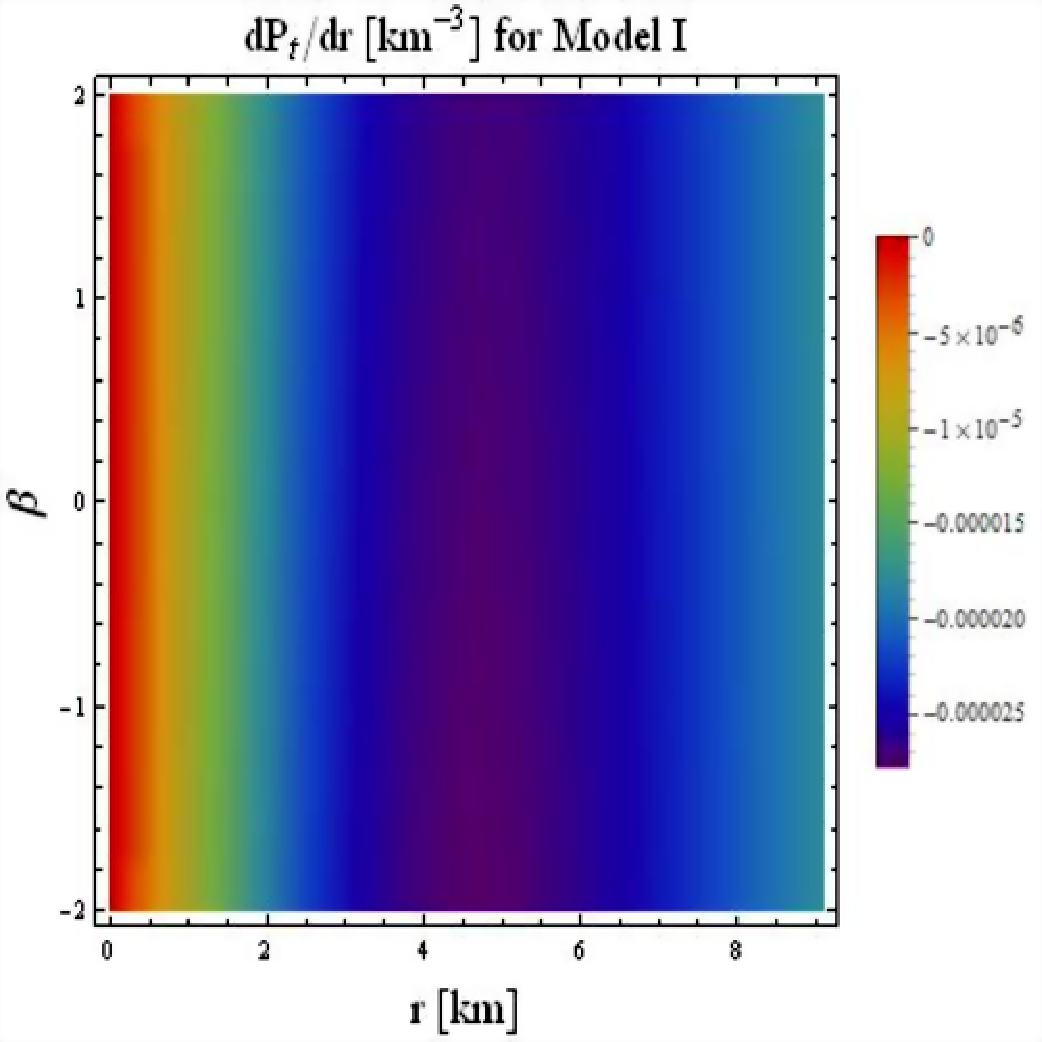,width=.4\linewidth}\epsfig{file=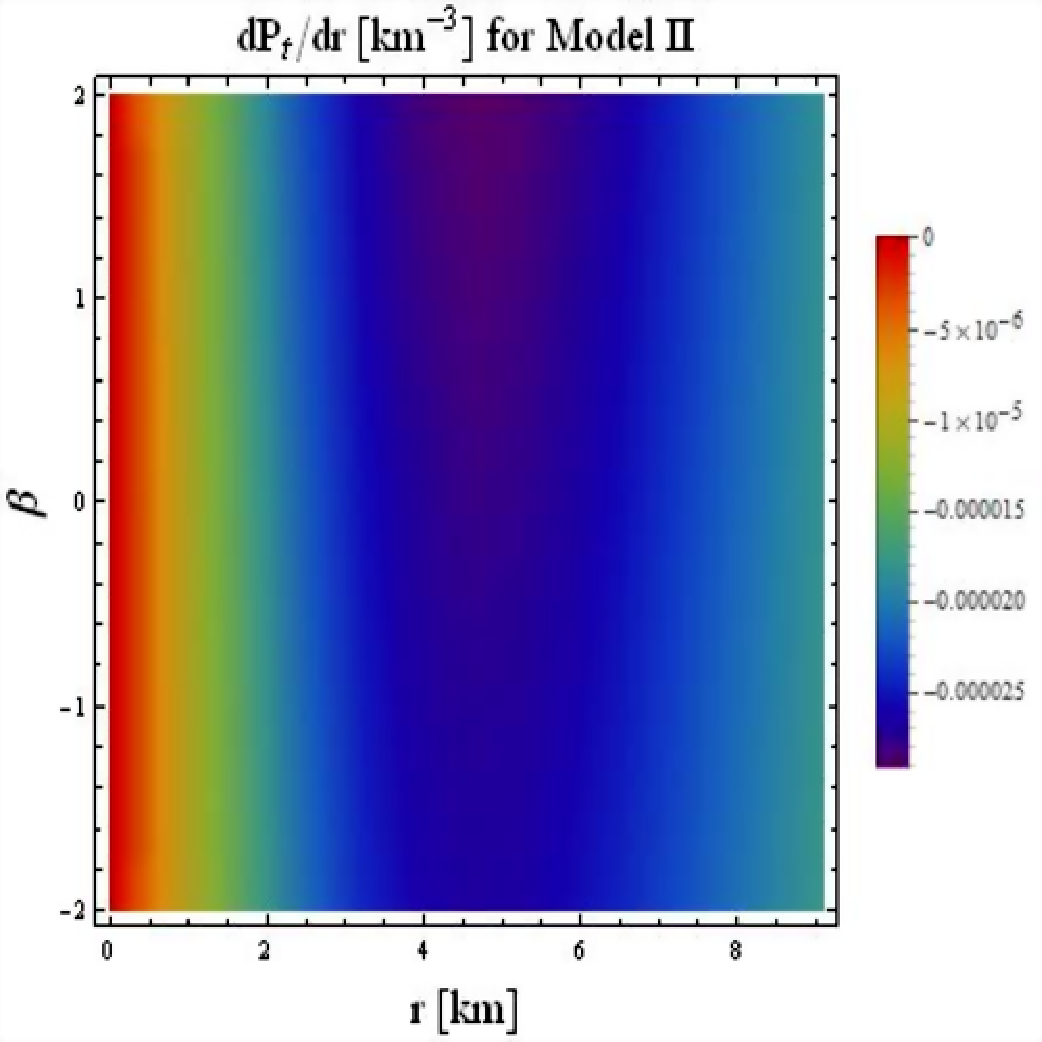,width=.4\linewidth}
\caption{First derivative of matter variables versus $\beta$ and
$r$.}
\end{figure}
\begin{figure}[h!]\center
\epsfig{file=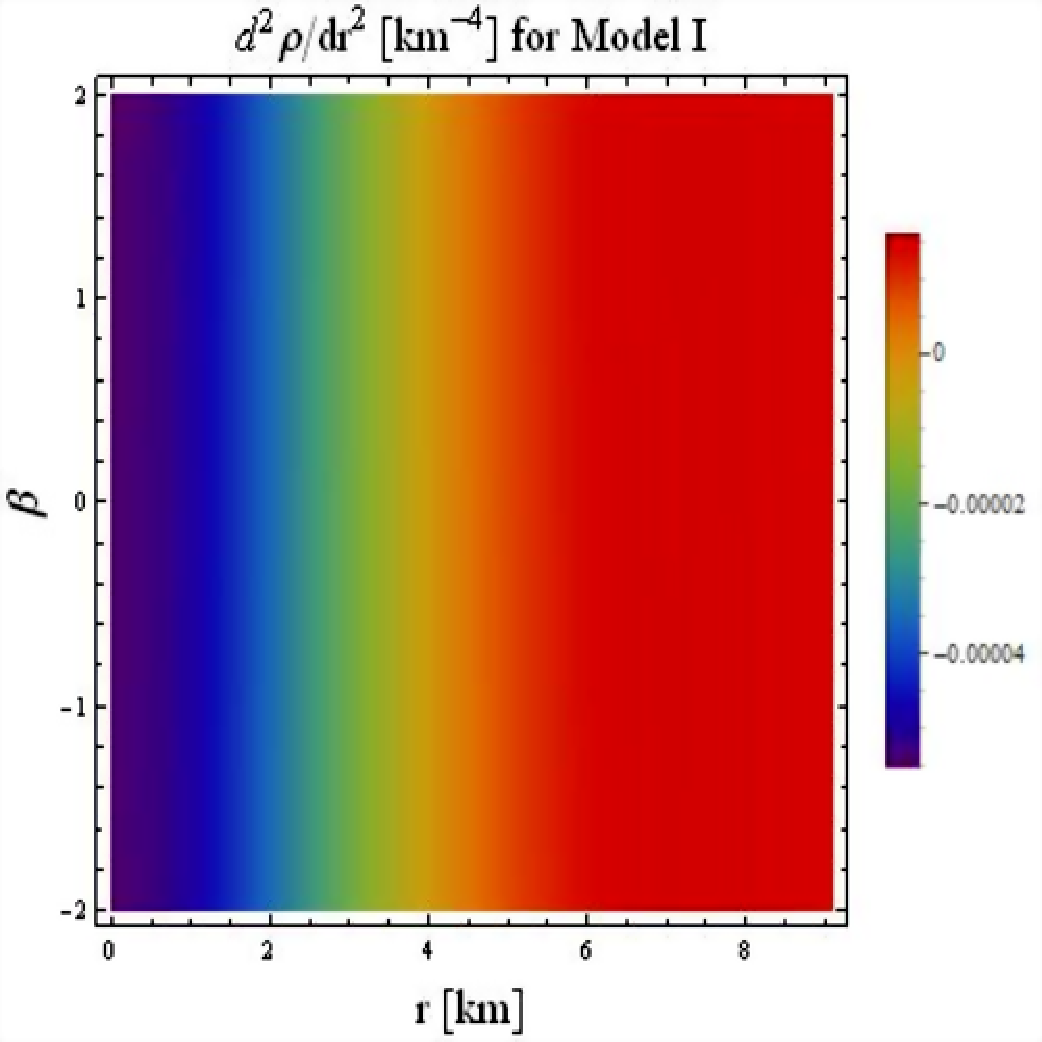,width=.4\linewidth}\epsfig{file=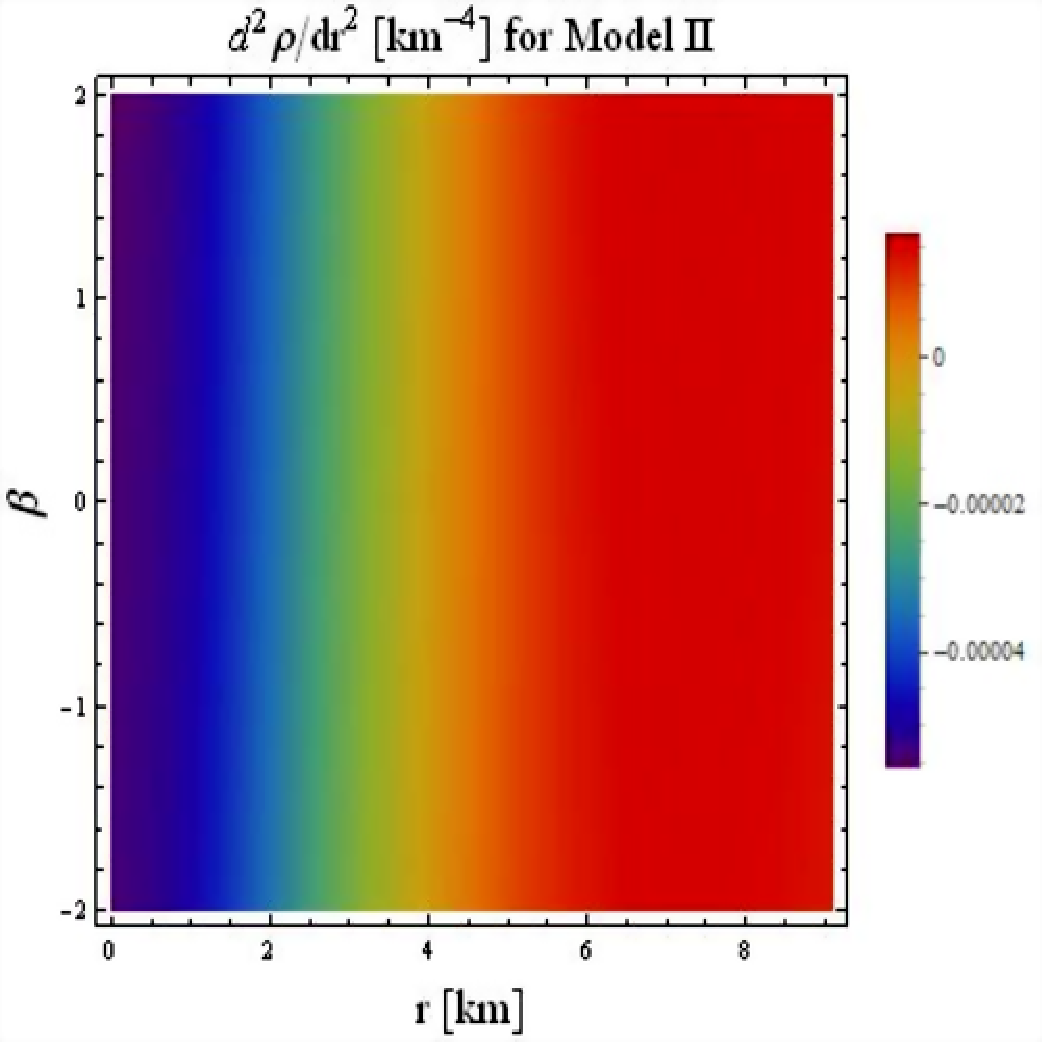,width=.4\linewidth}
\epsfig{file=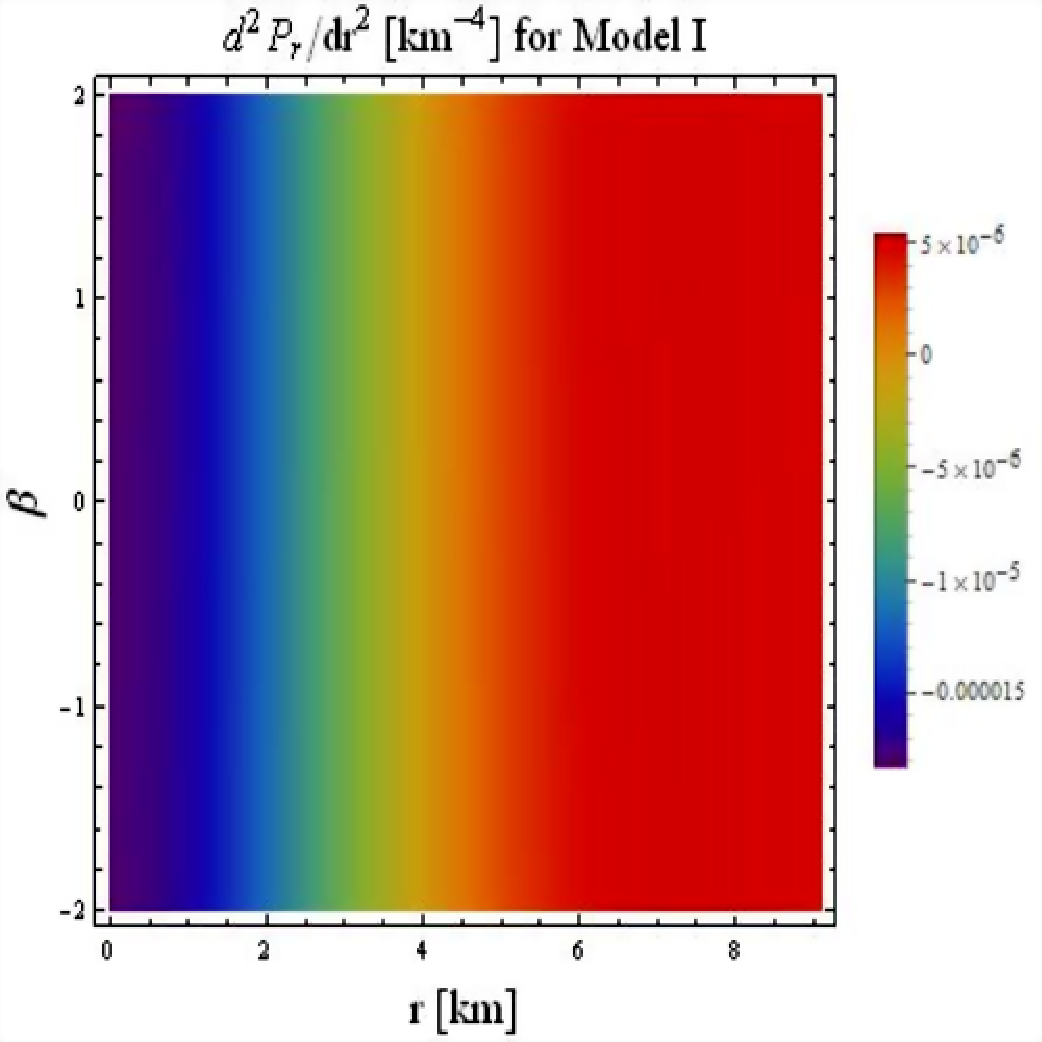,width=.4\linewidth}\epsfig{file=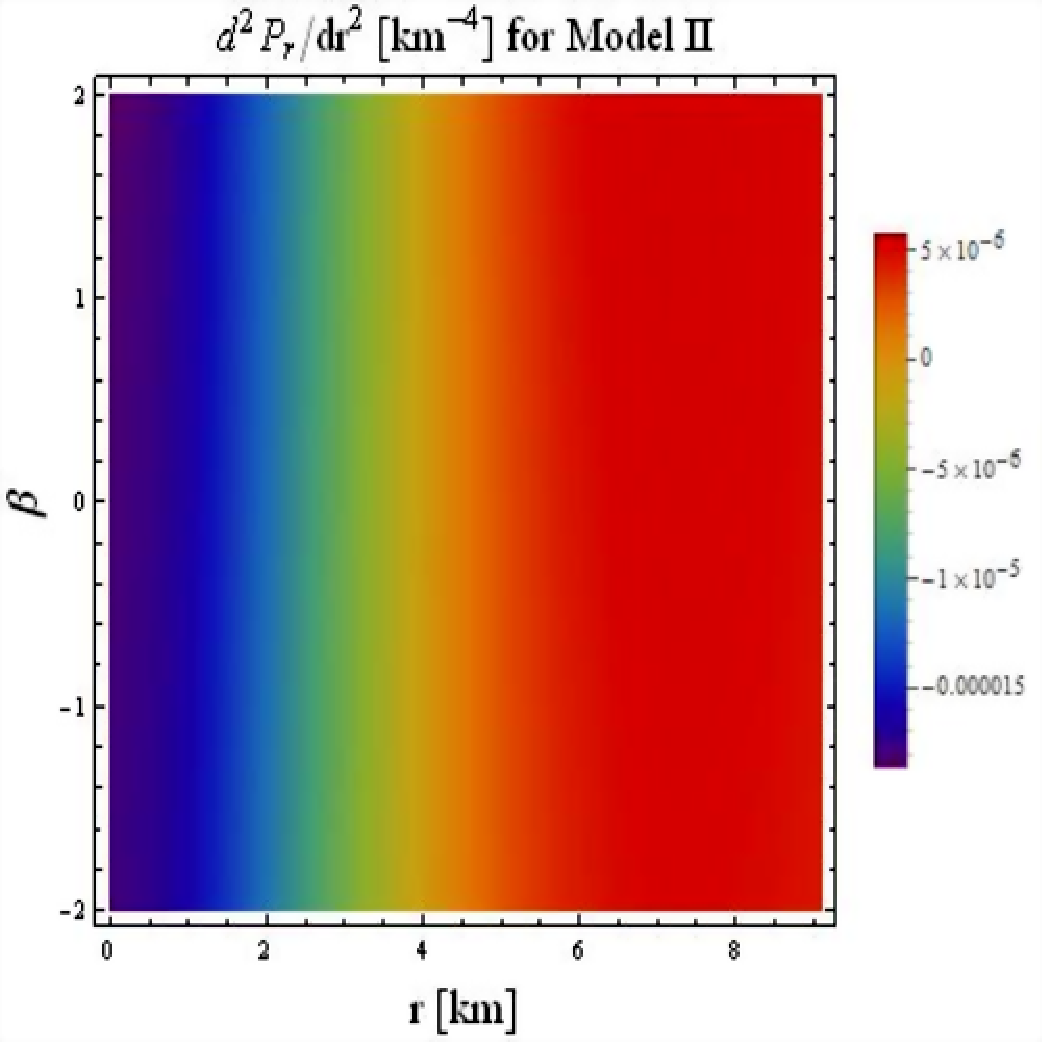,width=.4\linewidth}
\epsfig{file=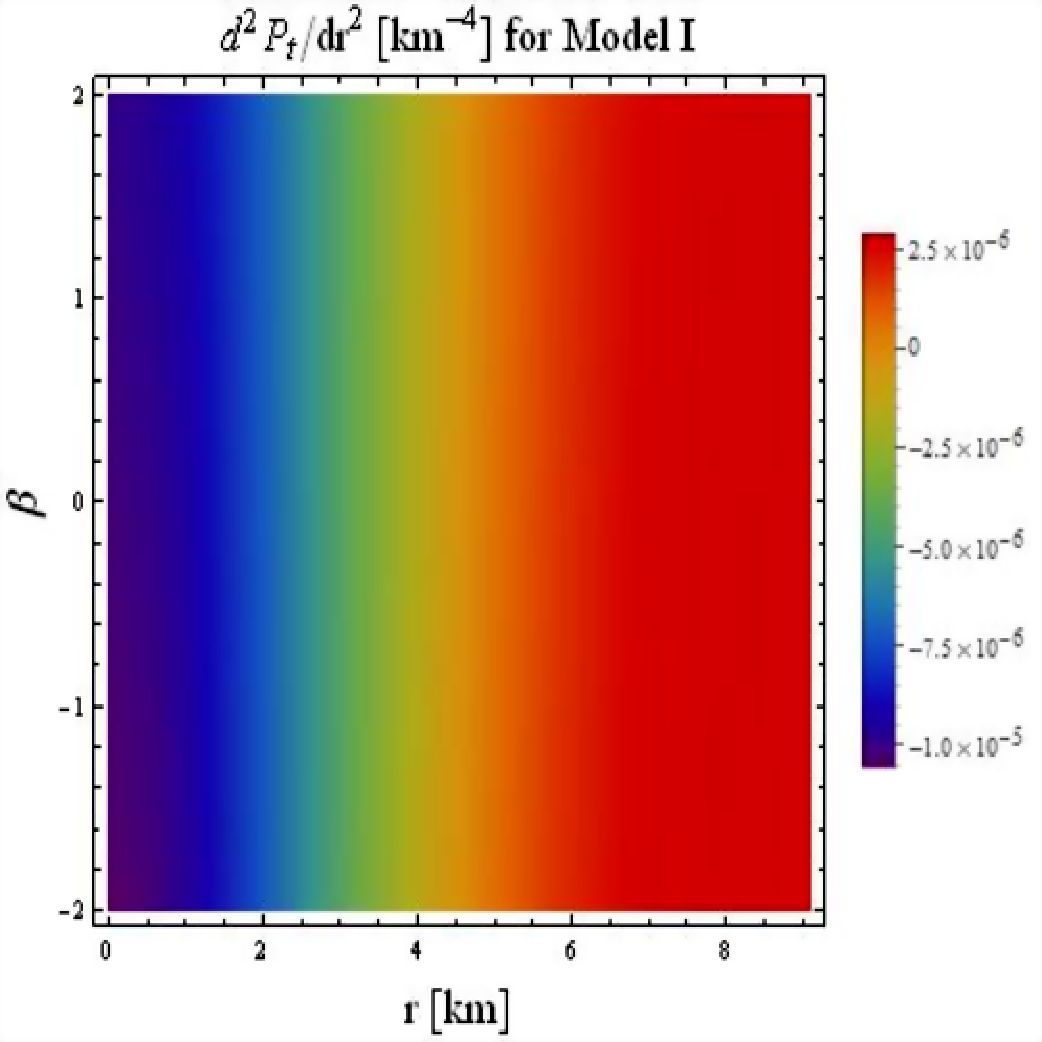,width=.4\linewidth}\epsfig{file=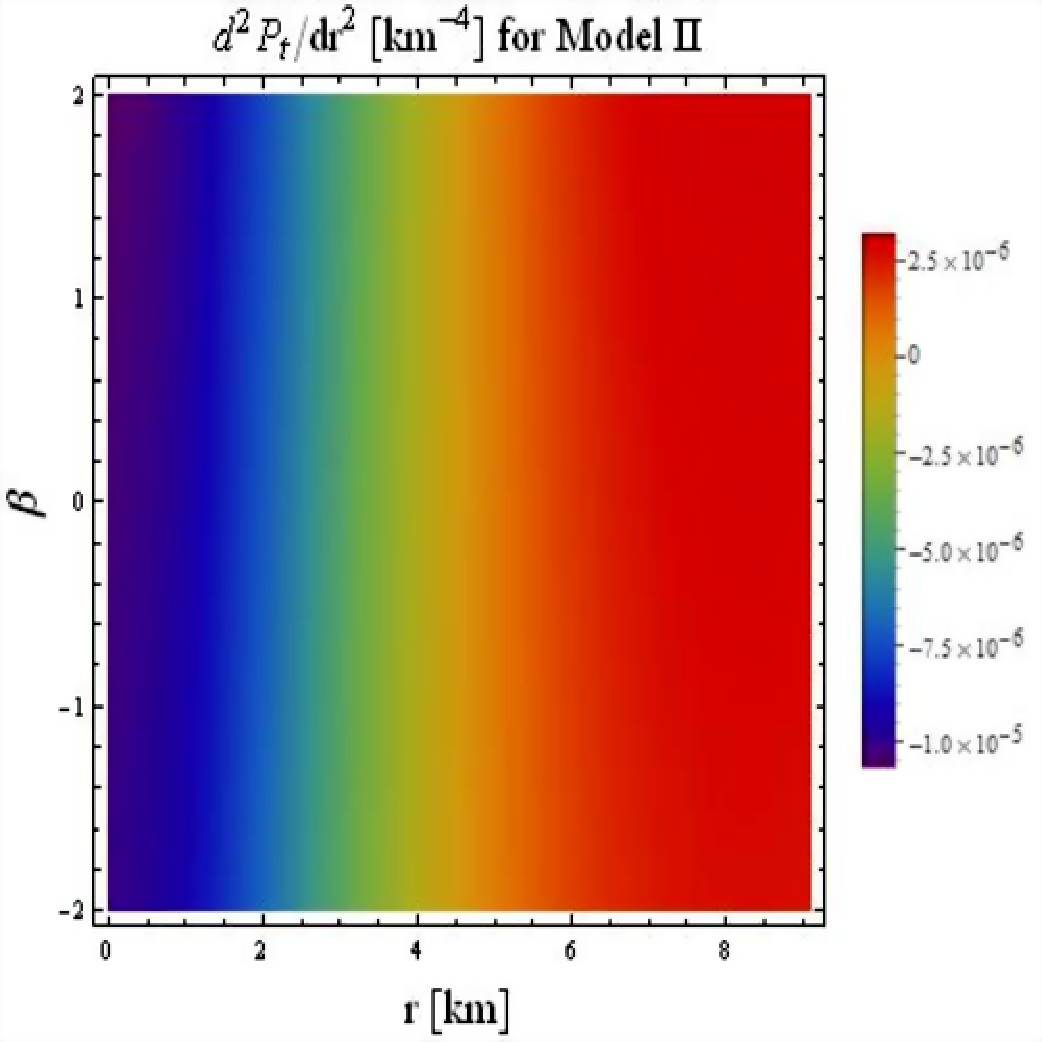,width=.4\linewidth}
\caption{Second derivative of matter variables versus $\beta$ and
$r$.}
\end{figure}

\subsection{Anisotropic Pressure}

The anisotropy is defined by $\Pi=P_t-P_r$. This subsection explores
the influence of this factor on the structural evolution of the
compact star under investigation through a graphical analysis of
this phenomenon. The characteristics of anisotropy can be elucidated
as follows.
\begin{itemize}
\item Anisotropy manifests as an outward expansion when the radial
pressure is lower than the tangential pressure.
\item Conversely, it manifests as an inward contraction when the tangential pressure
falls below the radial pressure.
\end{itemize}
The anisotropy corresponding to both models I and II is presented in
Appendix \textbf{B}. We observe this to be zero at the center from
Figure \textbf{5} and increasing outwards, exhibiting maximum value
at the boundary. It is also found that the compact structure
analogous to the model I possesses a little more anisotropy in
comparison with the other model.
\begin{figure}\center
\epsfig{file=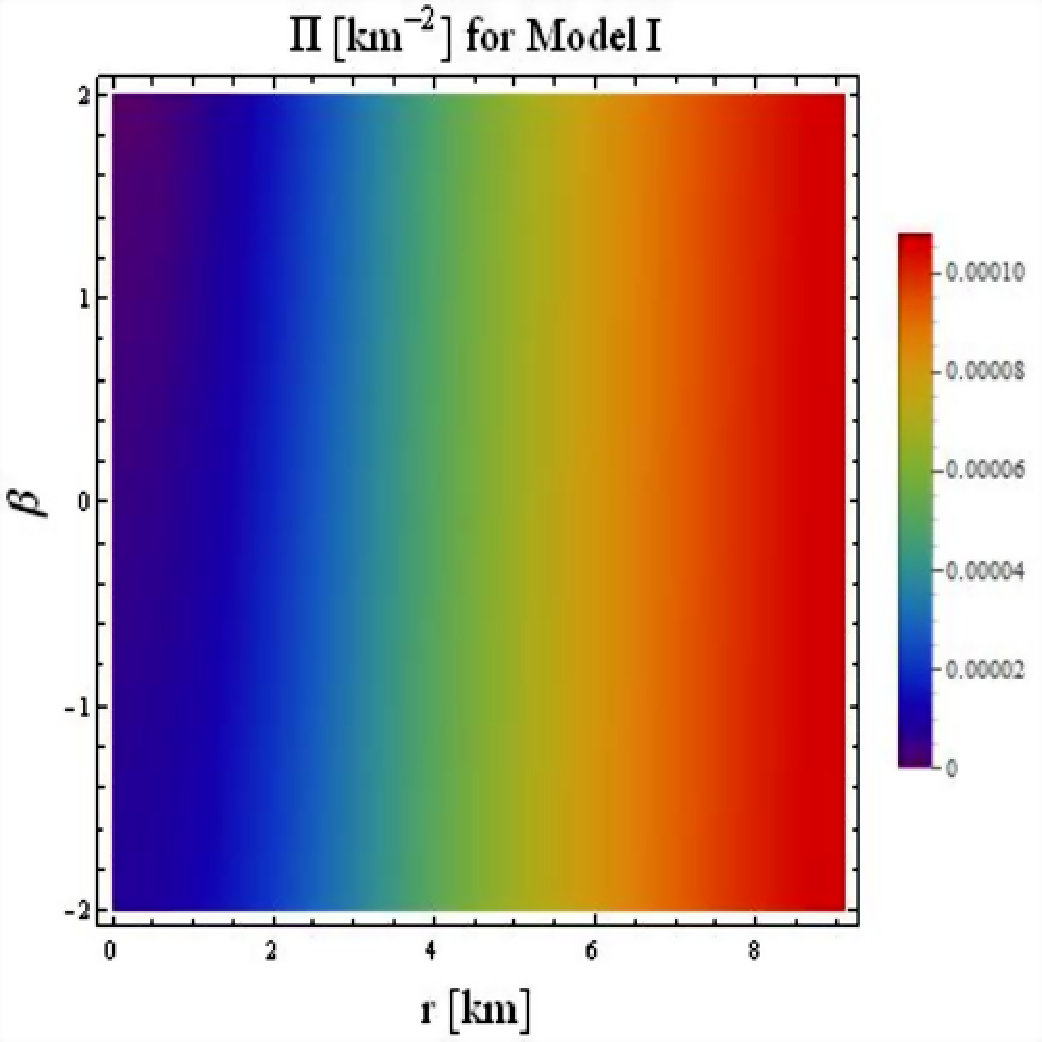,width=.4\linewidth}\epsfig{file=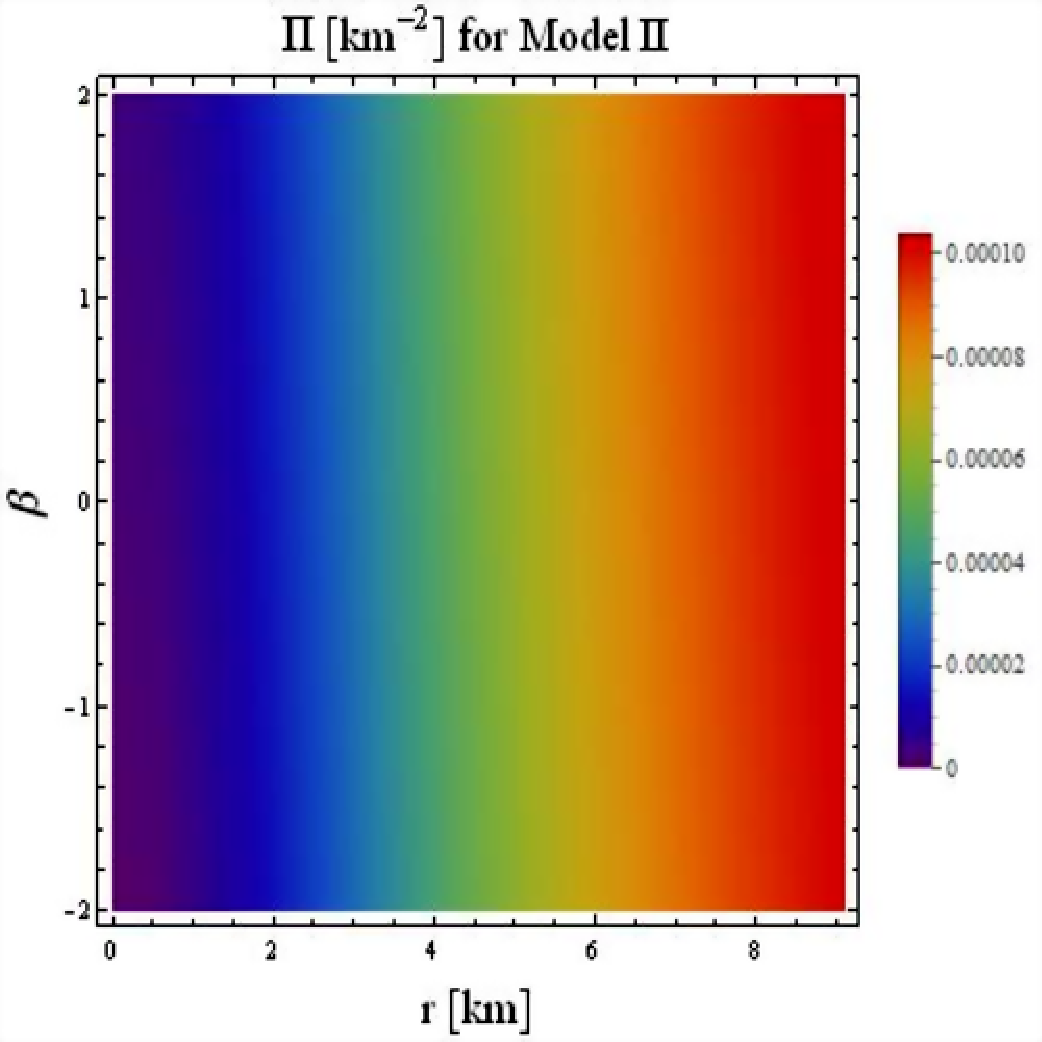,width=.4\linewidth}
\caption{Anisotropy versus $\beta$ and $r$.}
\end{figure}

\subsection{Mass, Compactness and Redshift Parameters}

We firstly present the mass function of a spherical structure in
terms of the geometric quantity \eqref{g12a} that remains same in GR
and modified theories. However, the following equation expresses the
spherical mass in terms of the energy density possessing modified
correction terms, thus, the impact of modified gravity can be
checked. This relation is defined by
\begin{equation}\label{g32}
m(r)=\frac{1}{2}\int_{0}^{\emph{R}}r^2\rho dr,
\end{equation}
where $\rho$ being the effective energy density (possessing the
correction terms). We already presented it in Eqs.\eqref{g14b} and
\eqref{g14e} analogous to both considered models. Figure
$\mathbf{6}$ (first row) implies that the mass function disappears
at $r=0$ and increases towards the spherical boundary for both
models. Notably, our model I generates a slightly massive structure
as compared to that produced by model II.
\begin{figure}[h!]\center
\epsfig{file=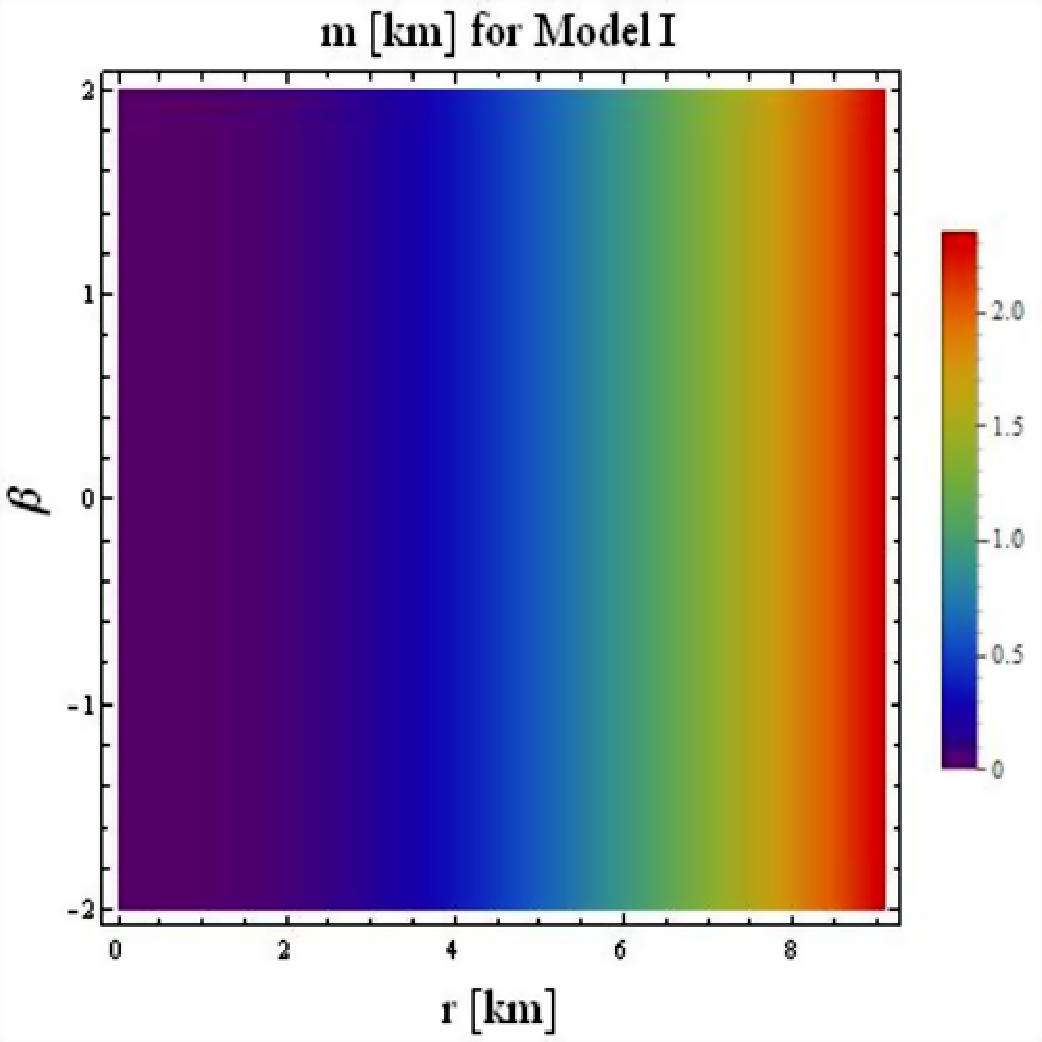,width=.4\linewidth}\epsfig{file=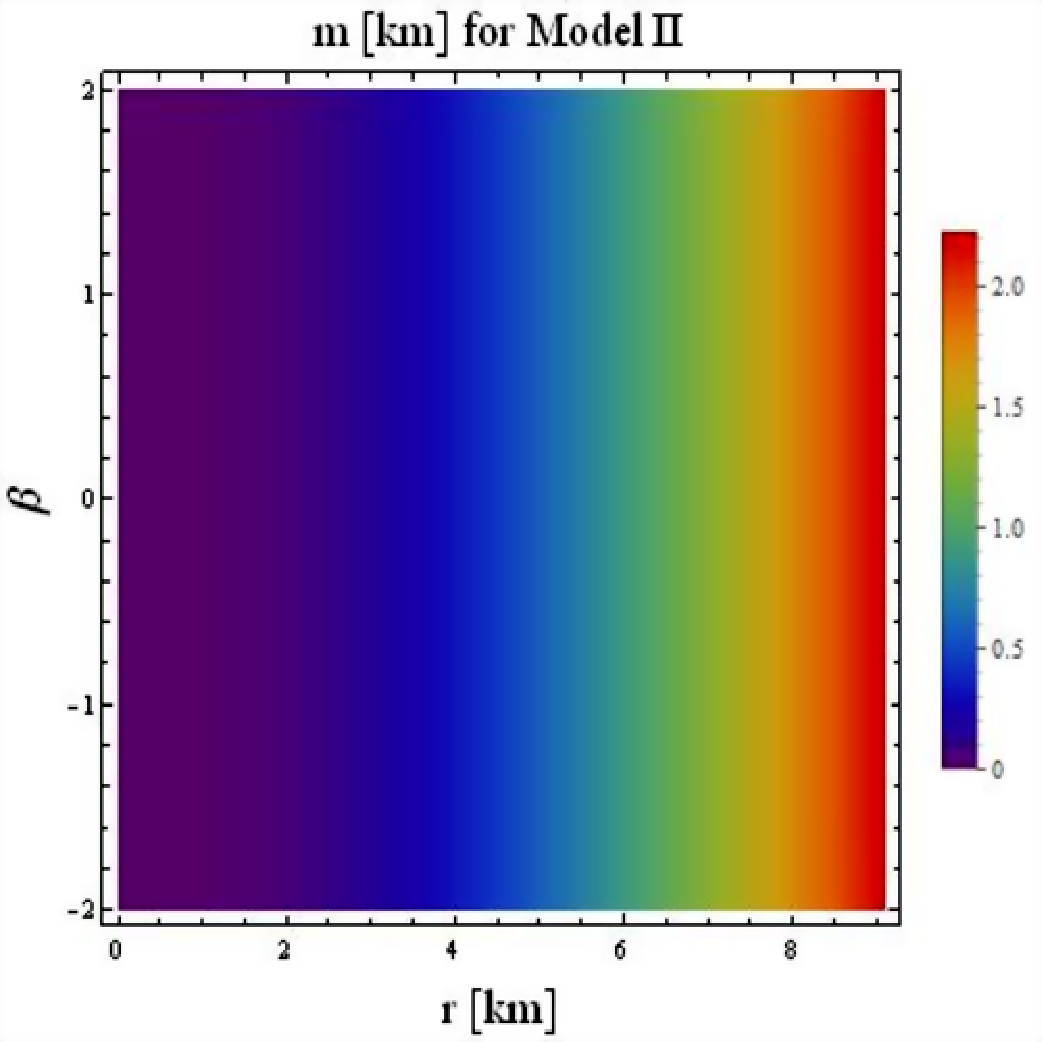,width=.4\linewidth}
\epsfig{file=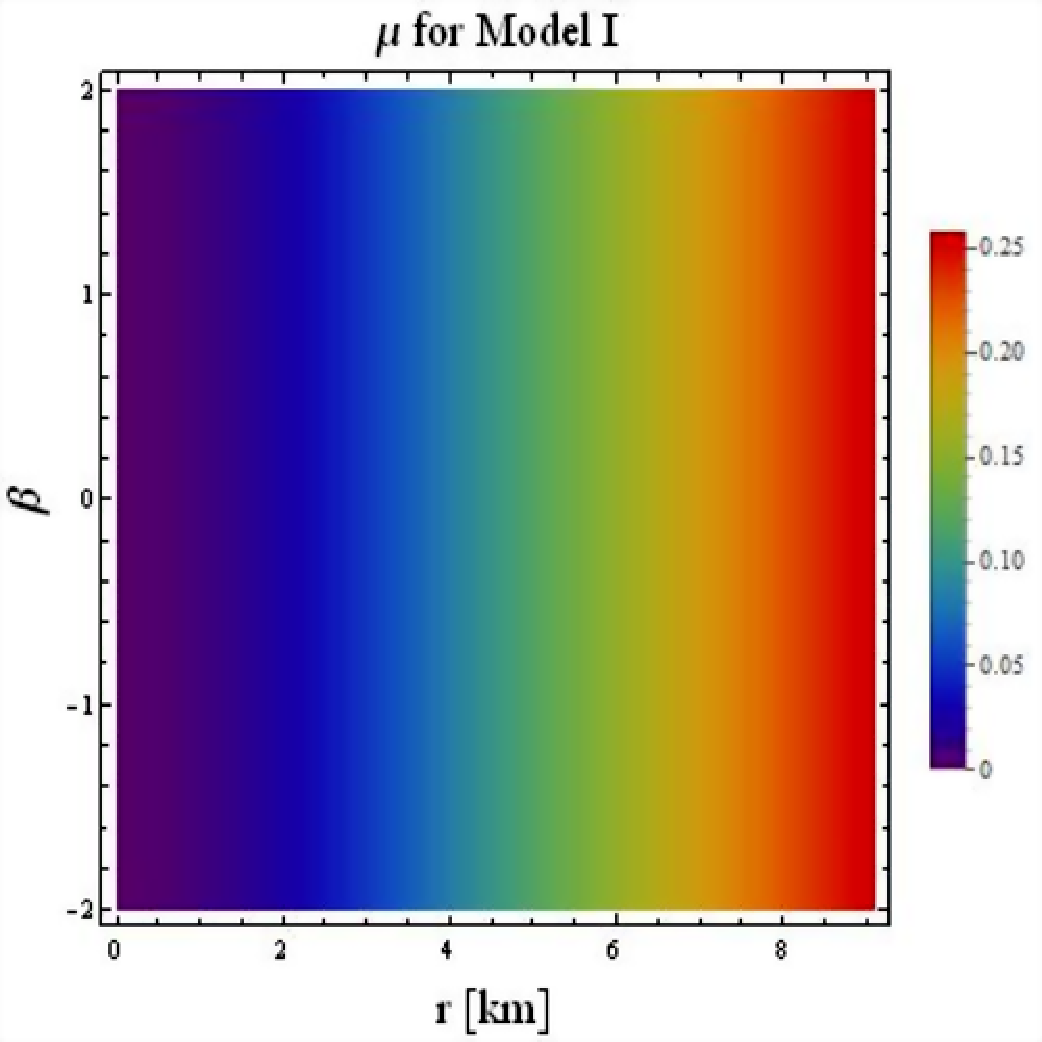,width=.4\linewidth}\epsfig{file=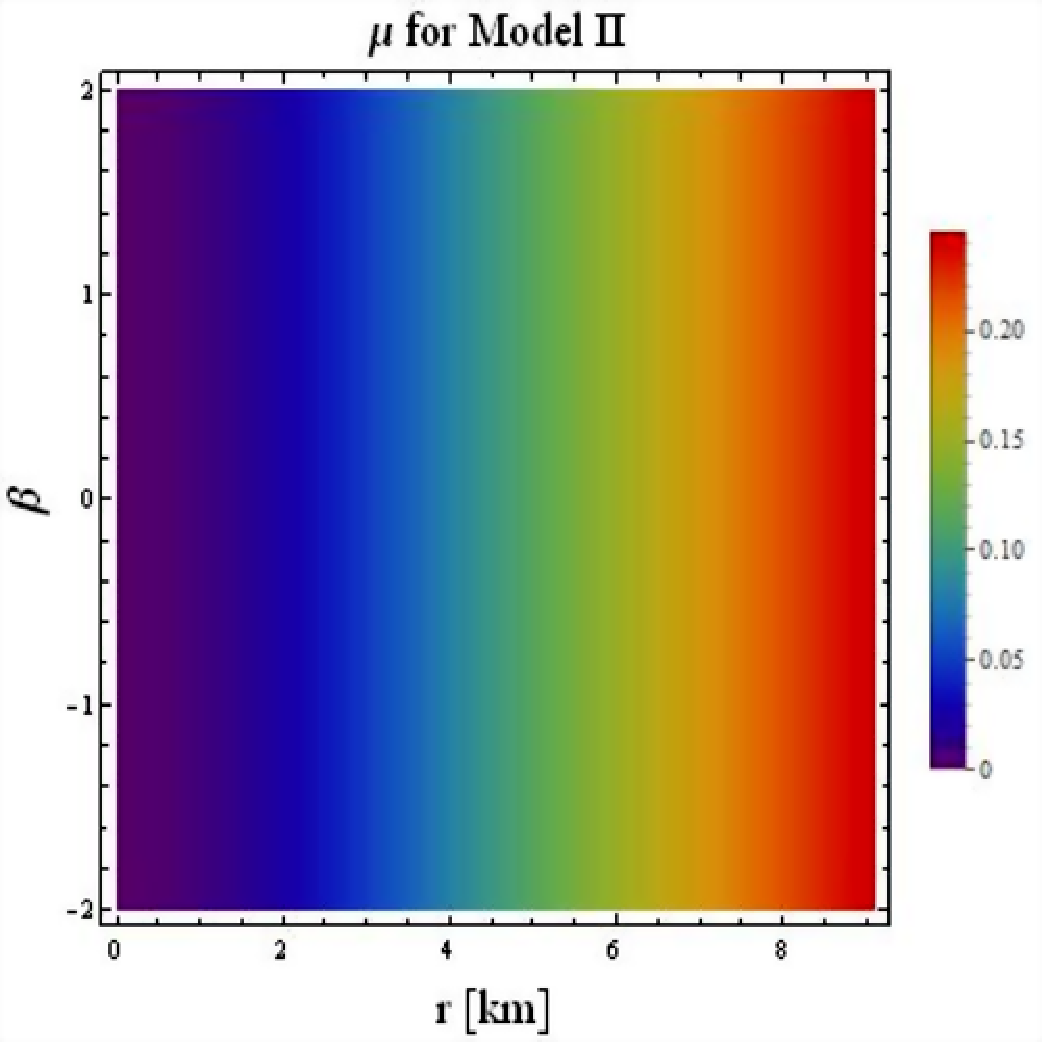,width=.4\linewidth}
\epsfig{file=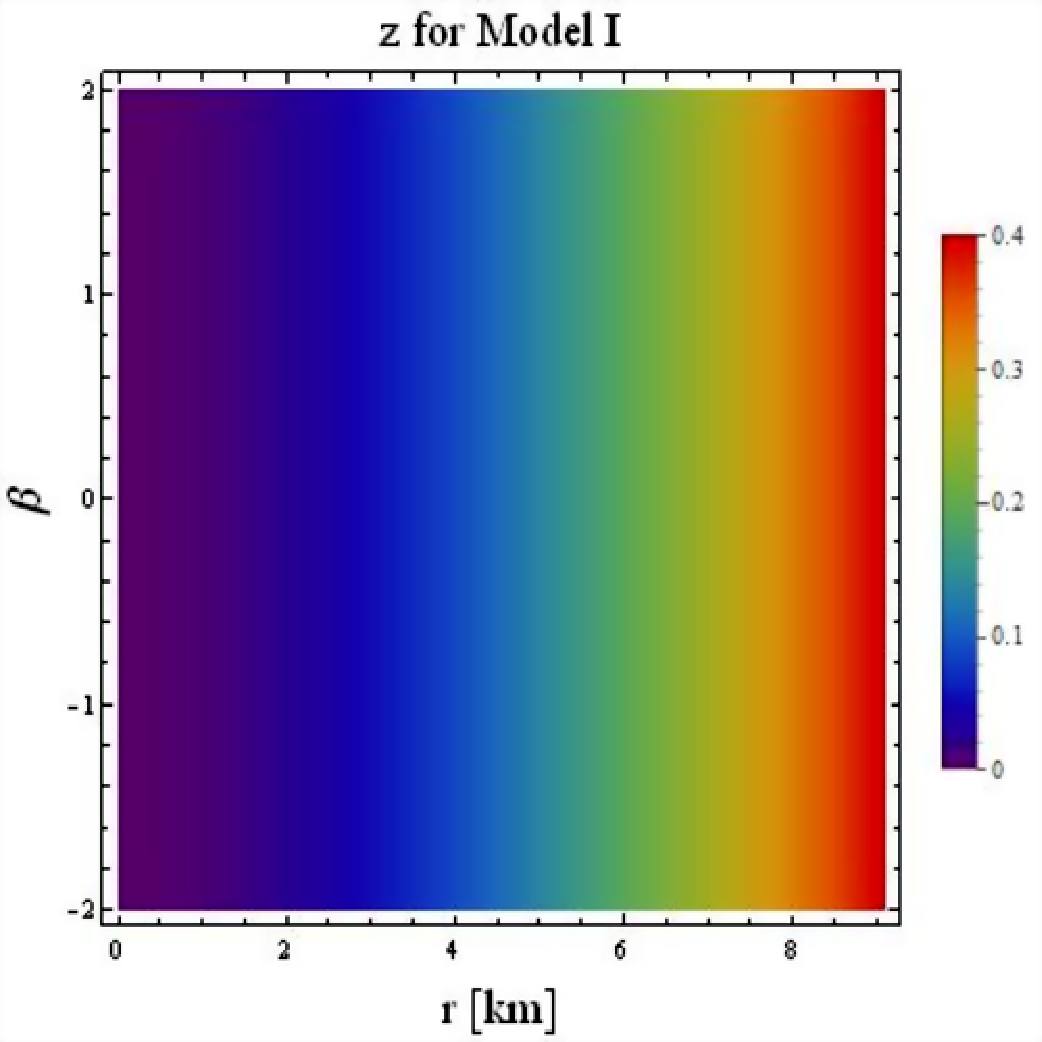,width=.4\linewidth}\epsfig{file=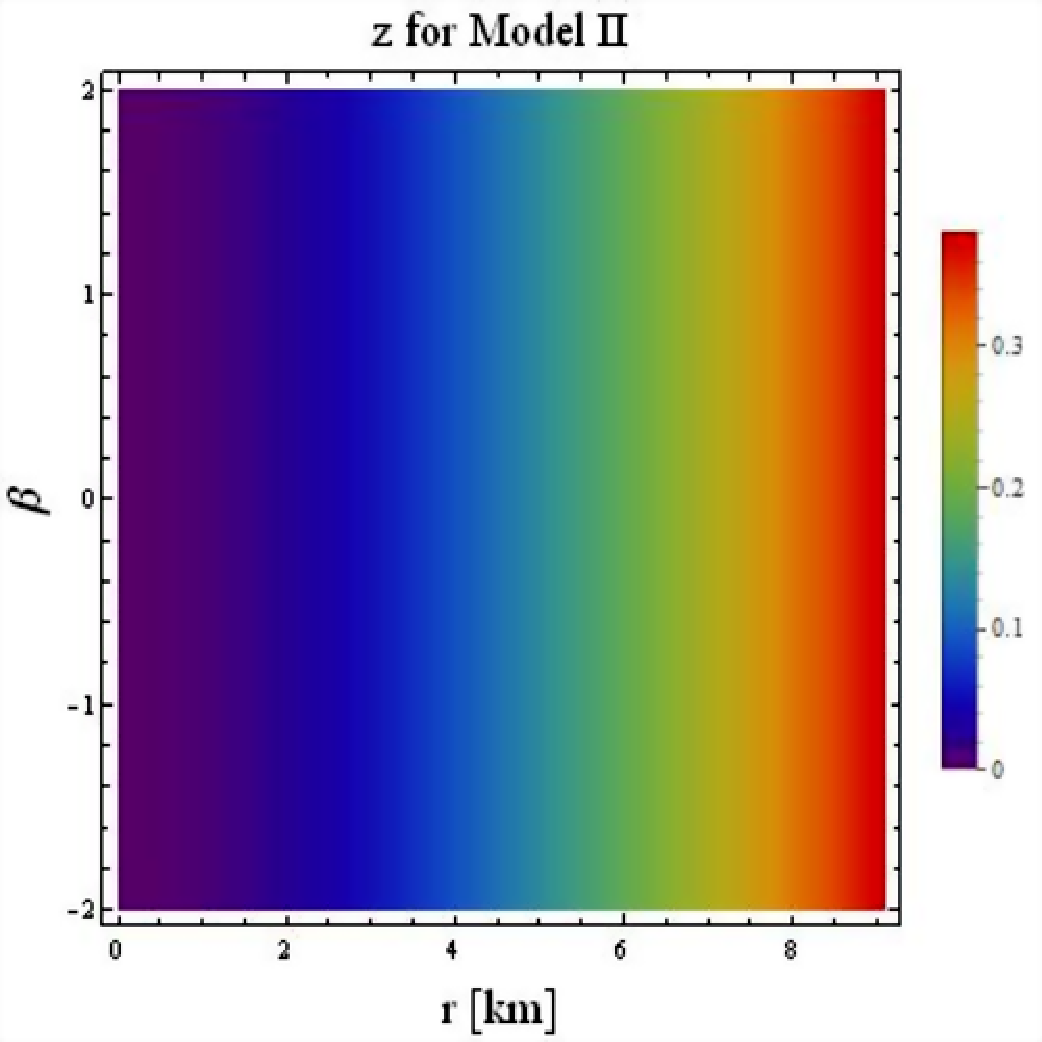,width=.4\linewidth}
\caption{Physical factors versus $\beta$ and $r$.}
\end{figure}

Numerous physical parameters, including compactness and surface
redshift, have been extensively examined in scientific literature to
gain insight into the evolution and properties of compact stellar
objects. Compactness, for instance, is defined as the ratio of a
star's mass to its radius. Buchdahl \cite{42a}, in his seminal work,
proposed a maximum value for compactness (represented by $\mu$)
within the framework of a celestial system, setting it at
$\frac{4}{9}$ through meticulous analysis that satisfies matching
criteria at the hypersurface ($r=\emph{R}$).

A massive celestial body, ensconced within a strong gravitational
field, emits electromagnetic radiations due to various nuclear
reactions occurring within its core. As these radiations traverse
space, their wavelengths can stretch, causing a phenomenon known as
redshift. The mathematical formulation of this parameter is
\begin{equation}\label{g35}
z(r)=\frac{1}{\sqrt{1-2\mu(r)}}-1.
\end{equation}
while, theoretical models with uniformly distributed matter within
such bodies have an upper limit of 2. A noteworthy development
emerged in the work of Ivanov \cite{42b} who extended this limit to
5.211 for the anisotropic distribution. The second and last rows of
Figure $\mathbf{6}$ confirm both above factors within their
suggested bounds.

\subsection{Equation of State Parameters}

Another essential aspect in assessing the relevancy of the
self-gravitating models is to explore the EoS parameters. These
equations establish the connection between the density and pressure
of the considered fluid distribution. Given that we are working
within the anisotropic framework, these equations take the form as
\begin{equation}\label{g361}
\omega_r=\frac{P_r}{\rho}, \quad \omega_t=\frac{P_t}{\rho}.
\end{equation}
It is crucial that these parameters must fall within the range of
$[0,1]$ for the viability of stellar matter configurations. The
information depicted in Figure \textbf{7} offers empirical
confirmation that both models I and II adhere to the prescribed
range for these parameters.
\begin{figure}[h!]\center
\epsfig{file=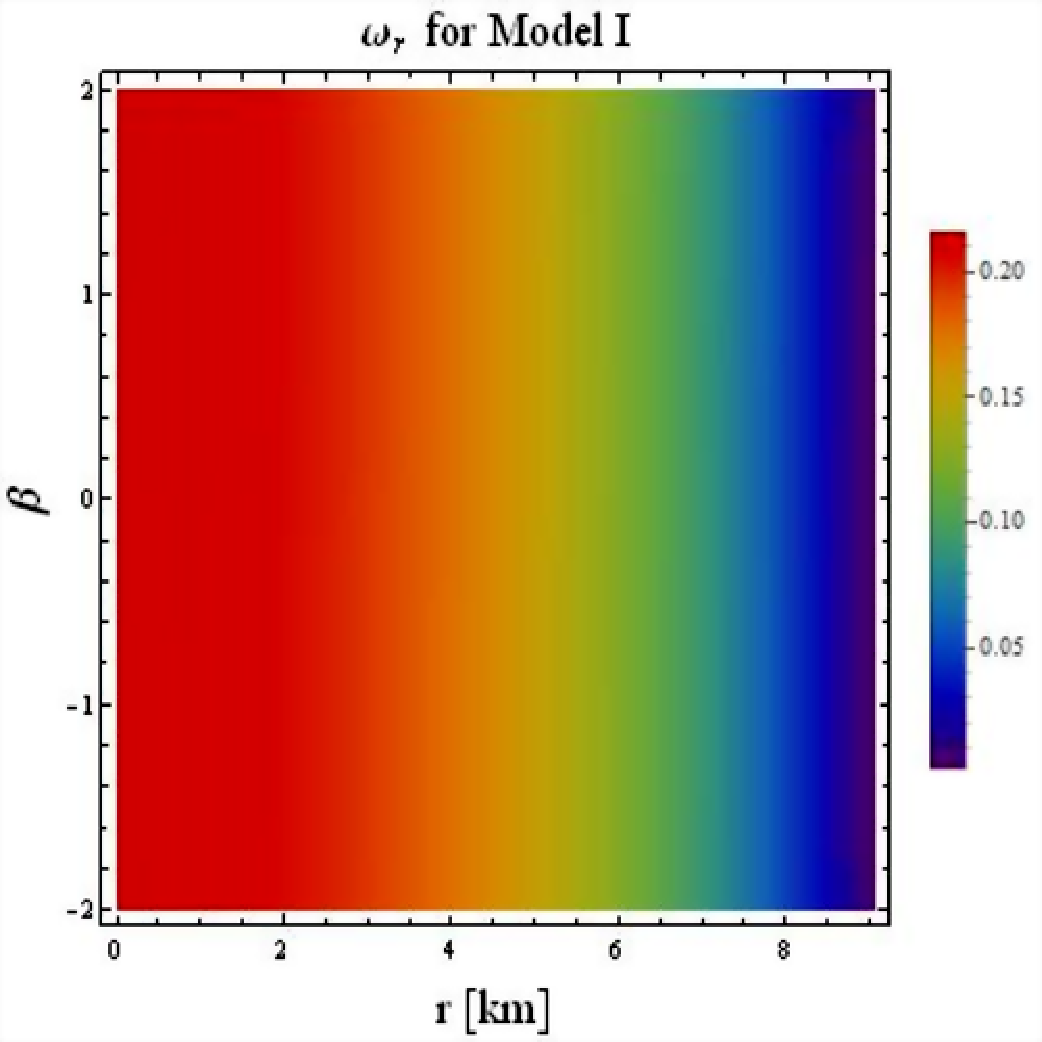,width=.4\linewidth}\epsfig{file=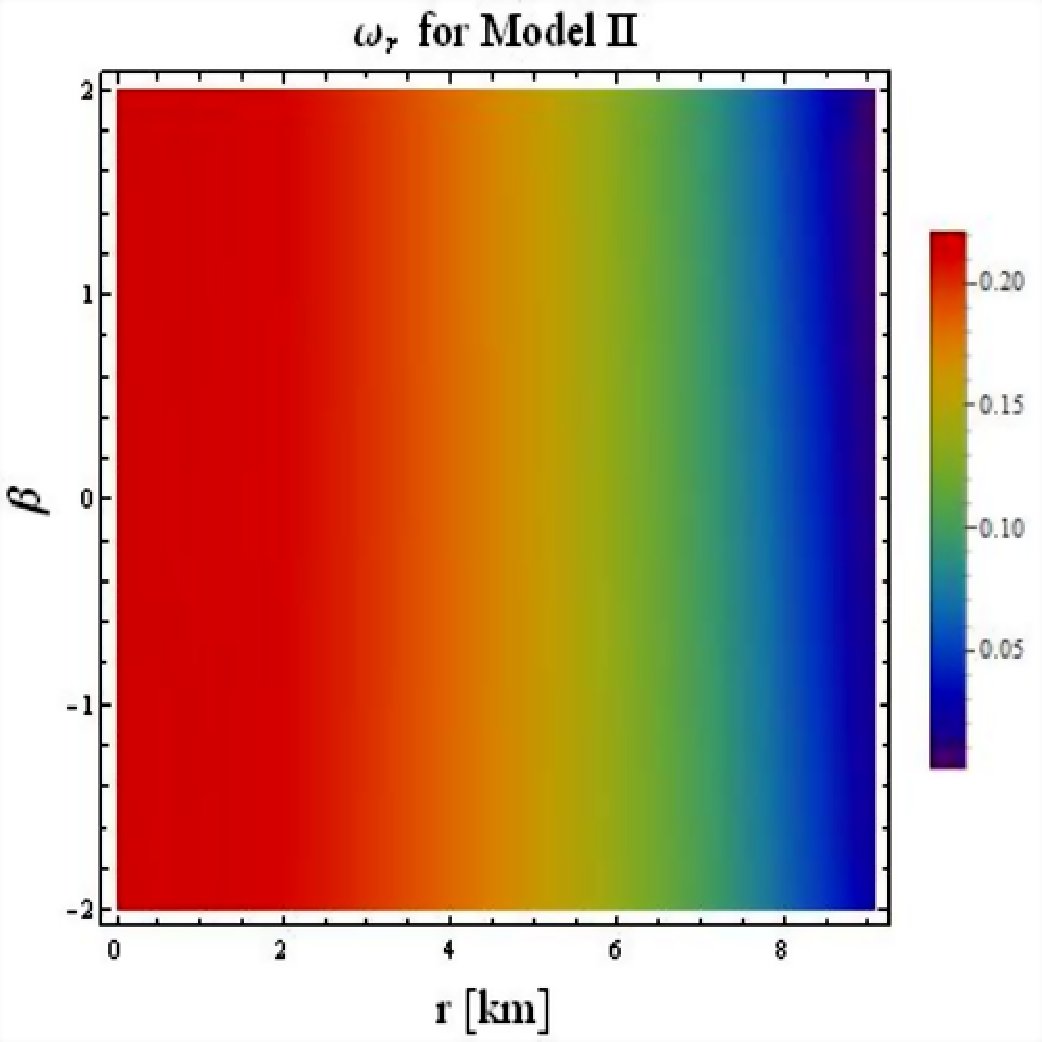,width=.4\linewidth}
\epsfig{file=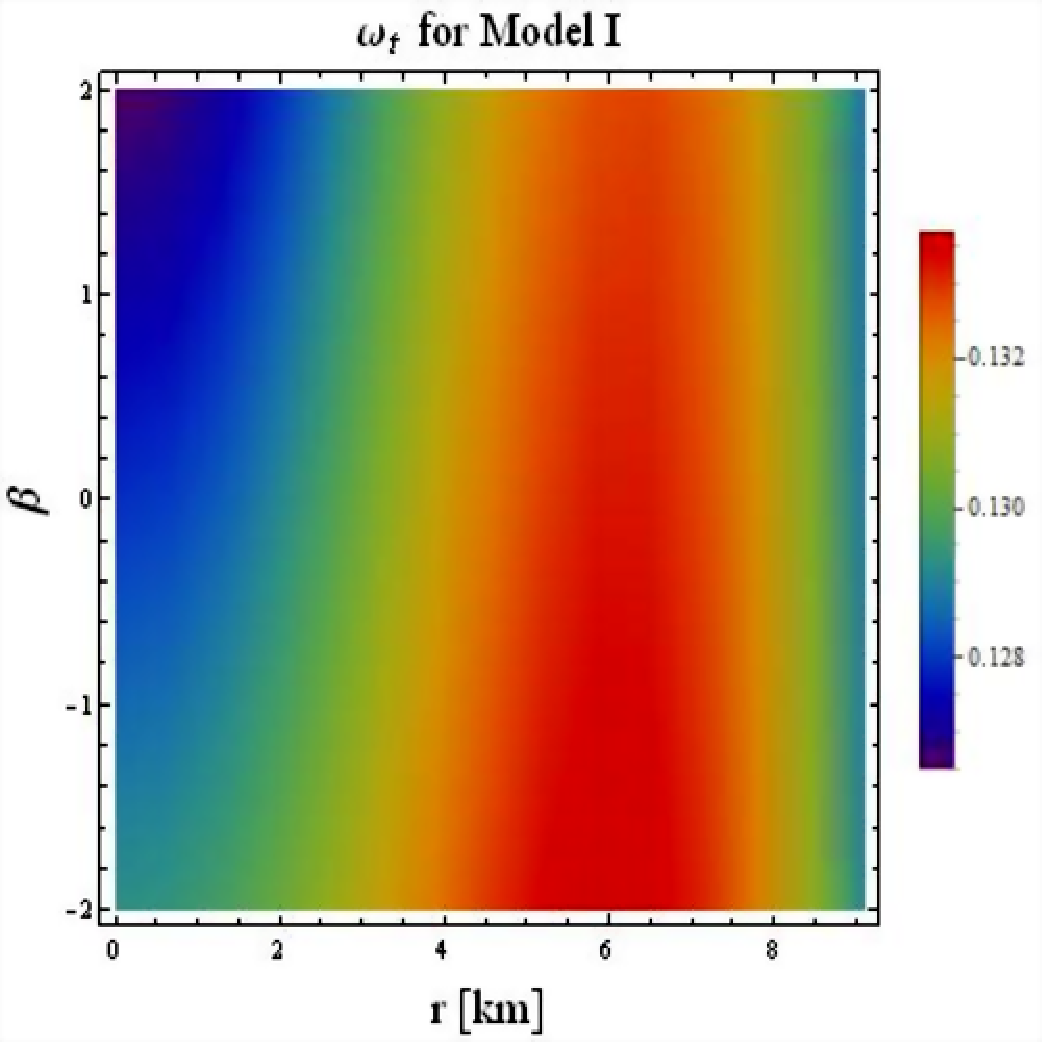,width=.4\linewidth}\epsfig{file=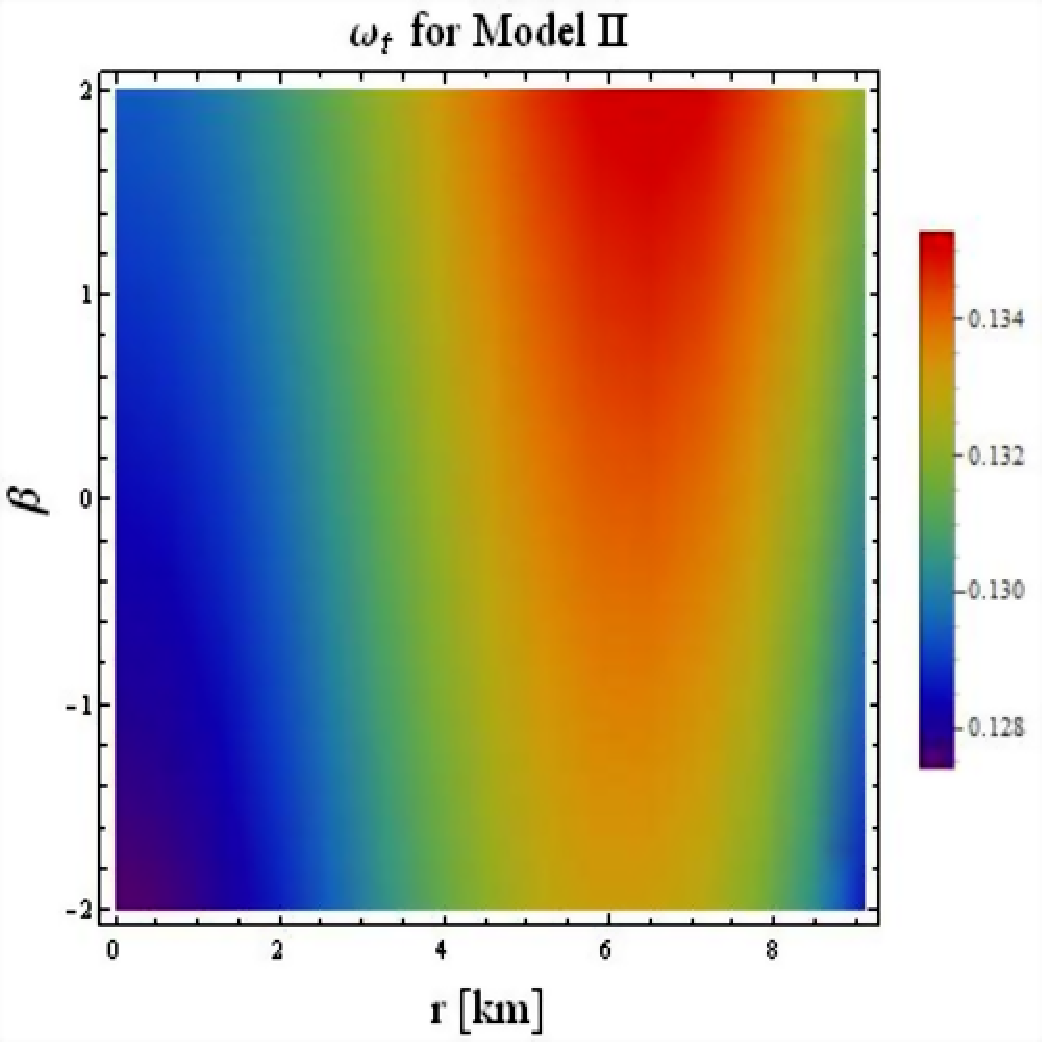,width=.4\linewidth}
\caption{Equation of state parameters versus $\beta$ and $r$.}
\end{figure}

\subsection{Energy Conditions}

The inner composition of celestial structures can encompass either
normal or exotic fluids. The existence of a normal fluid within a
compact star is determined by the energy conditions that rely on
various physical parameters, including pressure and energy density.
It is essential to consider these conditions when analyzing
astronomical systems in modified theories, as the correction terms
can have a significant impact on them. Consequently, meeting with
the following criteria ensures the formation of a realistic and
scientifically sound configuration
\begin{itemize}
\item Weak: $\rho+P_t \geq 0$, \quad $\rho \geq 0$, \quad $\rho+P_r \geq 0$,
\item Strong: $\rho+P_r+2P_t \geq 0$,
\item Null: $\rho+P_t \geq 0$, \quad $\rho+P_r \geq 0$,
\item Dominant: $\rho\pm P_t \geq 0$, \quad $\rho\pm P_r \geq 0$.
\end{itemize}
The graphical representation of these constraints can be observed in
Figures \textbf{8} and \textbf{9}, which correspond to models I and
II, respectively. All the plots illustrate consistently positive
behaviors. Consequently, this suggests that the solutions developed
for both $f(\mathcal{R},\mathcal{T},\mathcal{Q})$ models conform to
physical viability standards, indicating the presence of normal
matter within their respective interiors.
\begin{figure}[h!]\center
\epsfig{file=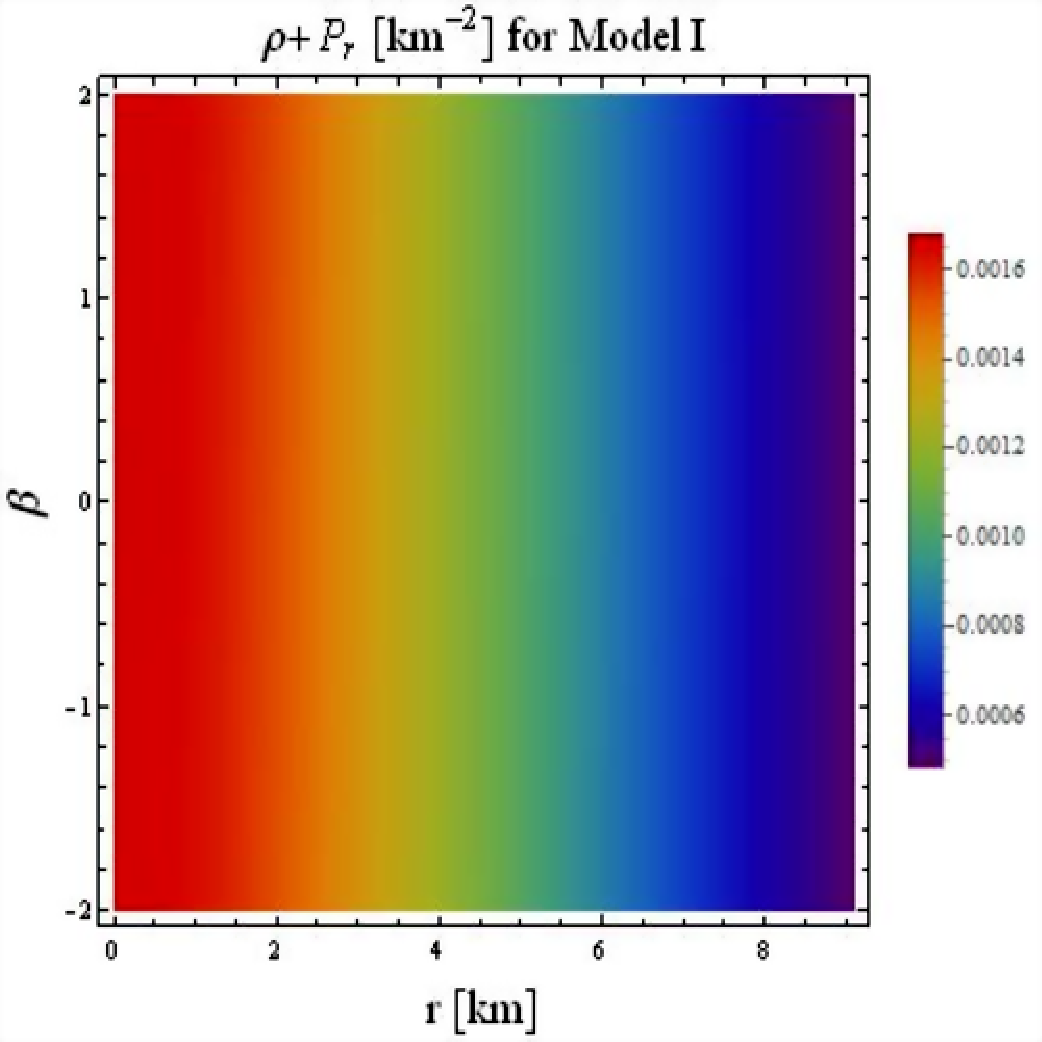,width=.4\linewidth}\epsfig{file=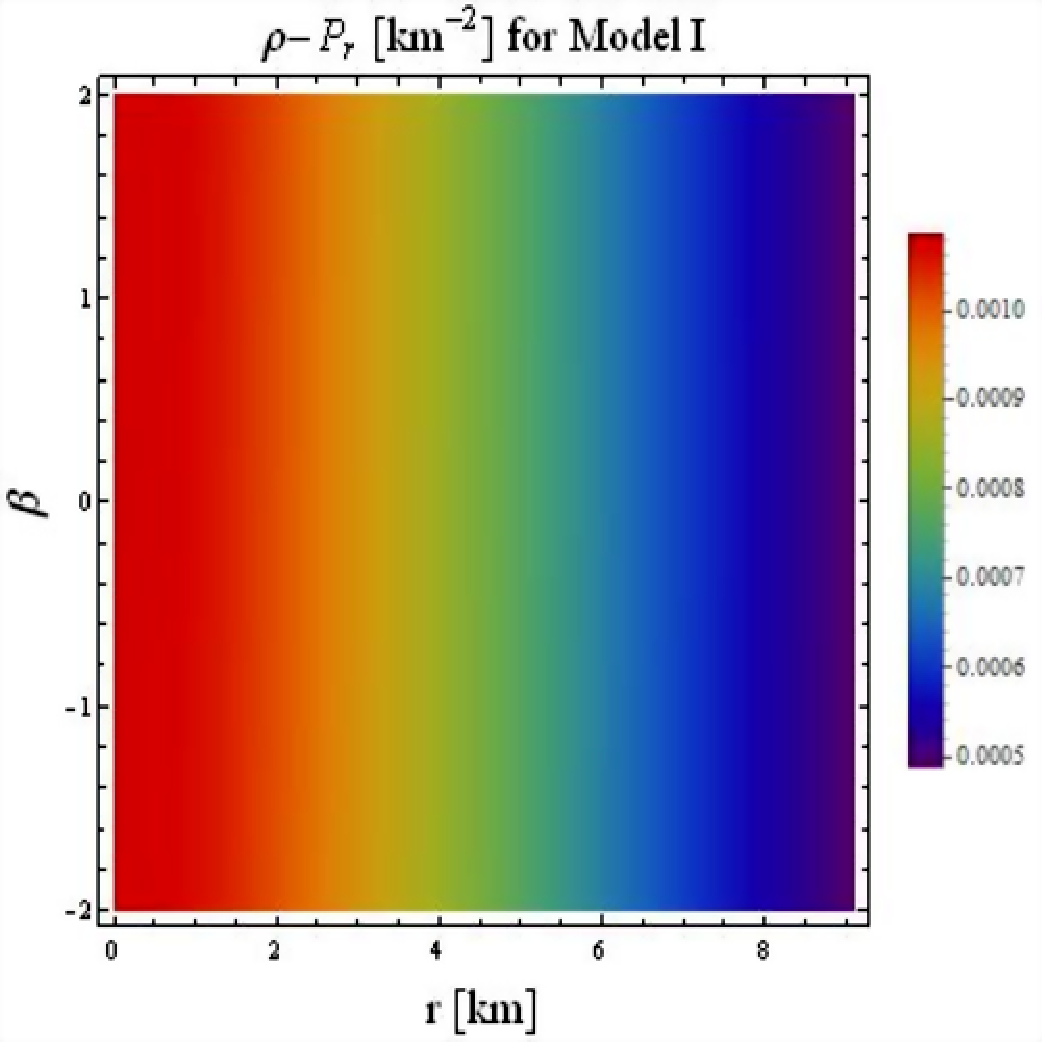,width=.4\linewidth}
\epsfig{file=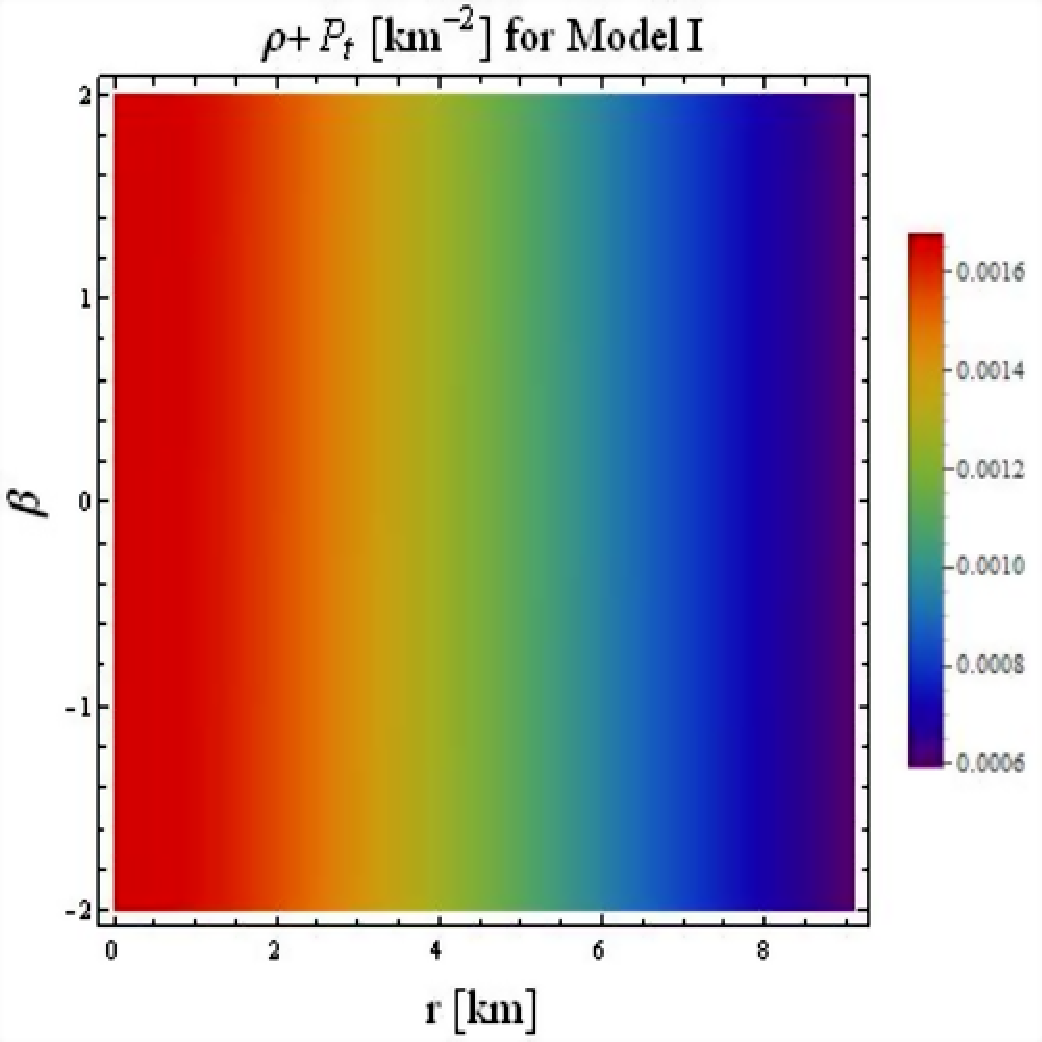,width=.4\linewidth}\epsfig{file=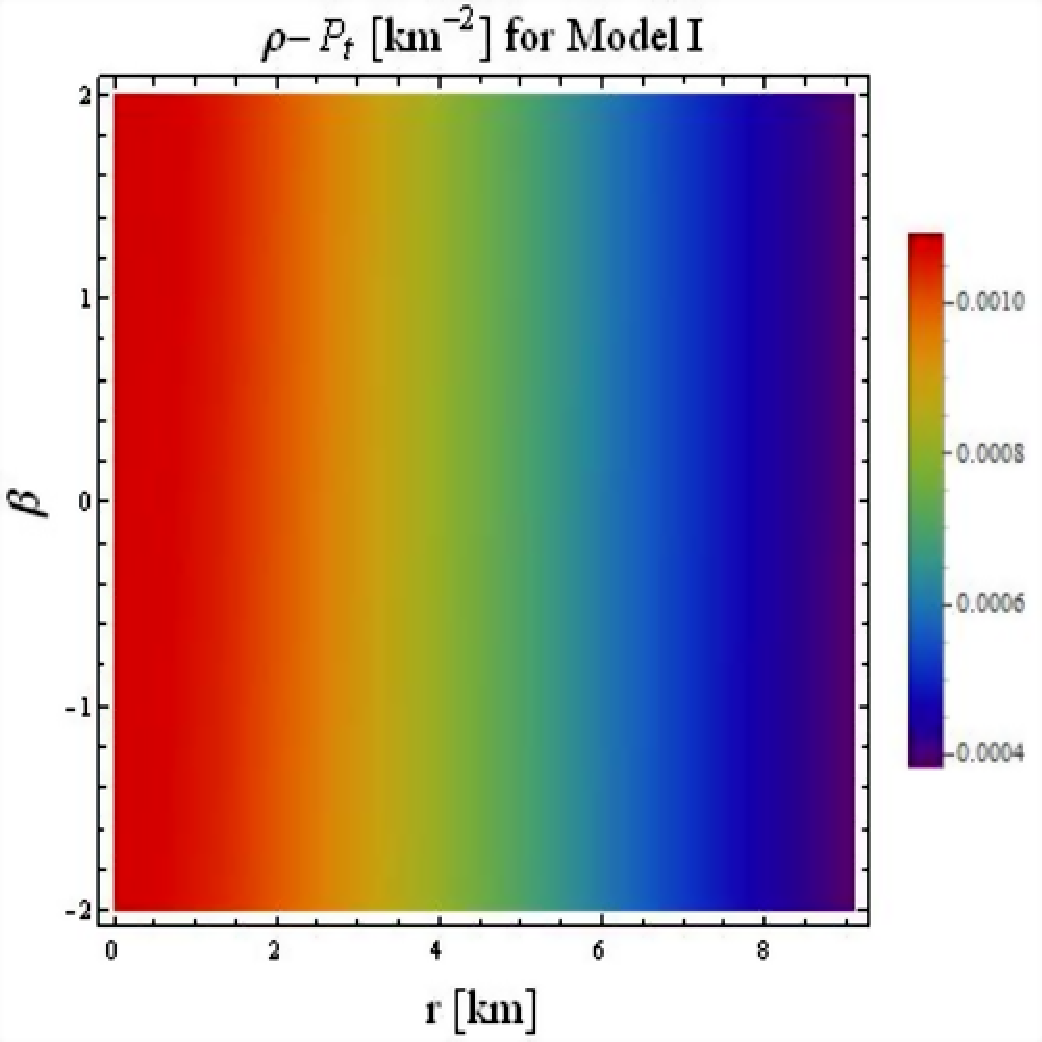,width=.4\linewidth}
\epsfig{file=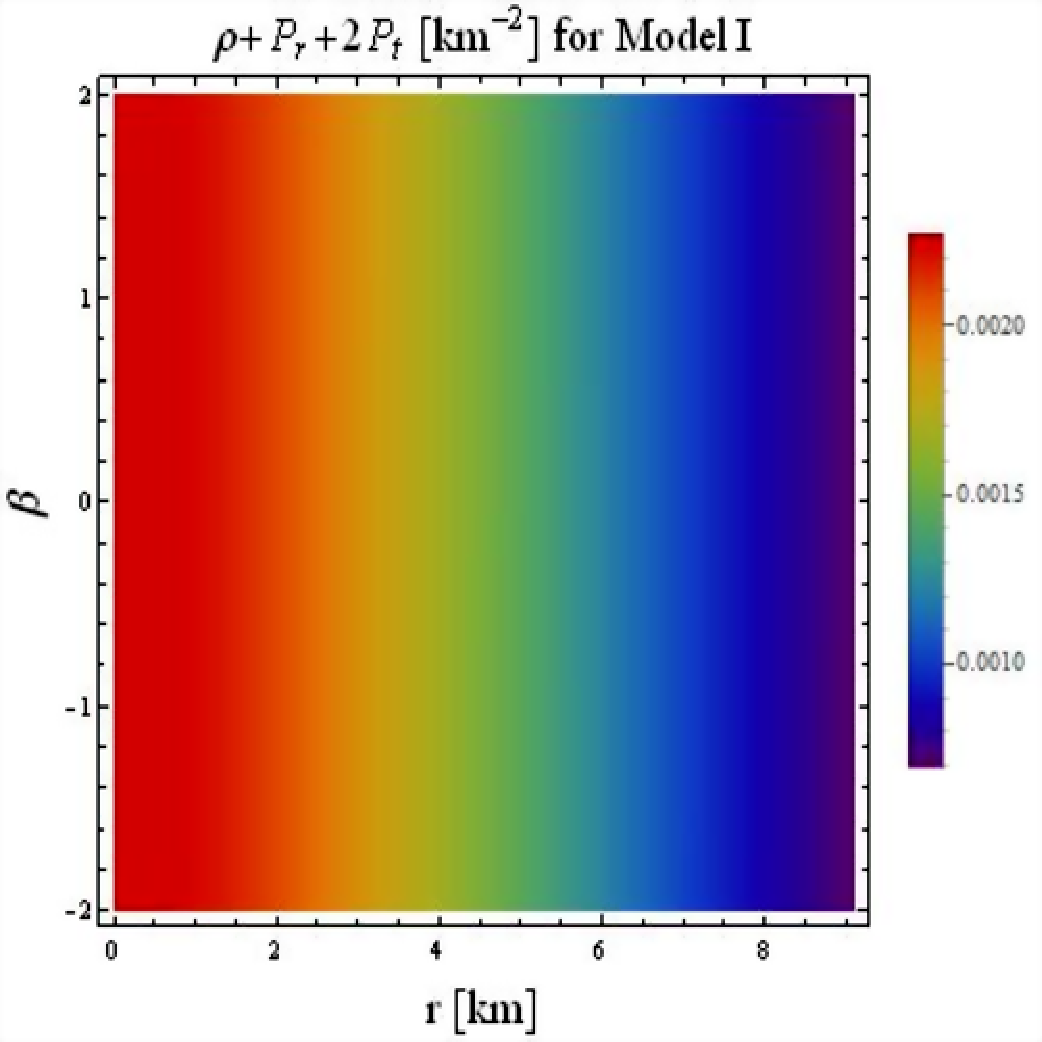,width=.4\linewidth} \caption{Energy conditions
versus $\beta$ and $r$.}
\end{figure}
\begin{figure}[h!]\center
\epsfig{file=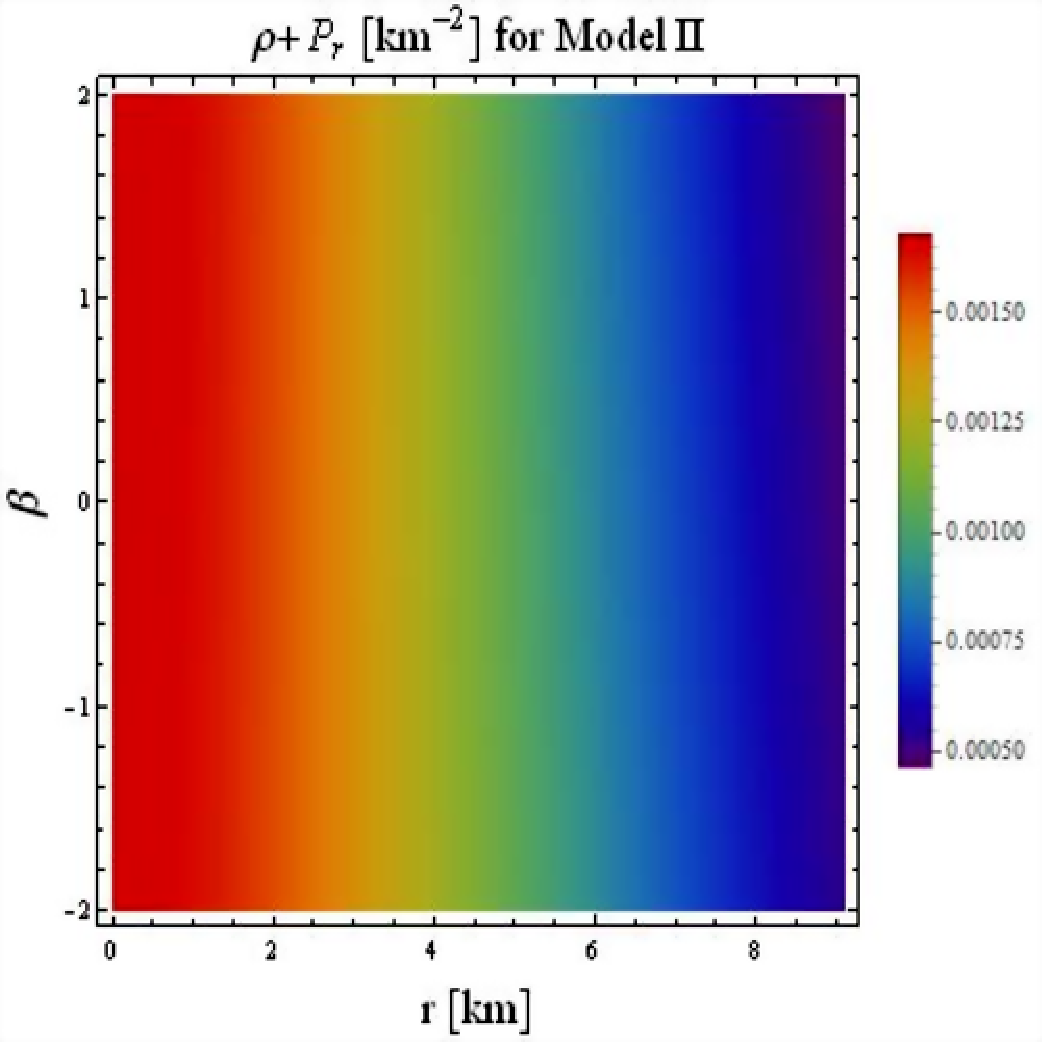,width=.4\linewidth}\epsfig{file=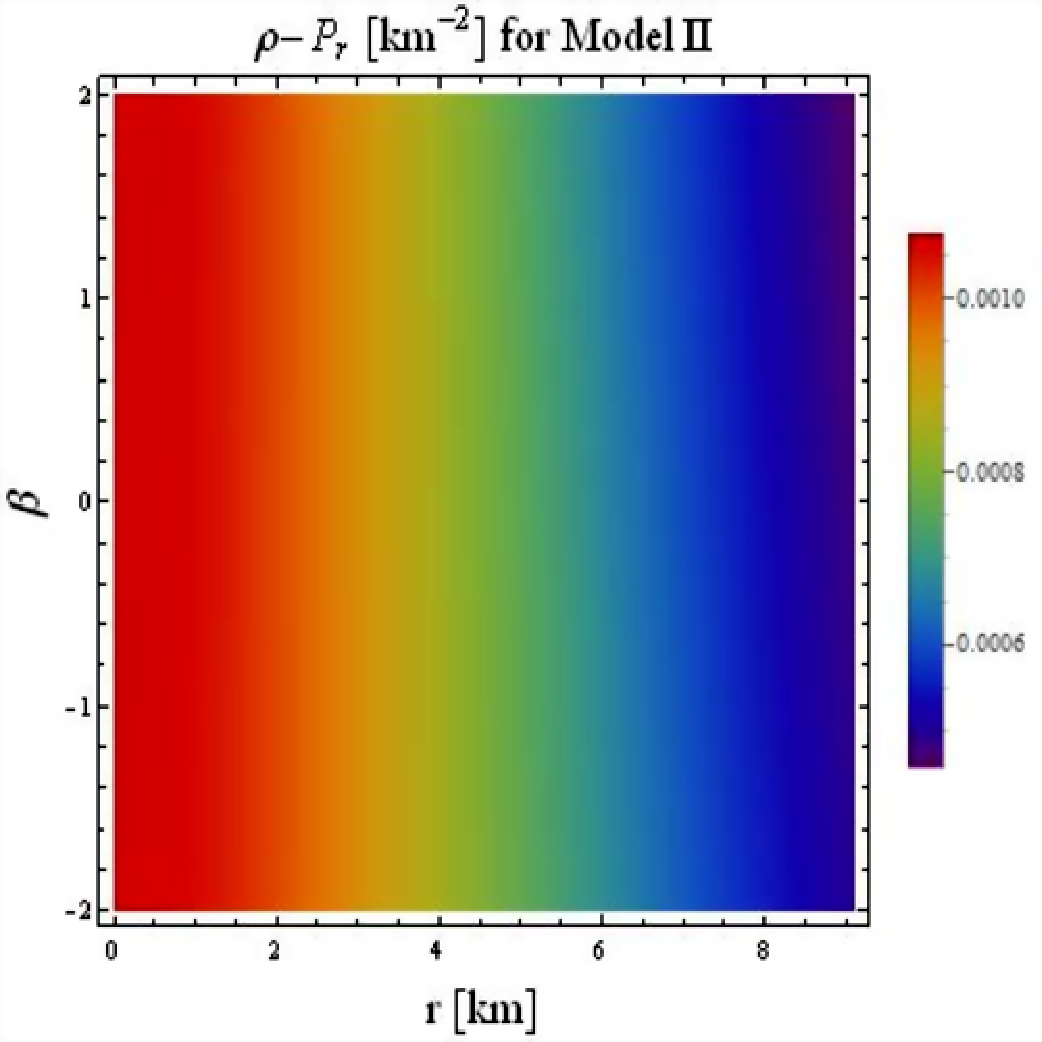,width=.4\linewidth}
\epsfig{file=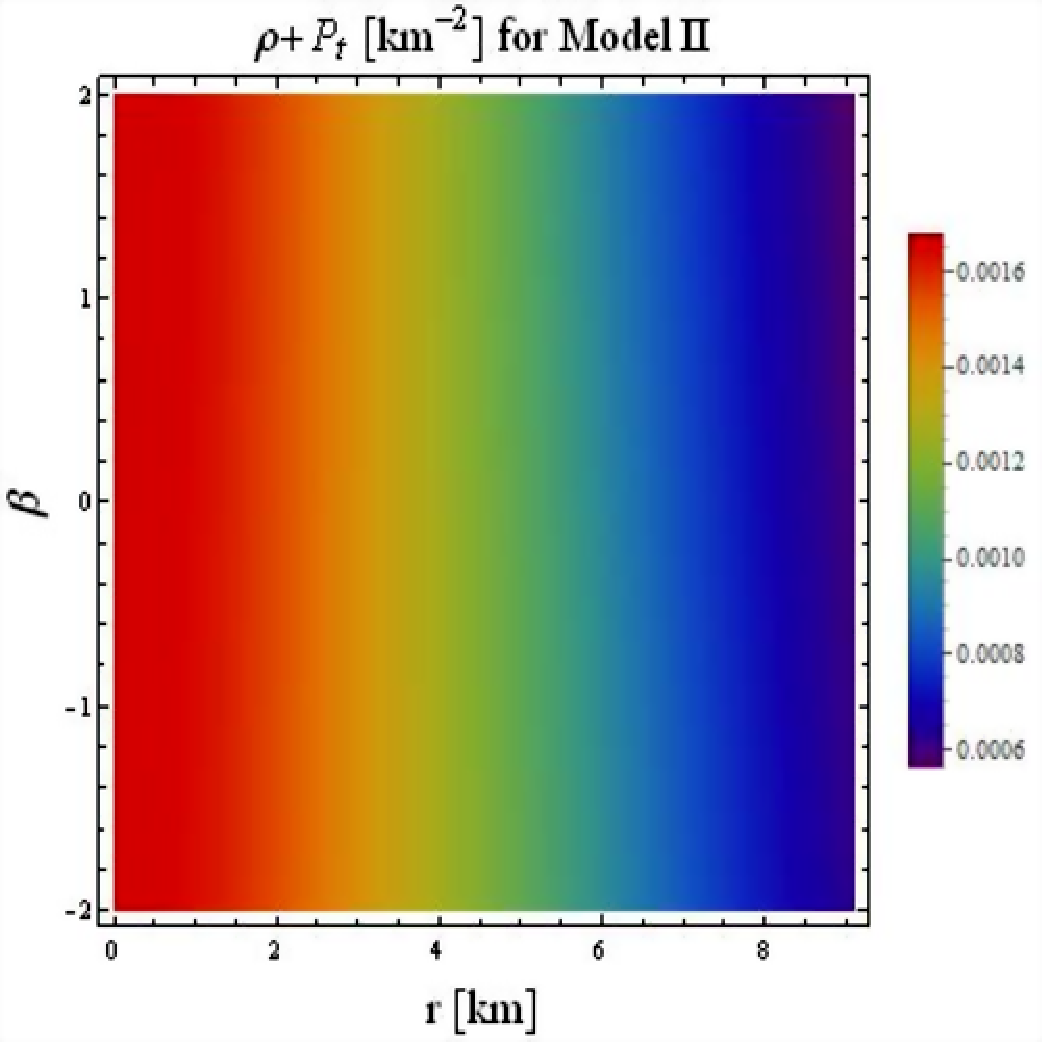,width=.4\linewidth}\epsfig{file=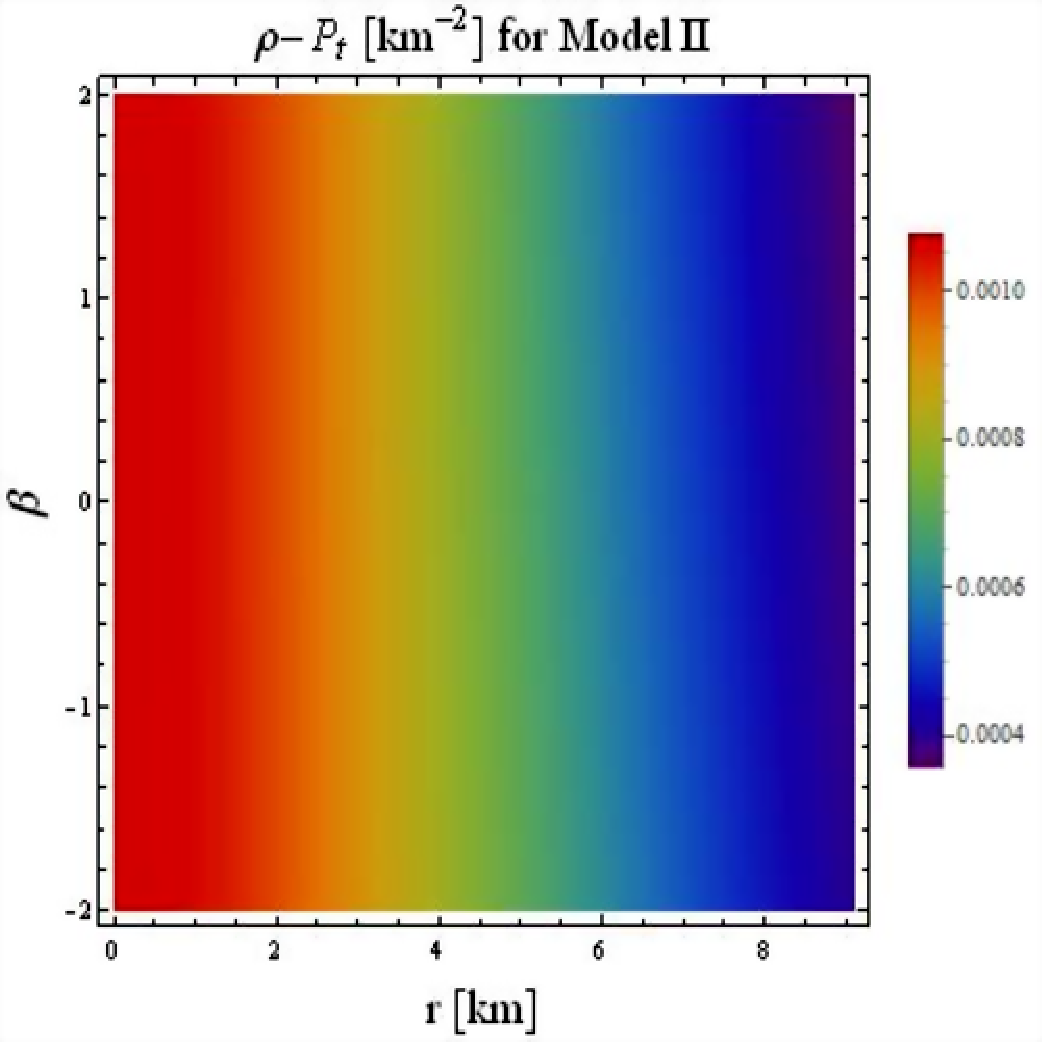,width=.4\linewidth}
\epsfig{file=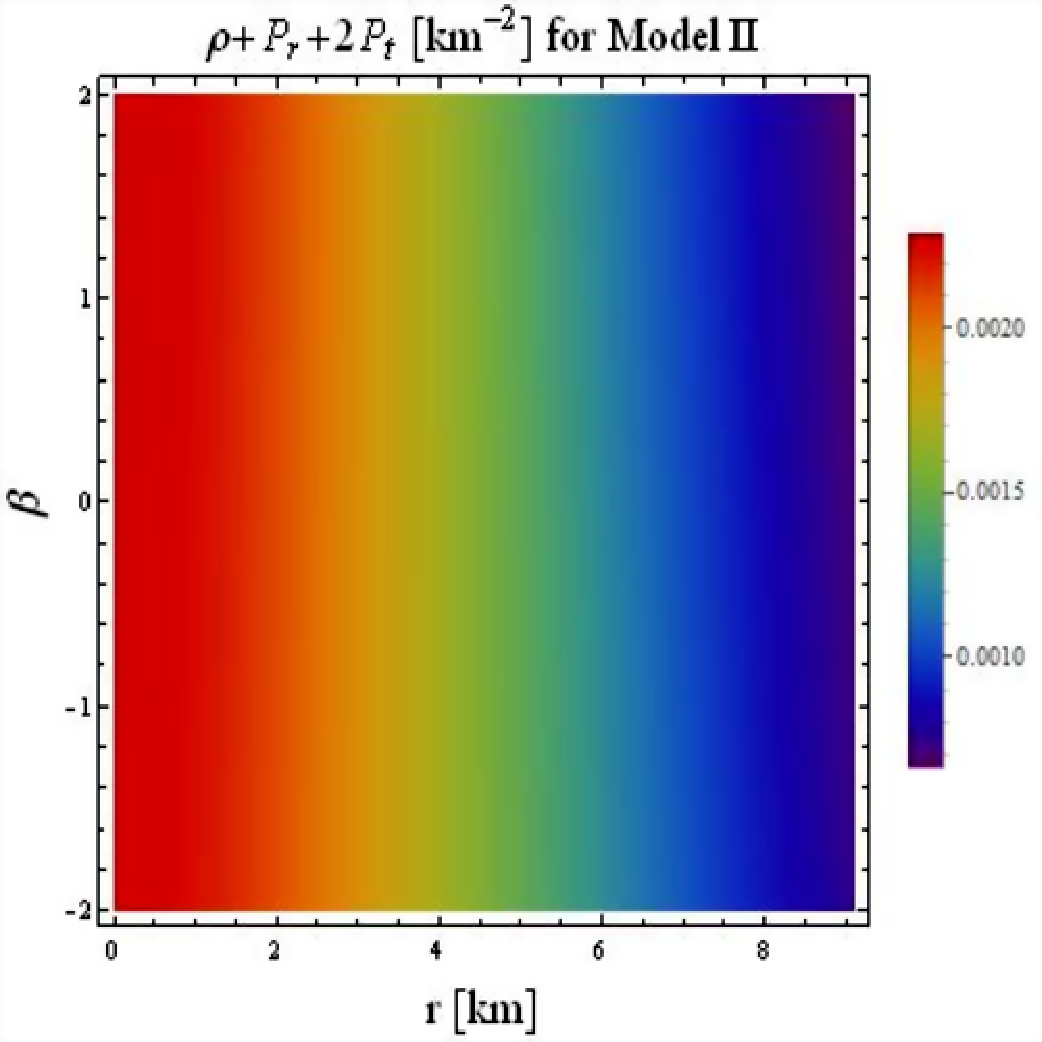,width=.4\linewidth} \caption{Energy conditions
versus $\beta$ and $r$.}
\end{figure}

\subsection{Tolman-Opphenheimer-Volkoff Equation}

The analysis of different fundamental forces is indeed necessary in
order to figure out the evolutionary pattern of a self-gravitating
matter distribution. We must verify these forces whether they are in
equilibrium or not \cite{37ccc,37ddd}. This phenomenon can be
studied by constructing the Tolman-Opphenheimer-Volkoff (TOV)
equation. In the following, we calculate this through Eq.\eqref{g4a}
with respect to both considered modified models, respectively, as
\begin{align}\nonumber
&\frac{dP_r}{dr}+\frac{A_1'}{2}\left(\rho
+P_r\right)-\frac{2}{r}\left(P_t-P_r\right)-\frac{2\beta
e^{-{A_2}}}{\beta\mathcal{R}+16\pi}\bigg[\frac{A_1'\rho}{8}\bigg(A_1'^2-A_1'{A_2}'+2A_1''+\frac{4A_1'}{r}\bigg)\\\nonumber
&-\frac{\rho'}{8}\bigg(A_1'^2-A_1'{A_2}'+2A_1''-\frac{4A_1'}{r}-\frac{8e^{A_2}}{r^2}+\frac{8}{r^2}\bigg)
+P_r\bigg(\frac{5A_1'^2{A_2}'}{8}-\frac{5A_1'{A_2}'^2}{8}+\frac{7A_1''{A_2}'}{4}\\\nonumber
&-A_1'A_1''+\frac{A_1'{A_2}''}{2}-\frac{5{A_2}'^2}{2r}-\frac{A_1'''}{2}+\frac{2{A_2}''}{r}+\frac{A_1'{A_2}'}{r}-\frac{{A_2}'}{r^2}
-\frac{A_1''}{r}+\frac{A_1'}{r^2}+\frac{2e^{A_2}}{r^3}-\frac{2}{r^3}\bigg)\\\nonumber
&-\frac{P'_r}{8}\bigg(A_1'^2-A_1'{A_2}'+2A_1''-\frac{4{A_2}'}{r}\bigg)+\frac{P_t}{r^2}\bigg({A_2}'-A_1'+\frac{2e^{A_2}}{r}
-\frac{2}{r}\bigg)-\frac{P'_t}{r}\bigg(\frac{{A_2}'}{2}-\frac{A_1'}{2}\\\label{g11}
&+\frac{e^{A_2}}{r}-\frac{1}{r}\bigg)\bigg]=0,
\end{align}
and
\begin{align}\nonumber
&\frac{dP_r}{dr}+\frac{A_1'}{2}\left(\rho
+P_r\right)-\frac{2}{r}\left(P_t-P_r\right)-\frac{2\beta}{\beta\mathcal{R}^2+16\pi}
\bigg[\rho\bigg\{\frac{e^{-A_1-{A_2}}A_1'\mathcal{R}\mathcal{R}_{00}}{2}-e^{-2{A_2}}\\\nonumber
&\times\mathcal{R}'\bigg(\frac{A_1'}{r}-\frac{e^{{A_2}}}{r^2}+\frac{1}{r^2}\bigg)\bigg\}-\rho'\bigg\{\frac{e^{-A_1-{A_2}}\mathcal{R}\mathcal{R}_{00}}{2}
-e^{-2{A_2}}\mathcal{R}\bigg(\frac{A_1'}{r}-\frac{e^{{A_2}}}{r^2}+\frac{1}{r^2}\bigg)\bigg\}\\\nonumber
&+P_r\bigg\{\mathcal{R}'\mathcal{R}^{11}+\mathcal{R}(\mathcal{R}^{11})'-e^{-{A_2}}\mathcal{R}\mathcal{R}'+\frac{e^{-2{A_2}}{A_2}'
\mathcal{R}\mathcal{R}_{11}}{2}\bigg\}-\frac{\mathcal{R}\mathcal{R}_{22}}{e^{{A_2}}}\bigg\{\frac{P'_{t}}{r^2}-\frac{2P_{t}}{r^3}\bigg\}\\\label{g11a}
&+P'_r\bigg\{\mathcal{R}\mathcal{R}^{11}-\frac{e^{-2{A_2}}\mathcal{R}\mathcal{R}_{11}}{2}\bigg\}\bigg]=0.
\end{align}
The non-zero Ricci tensors $\mathcal{R}_{00},~\mathcal{R}_{11}$ and
$\mathcal{R}_{22}$ are provided in Appendix \textbf{A}. The compact
form of the above TOV equations is
\begin{equation}\label{g36a}
f_h+f_a+f_x=0,
\end{equation}
with $f_h$ and $f_a$ being hydrostatic and anisotropic forces,
respectively, given by
\begin{equation}\nonumber
f_h=-\frac{dP_r}{dr},\quad f_a=\frac{2}{r}\big(P_t-P_r\big).
\end{equation}
Moreover, we express the third force as $f_x=f_g+f_e$, where $f_g$
and $f_e$ are the gravitational and additional forces appearing due
to modified theory, respectively. The force $f_x$ contains all of
the remaining terms of Eqs.\eqref{g11} and \eqref{g11a} multiplied
by a negative sign. Figure \textbf{10} guarantees that the developed
solutions support the hydrostatic equilibrium in the interior of 4U
1820-30 compact star.
\begin{figure}\center
\epsfig{file=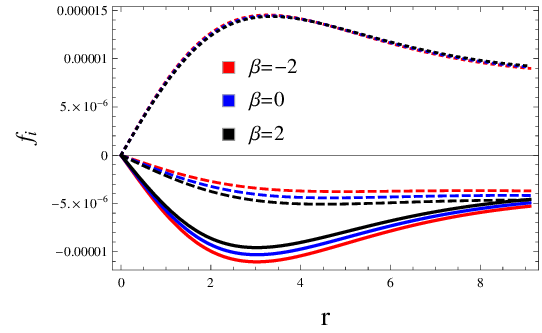,width=.4\linewidth}\epsfig{file=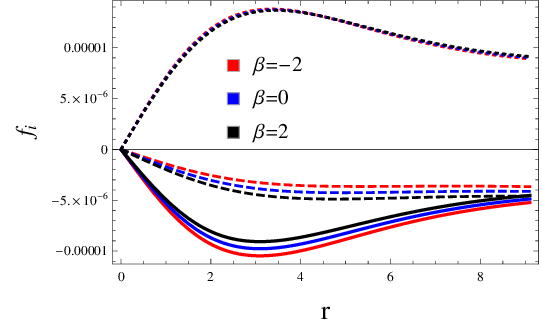,width=.4\linewidth}
\caption{Variation in $f_a$ (solid), $f_h$ (dashed) and $f_x$
(dotted) versus $r$.}
\end{figure}

\subsection{Stability Analysis of 4U 1820-30 Compact Star}

\subsubsection{Causality and Cracking}

In the vast realm of cosmic phenomena, particular attention has been
devoted to celestial bodies and gravitational models that adhere to
stability criteria. Numerous methodologies have been enlisted in the
scientific literature to scrutinize the stability of these models.
One essential condition that arises in these investigations is the
causality criterion \cite{42d}, which stipulates that the speed of
light within any given medium must be less than the speed of light
in a stable structure. This condition is often expressed
mathematically as $0 < v_{s}^{2} < 1$, and for anisotropic
configurations, it takes the form of $0 < v_{st}^{2} < 1$
(tangential) and $0 < v_{sr}^{2} < 1$ (radial component). They are
defined as
\begin{equation}
v_{st}^{2}=\frac{dP_{t}}{d\rho}, \quad
v_{sr}^{2}=\frac{dP_{r}}{d\rho}.
\end{equation}
Herrera \cite{42e} introduced the notion of cracking and put forth
an inequality based on the previously defined parameters as $0 <
\mid v_{st}^{2}-v_{sr}^{2} \mid < 1$, which is applicable only to
those objects that are in a stable state. In Figure \textbf{11}, we
observe graphical representations of the causality condition as well
as cracking approach. The satisfaction of both these conditions
ensures the stability of our resulting models for all $\beta$ in the
considered range.
\begin{figure}[h!]\center
\epsfig{file=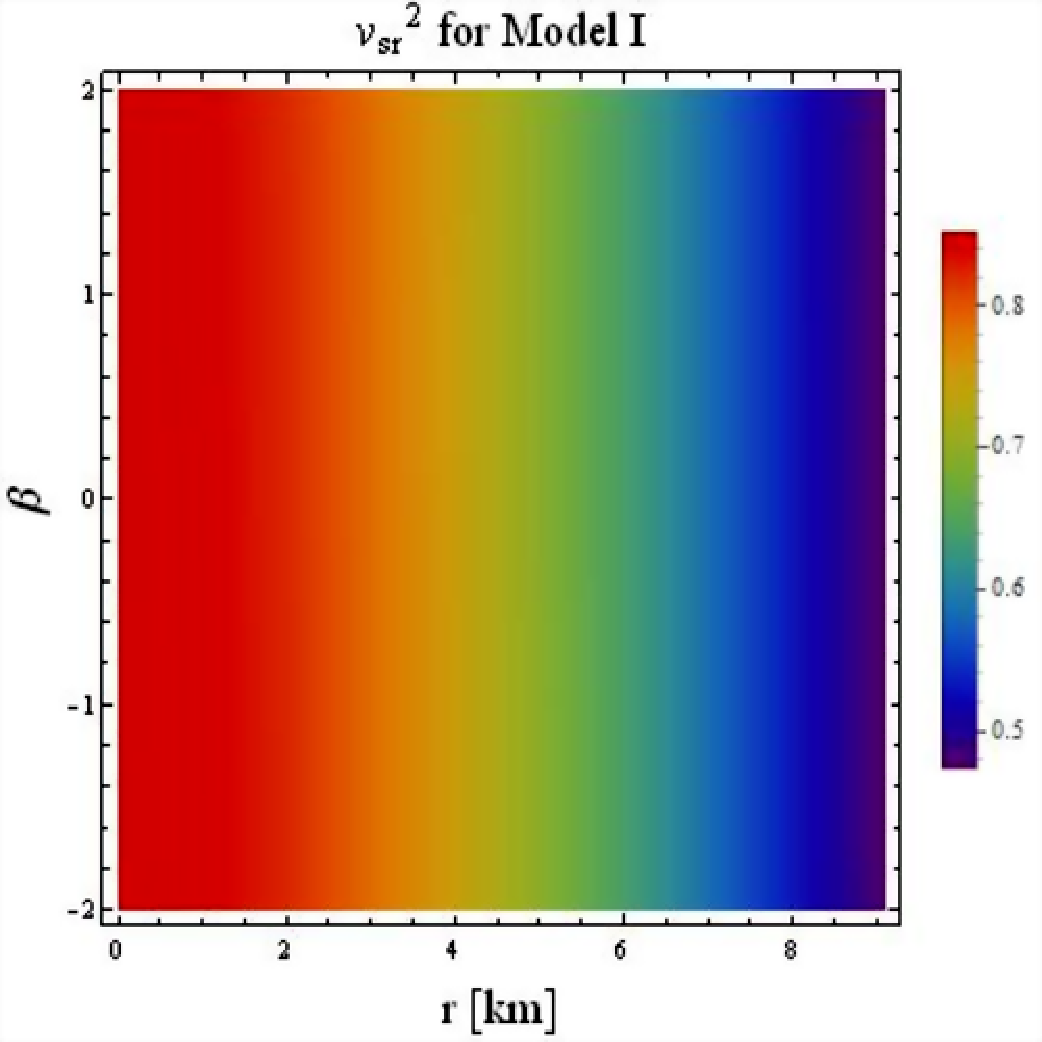,width=.4\linewidth}\epsfig{file=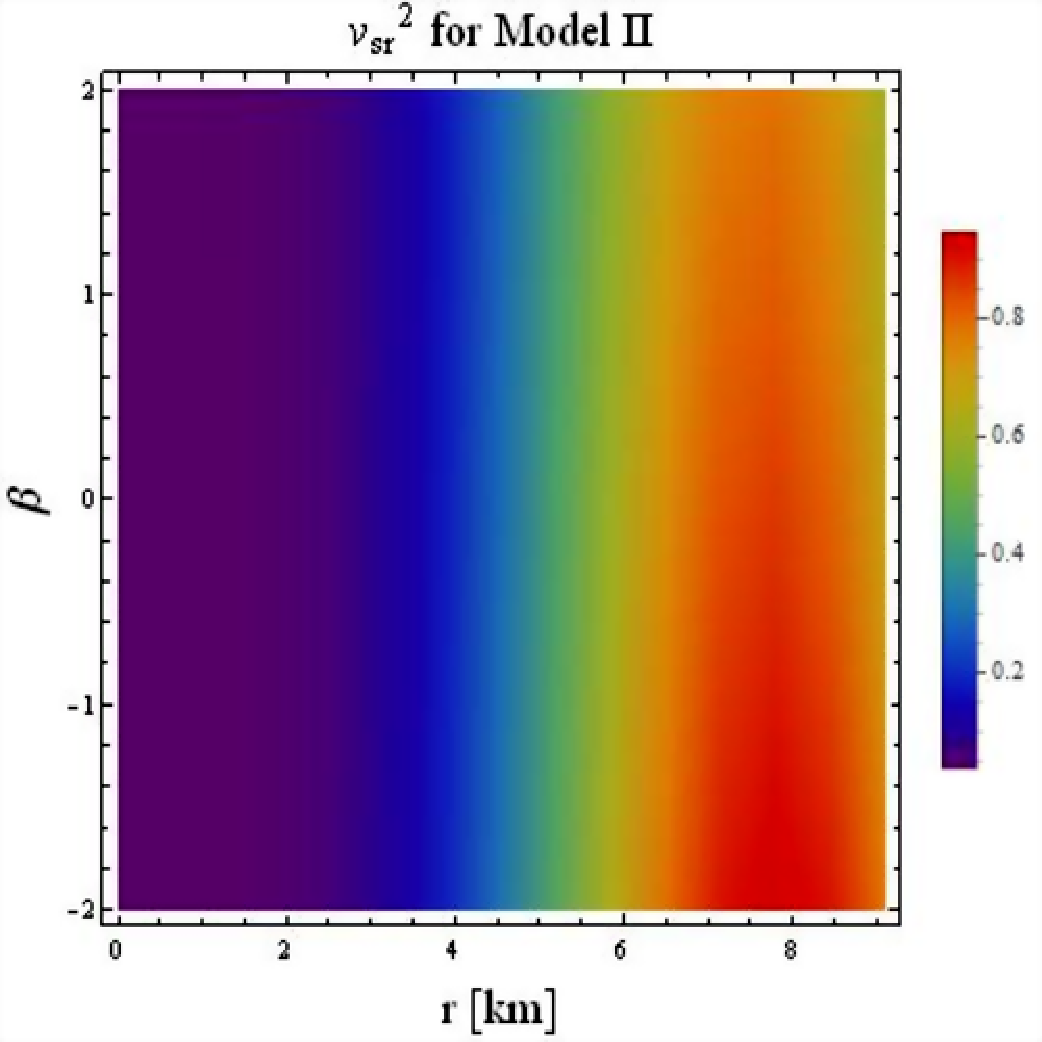,width=.4\linewidth}
\epsfig{file=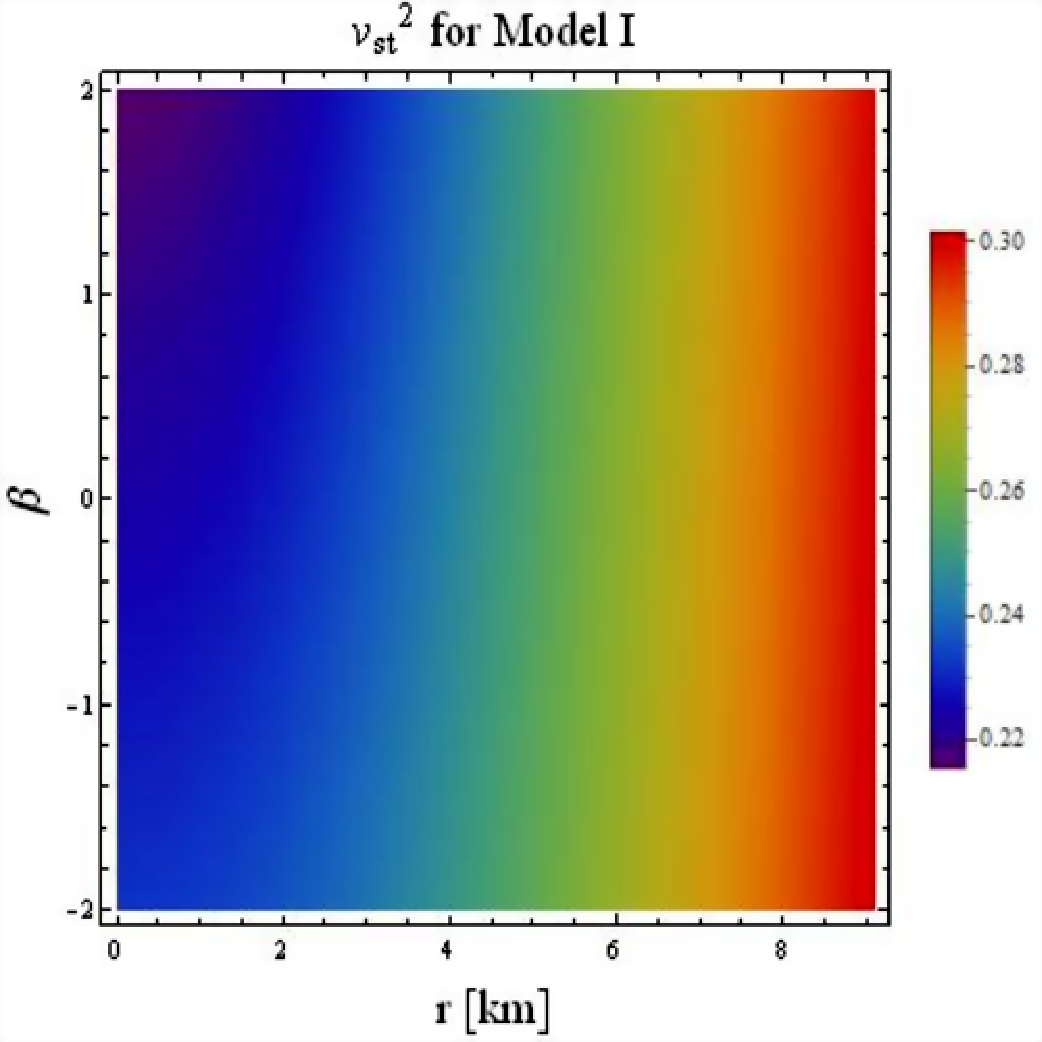,width=.4\linewidth}\epsfig{file=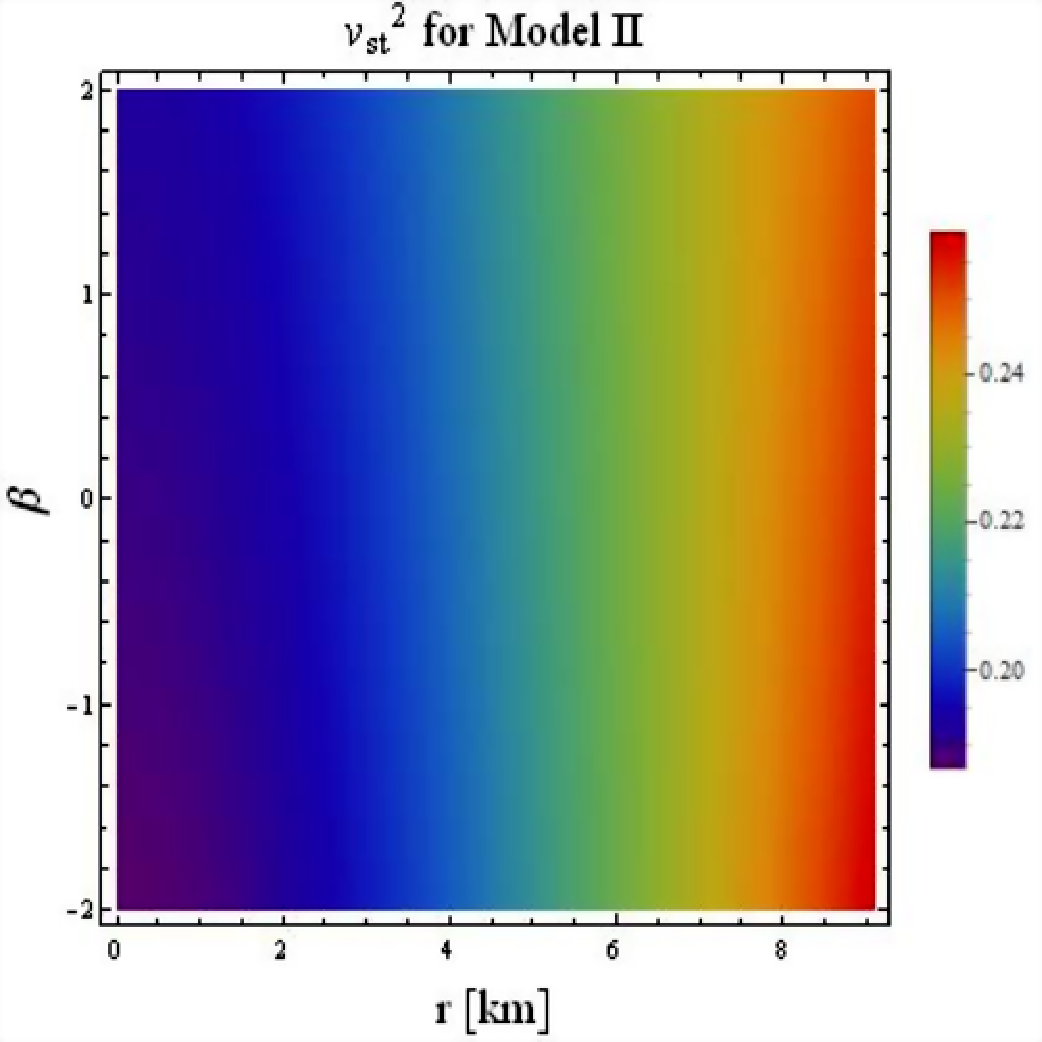,width=.4\linewidth}
\epsfig{file=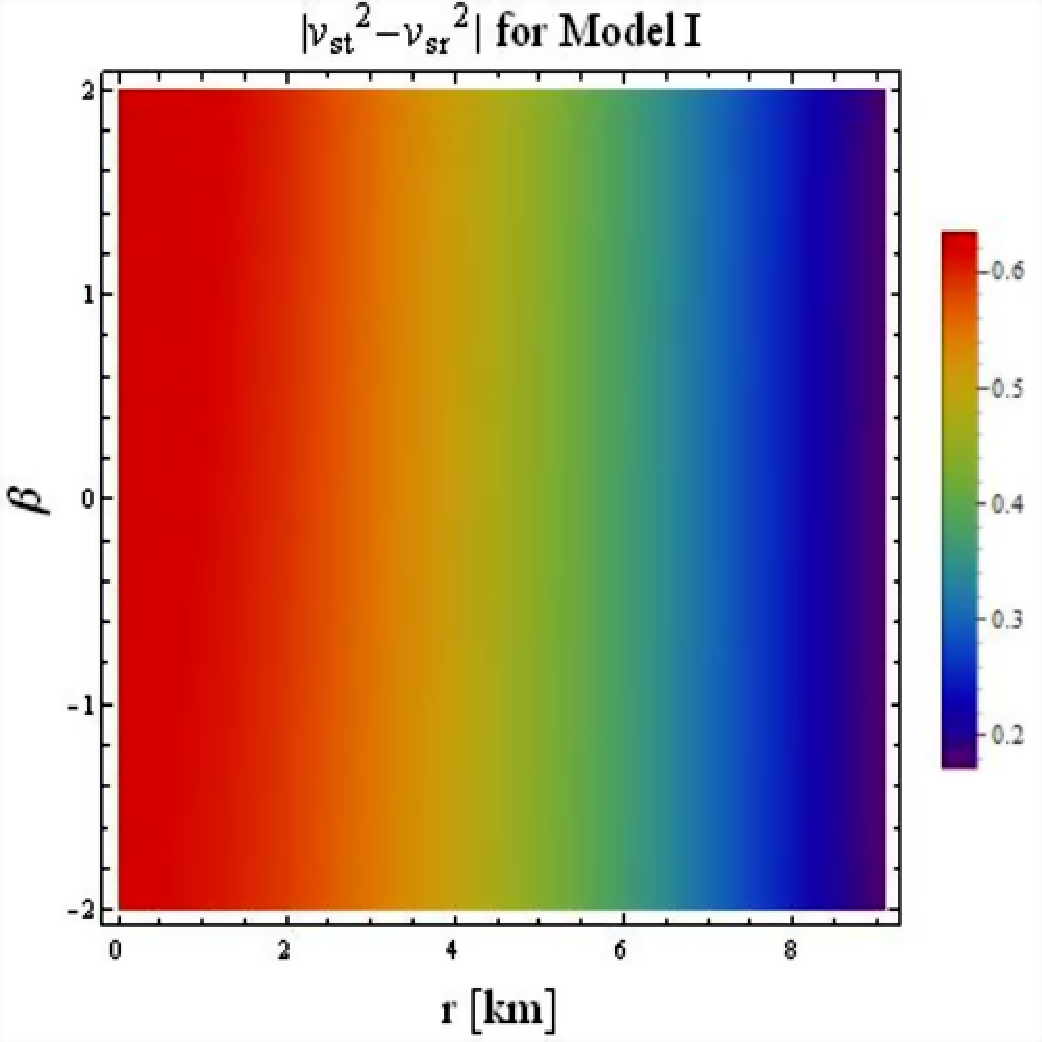,width=.4\linewidth}\epsfig{file=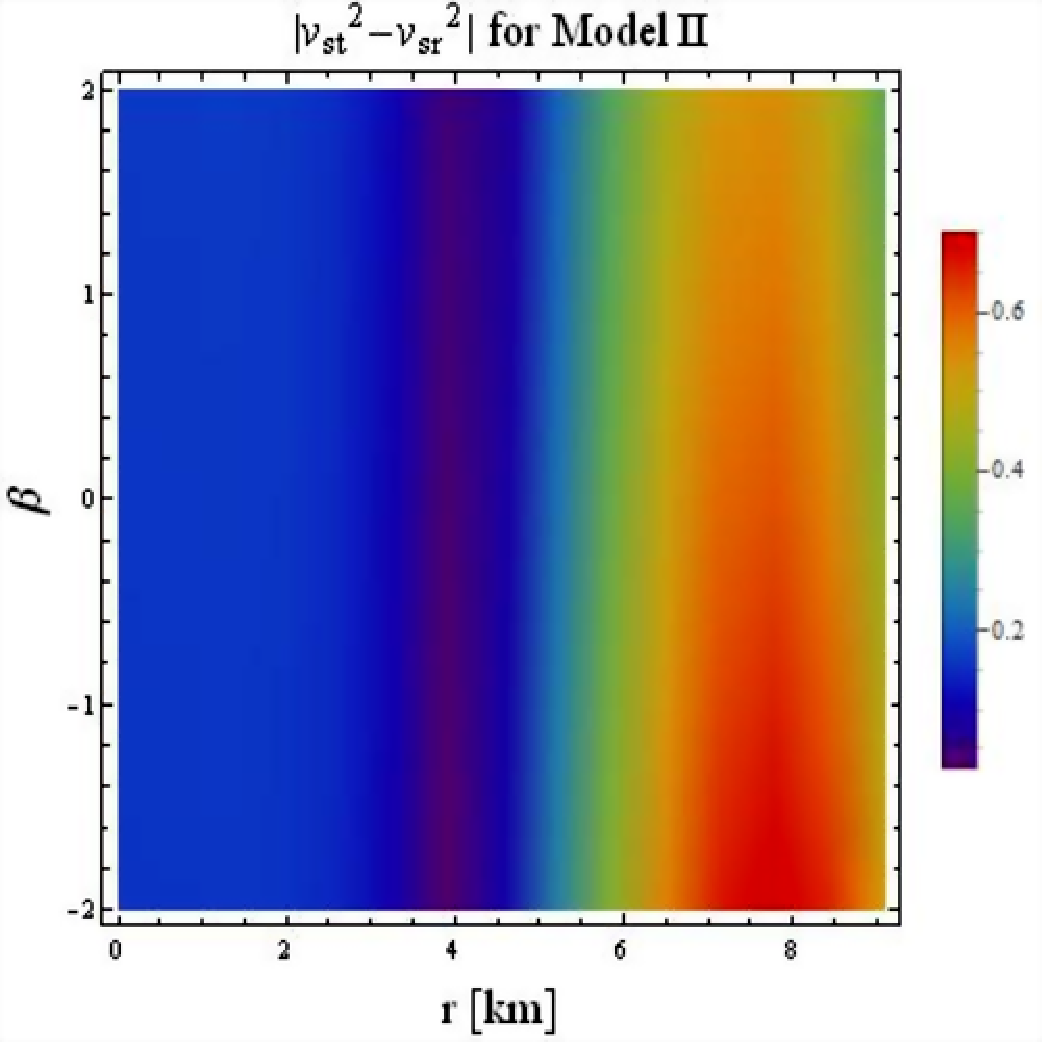,width=.4\linewidth}
\caption{Stability analysis versus $\beta$ and $r$.}
\end{figure}

\subsubsection{Adiabatic Index}

The adiabatic index, denoted by $\Gamma$, serves as a valuable tool
for computing the stability of compact objects. This criterion has
been widely utilized in the examination of self-gravitating
structures, revealing a noteworthy findings. It has been determined
that, in the context of stable objects, the lower limit of this
index consistently remains at a value of $\frac{4}{3}$ \cite{42f}.
Here, $\Gamma$ is divided into two components for the case of
anisotropic fluid. They are defined as
\begin{equation}\label{g62}
\Gamma_r=\frac{\rho+P_{r}}{P_{r}} \left(\frac{dP_{r}}{d\rho}\right),
\quad \Gamma_t=\frac{\rho+P_{t}}{P_{t}}
\left(\frac{dP_{t}}{d\rho}\right).
\end{equation}
Figure $\mathbf{12}$ shows the stable behavior of both resulting
models as they are consistent with the desired behavior of this
index.
\begin{figure}[h!]\center
\epsfig{file=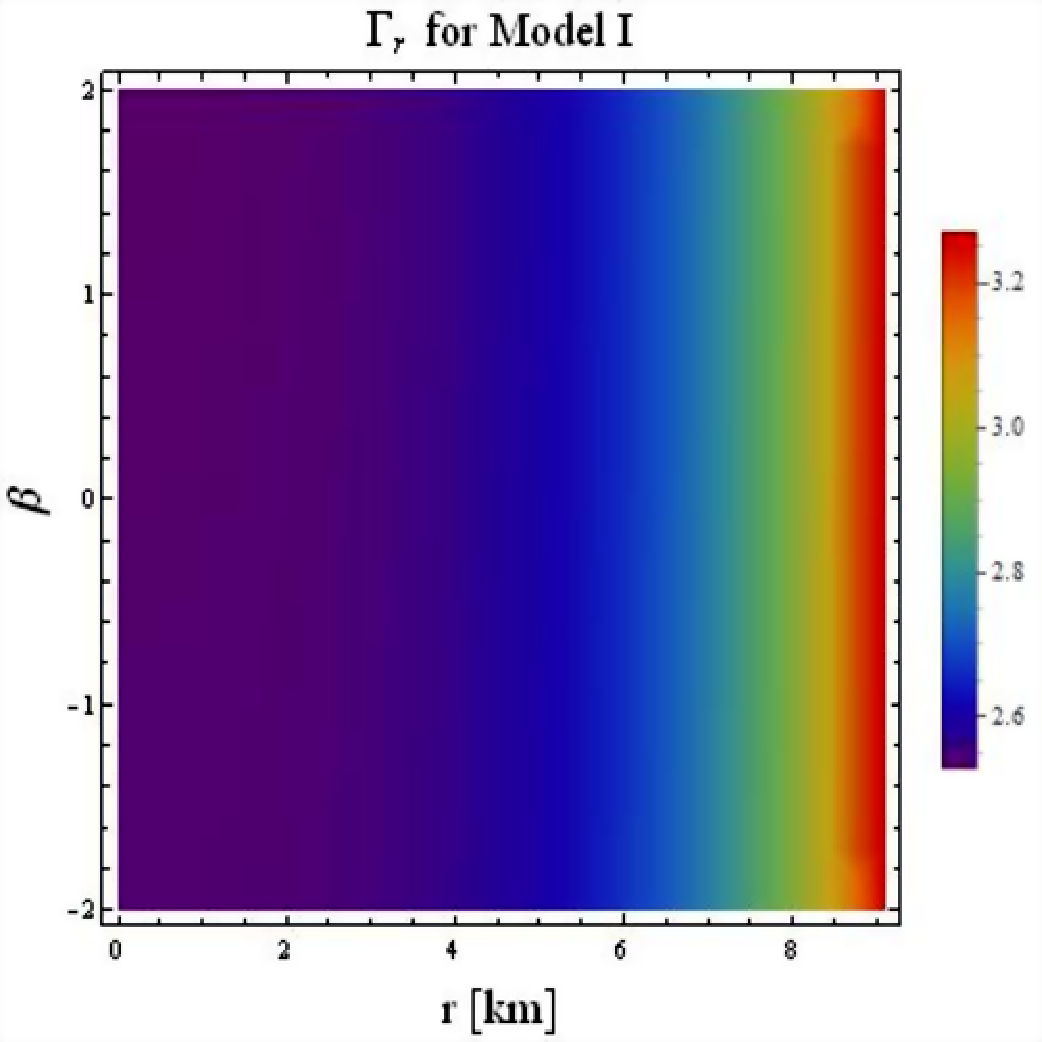,width=.4\linewidth}\epsfig{file=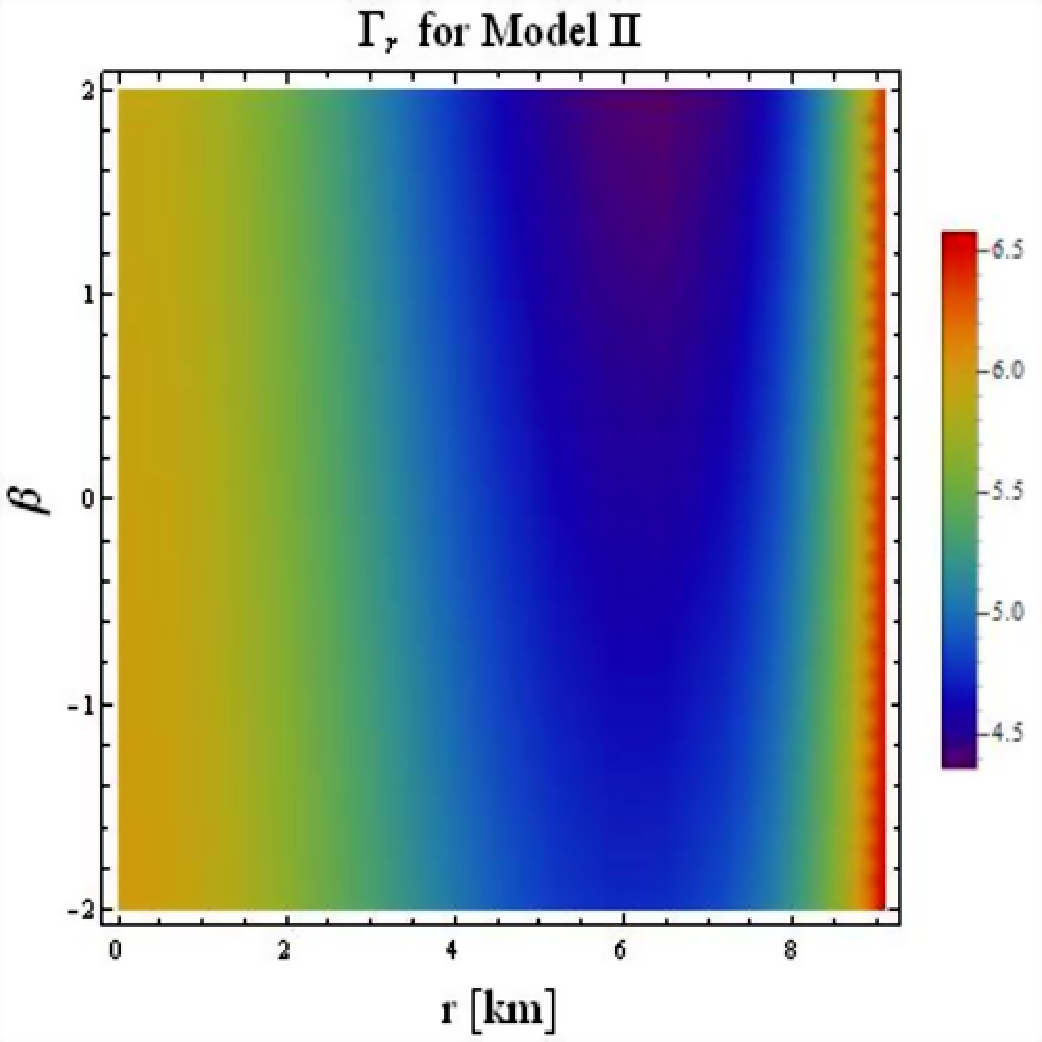,width=.4\linewidth}
\epsfig{file=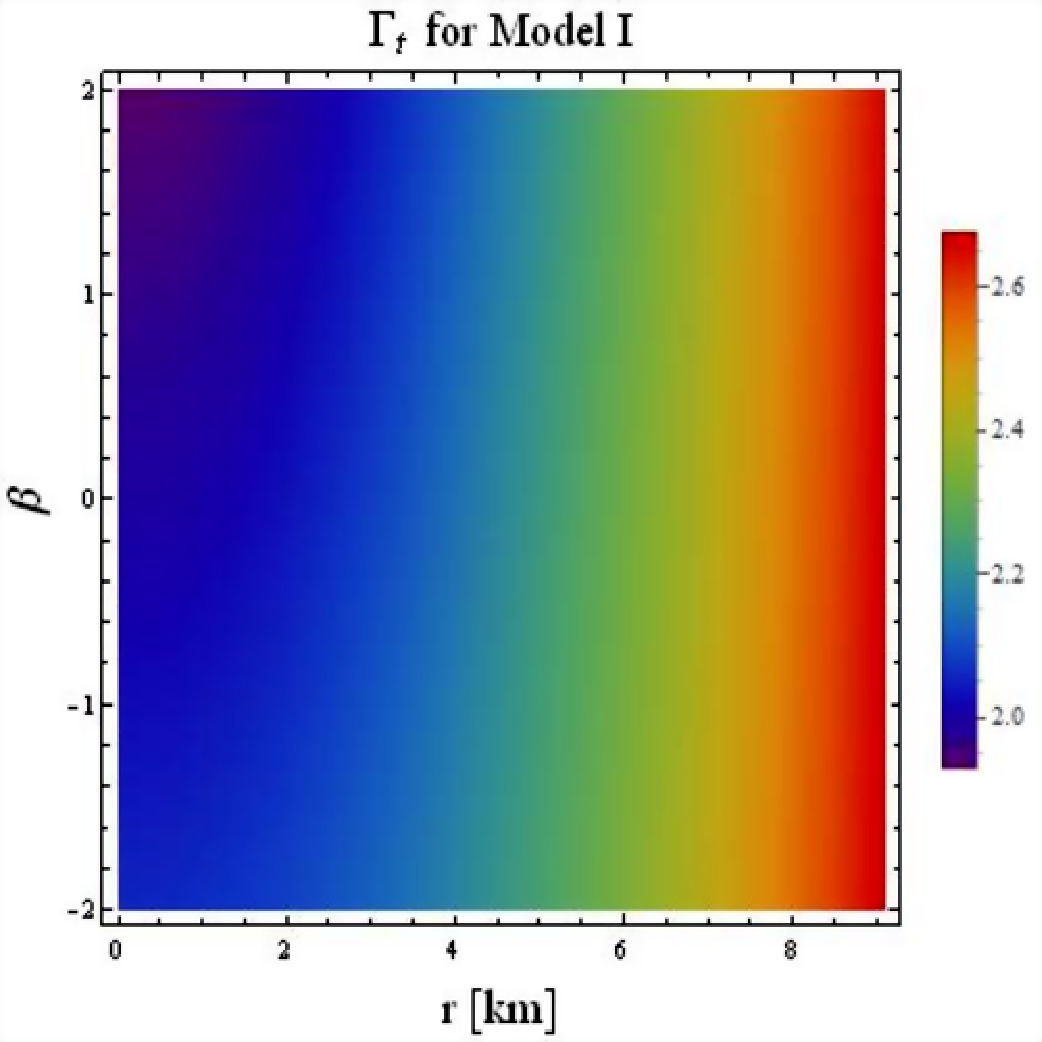,width=.4\linewidth}\epsfig{file=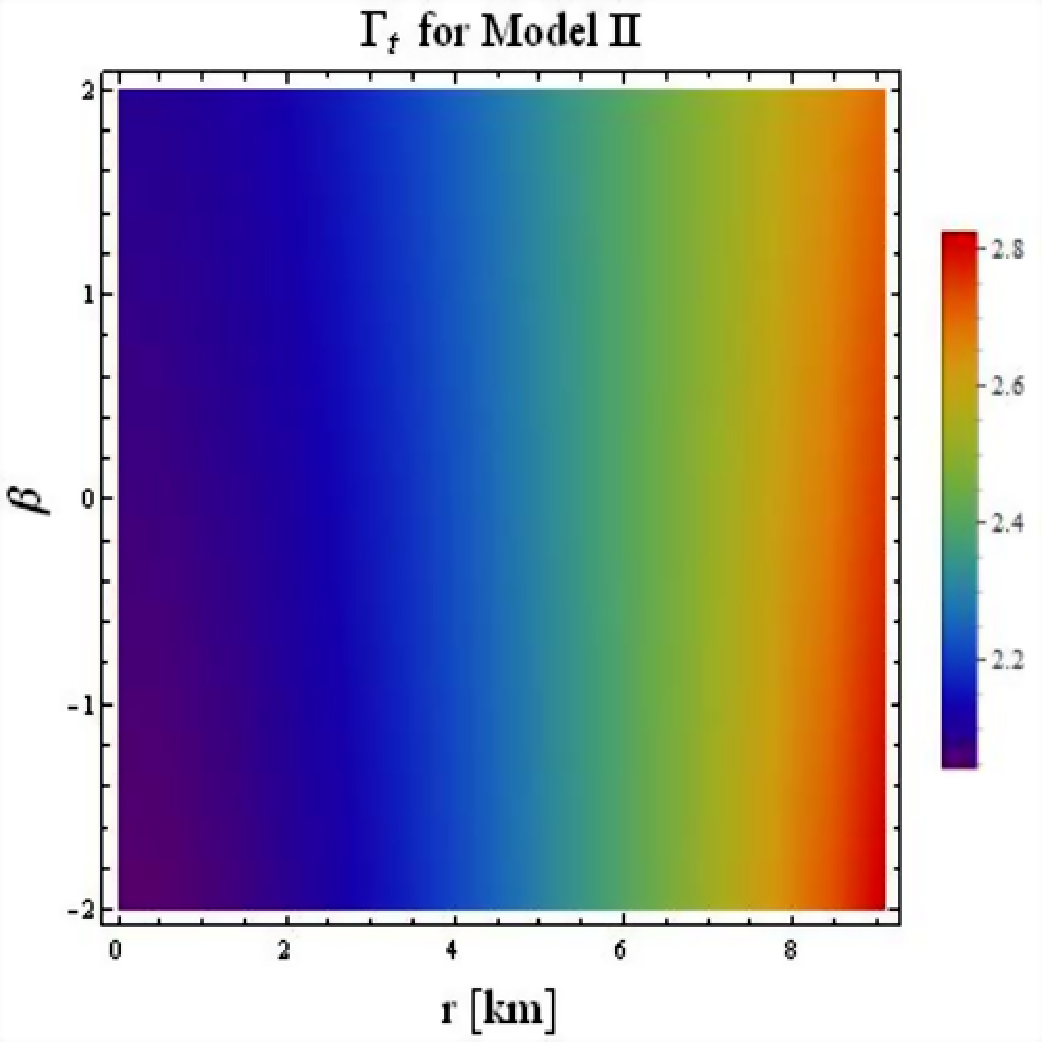,width=.4\linewidth}
\caption{Adiabatic index versus $\beta$ and $r$.}
\end{figure}

\section{Estimating the Model Parameter}

This section calculates the values of the model parameter for eight
different compact stars which are consistent with their observed
data such as radii and masses. We already discussed that there are
two fundamental forms of the matching conditions. The second form
states that the radial pressure in any compact fluid configurations
becomes zero at the boundary surface, i.e., $P_r{_=^\Sigma}0$. When
we apply this condition on Eqs.\eqref{g14c}, the resulting
expression becomes
\begin{align}\nonumber
&3\mathrm{M}(\emph{R}-2\mathrm{M})^2+4\mathrm{B_c}\big\{11\beta\mathrm{M}^3-\mathrm{M}^2\big(8\pi\emph{R}^3+11\beta\emph{R}\big)\\\label{g26}
&+3\mathrm{M}\emph{R}^2\big(\beta+4\pi\emph{R}^2\big)-4\pi\emph{R}^5\big\}=0,
\end{align}
and Eq.\eqref{g14f} takes the form after combining with the
vanishing radial pressure constraint
\begin{align}\nonumber
&\mathrm{B_c}\big[2\beta\mathrm{M}^5\big(4\emph{R}^4+189\emph{R}^2-198\big)+\beta\mathrm{M}^4\big(487-20\emph{R}^4
-634\emph{R}^2\big)\emph{R}+2\mathrm{M}^3\big\{4\pi\emph{R}^6\\\nonumber
&+\beta\big(8\emph{R}^4+184\emph{R}^2-91\big)\emph{R}^2\big\}-\mathrm{M}^2\big\{20\pi\emph{R}^7
+\beta\emph{R}^3\big(4\emph{R}^4+85\emph{R}^2-19\big)\big\}\\\label{g27}
&+2\mathrm{M}\big(8\pi\emph{R}^8+3\beta\emph{R}^6\big)-4\pi\emph{R}^9\big]-4\big[3\mathrm{M}\emph{R}^3(\mathrm{M}
-\emph{R})(\emph{R}-2\mathrm{M})^2\big]=0.
\end{align}
We use Eqs.\eqref{g26} and \eqref{g27} to construct Tables
\textbf{1} and \textbf{2}, providing the values of $\beta$ for
different compact stellar models corresponding to three different
choices of the bag constant.
\begin{table}[H]
\scriptsize \centering \caption{Values of model parameter
corresponding to different choices of the bag constant for model I.}
\label{Table1} \vspace{+0.1in} \setlength{\tabcolsep}{0.68em}
\begin{tabular}{cccccc}
\hline\hline  Values of $\mathrm{B_c}$ & & & $75~MeV/fm^3$ &
$85~MeV/fm^3$ & $95~MeV/fm^3$
\\\hline Star Models & Mass ($\mathrm{M}_{\bigodot}$) & Radius (km) & $\beta$ & $\beta$ & $\beta$
\\\hline Cen X-3 & 1.49 & 9.51 & -151.814 & 149.225 & 386.888
\\\hline
SMC X-4 & 1.29 & 8.83 & -401.983 & -98.228 & 141.578
\\\hline
Her X-1 & 0.85 & 8.1 & -826.024 & -525.799 & -288.779
\\\hline
4U 1820-30 & 1.58 & 9.1 & -459.042 & -164.623 & 67.813
\\\hline
4U 1608-52 & 1.74 & 9.3 & -427.304 & -140.648 & 85.659
\\\hline
PSR J 1614 2230 & 1.97 & 10.3 & 5.949 & 289.672 & 513.663
\\\hline
PSR J 1903+327 & 1.67 & 9.82 & -99.141 & 196.936 & 430.68
\\\hline
SAX J 1808.4-3658 & 0.9 & 7.95 & -496.977 & -189.679 & 52.925  \\
\hline\hline
\end{tabular}
\end{table}
\begin{table}[H]
\scriptsize \centering \caption{Values of model parameter
corresponding to different choices of the bag constant for model
II.} \label{Table2} \vspace{+0.1in} \setlength{\tabcolsep}{0.68em}
\begin{tabular}{cccccc}
\hline\hline  Values of $\mathrm{B_c}$ & & & $75~MeV/fm^3$ &
$85~MeV/fm^3$ & $95~MeV/fm^3$
\\\hline Star Models & Mass ($\mathrm{M}_{\bigodot}$) & Radius (km) & $\beta$ & $\beta$ & $\beta$
\\\hline Cen X-3 & 1.49 & 9.51 & 3.608 & -3.547 & -9.195
\\\hline
SMC X-4 & 1.29 & 8.83 & 12.081 & 2.952 & -4.255
\\\hline
Her X-1 & 0.85 & 8.1 & 16.075 & -0.846 & -14.205
\\\hline
4U 1820-30 & 1.58 & 9.1 & 10.449 & 3.747 & -1.544
\\\hline
4U 1608-52 & 1.74 & 9.3 & 8.543 & 2.812 & -1.713
\\\hline
PSR J 1614 2230 & 1.97 & 10.3 & -0.095 & -4.631 & -8.211
\\\hline
PSR J 1903+327 & 1.67 & 9.82 & 1.996 & -3.965 & -8.671
\\\hline
SAX J 1808.4-3658 & 0.9 & 7.95 & 25.646 & 9.788 & -2.731  \\
\hline\hline
\end{tabular}
\end{table}

\section{Conclusions}

This study focuses on the possible existence of a specific compact
object, 4U 1820-30, within the framework of
$f(\mathcal{R},\mathcal{T},\mathcal{R}_{\alpha\gamma}\mathcal{T}^{\alpha\gamma})$
theory. Two different modified models have been employed to explore
the implications of the non-minimal interaction between matter and
geometry, while considering a specific range of the coupling
parameter $\beta$. We have derived the modified field equations
\eqref{g8}-\eqref{g8b} and \eqref{g8c}-\eqref{g8e} corresponding to
both models and observed them as the under-determined systems. The
Finch-Skea metric potentials, which satisfy the criteria \cite{41j},
have been adopted to compute solutions for these sets. Additionally,
we have utilized the MIT bag model to describe the internal
distribution of strange stars. Within the Finch-Skea spacetime,
characterized by three unknowns $(d_1,d_2,d_3)$, we have used the
matching conditions at the hypersurface to calculate them in
relation with the mass and radius of a self-gravitating star.

The graphical representation of the matter variables (Figure
\textbf{2}) ensures the validation of the obtained solutions
\eqref{g14b}-\eqref{g14c} and \eqref{g14e}-\eqref{g14f}. Examining
the mass distribution within the strange star reveals a consistent
outward increasing trend (Figure \textbf{6}). We have also observed
that model I exhibits a slightly denser interior in comparison to
model II for the considered values of $\beta$. Furthermore, the
plots of compactness and redshift fall within the acceptable limits.
The EoS parameters (Figure \textbf{7}) ensure the feasibility of the
developed models. The energy bounds maintain positive values
everywhere in the interior, confirming the physical viability of
both of our proposed solutions (Figures \textbf{8} and \textbf{9}).
Moreover, the TOV equations \eqref{g11} and \eqref{g11a} have been
plotted (Figure \textbf{10}) and it has been shown that the obtained
models agree with the hydrostatic equilibrium condition in the
presence of modified corrections.

Finally, we have employed three distinct criteria to assess the
stability within this framework. Their plots demonstrated the
stability of the solutions \eqref{g14b}-\eqref{g14c} and
\eqref{g14e}-\eqref{g14f}, as supported by the observations (Figures
\textbf{11} and \textbf{12}), in agreement with \cite{27a,38}.
Notably, it becomes apparent that our solutions exhibit superior
efficacy in comparison to those proposed in \cite{25a}, suggesting
that the additional force within this theoretical framework may
yield more favorable results for the considered parametric values.
Further, it is important to note here that there are several past
related works on compact stars. In the following, we highlight two
main differences which set our work different from them. Firstly, we
have used the disappearing radial pressure condition at the
spherical boundary to estimate the values of the model parameter
that correspond to the observed data (masses and radii) of eight
different compact stars. Tables \textbf{1} and \textbf{2} present
these parametric values for models I and II, respectively, for
different acceptable values of the bag constant. It is found that
$\beta$ takes both negative and positive values, indicating itself
to be an arbitrary parameter. Secondly, in the recent works, the
compact interiors have not been explored for our model II. We have
adopted this model for the very first time in this article and
presented a brief comparison with the results corresponding to the
first model. Our findings can ultimately be reduced to GR when
$\beta=0$.

\section*{Appendix A}

The values of $\tau_i^{'s}$ displayed in Eqs.\eqref{g8c}-\eqref{g8e}
are
\begin{align}\nonumber
\tau_1&=\frac{e^{-{A_2}}}{4}\bigg(A_1'^2-A_1'{A_2}'+2A_1''+\frac{4A_1'}{r}\bigg),\\\nonumber
\tau_2&=\frac{e^{-{A_2}}}{4}\bigg(A_1'{A_2}'-A_1'^2-2A_1''+\frac{4{A_2}'}{r}\bigg),\\\nonumber
\tau_3&=e^{-{A_2}}\bigg(\frac{{A_2}'}{r}-\frac{A_1'}{r}+\frac{2e^{A_2}}{r^2}-\frac{2}{r^2}\bigg),\\\nonumber
\tau_4&=\frac{e^{-{A_2}}}{4}\bigg(A_1'{A_2}'^2-A_1'^2{A_2}'-3A_1''{A_2}'-A_1'{A_2}''+2A_1'A_1''+2A_1'''-\frac{4A_1'}{r^2}\\\nonumber
&-\frac{4A_1'{A_2}'}{r}+\frac{4A_1''}{r}\bigg),\\\nonumber
\tau_5&=\frac{e^{-{A_2}}}{4}\bigg(A_1'^2{A_2}'-A_1'{A_2}'^2+3A_1''{A_2}'+A_1'{A_2}''-2A_1'A_1''-2A_1'''-\frac{4{A_2}'}{r^2}\\\nonumber
&-\frac{4{A_2}'^2}{r}+\frac{4{A_2}''}{r}\bigg),\\\nonumber
\tau_6&=e^{-{A_2}}\bigg(\frac{A_1'{A_2}'}{r}-\frac{4{A_2}'e^{A_2}}{r^2}-\frac{{A_2}'^2}{r}+\frac{{A_2}''}{r}+\frac{{A_2}'}{r^2}-\frac{A_1''}{r}
+\frac{A_1'}{r^2}-\frac{4e^{A_2}}{r^3}+\frac{4}{r^3}\bigg),\\\nonumber
\tau_7&=\frac{e^{-{A_2}}}{4}\bigg(A_1'^2{A_2}'^2-A_1'{A_2}'^3+4A_1''{A_2}'^2+3A_1'{A_2}'{A_2}''-5A_1'''{A_2}'
-4A_1''{A_2}''\\\nonumber
&-A_1'^2{A_2}''-A_1'{A_2}'''-4A_1'A_1''{A_2}'+2A_1''^2+2A_1'A_1'''+2A_1''''+\frac{4A_1'{A_2}'^2}{r}\\\nonumber
&-\frac{8A_1''{A_2}'}{r}-\frac{4A_1'{A_2}''}{r}+\frac{4A_1'''}{r}-\frac{8A_1''}{r^2}+\frac{8A_1'}{r^3}\bigg),\\\nonumber
\tau_8&=\frac{e^{-{A_2}}}{4}\bigg(A_1'{A_2}'^3-A_1'^2{A_2}'^2-4A_1''{A_2}'^2-3A_1'{A_2}'{A_2}''+4A_1'A_1''{A_2}'+5A_1'''{A_2}'\\\nonumber
&+A_1'^2{A_2}''+A_1'{A_2}'''+4A_1''{A_2}''-2A_1''^2-2A_1'A_1'''-2A_1''''+\frac{4{A_2}'^3}{r}-\frac{12{A_2}'{A_2}''}{r}\\\nonumber
&+\frac{8{A_2}'^2}{r^2}+\frac{4{A_2}'''}{r}-\frac{8{A_2}''}{r^2}+\frac{8{A_2}'}{r^3}\bigg),\\\nonumber
\tau_9&=e^{-{A_2}}\bigg(\frac{{A_2}'^3}{r}-\frac{A_1'{A_2}'^2}{r}-\frac{3{A_2}'{A_2}''}{r}+\frac{4A_1''{A_2}'}{r}-\frac{2A_1'{A_2}'}{r^2}
-\frac{6{A_2}'}{r^3}+\frac{A_1'{A_2}''}{r}\\\nonumber
&-\frac{{A_2}''e^{A_2}}{r^2}+\frac{8{A_2}'e^{A_2}}{r^3}+\frac{{A_2}'''}{r}-\frac{A_1'''}{r}+\frac{2A_1''}{r^2}-\frac{2A_1'}{r^3}+\frac{12e^{A_2}}{r^4}
-\frac{12}{r^4}\bigg).
\end{align}
The Ricci scalar and non-zero components of Ricci tensor for metric
\eqref{g6} are
\begin{align}\nonumber
\mathcal{R}&=\frac{1}{2e^{{A_2}}}\bigg(A_1'^2-{A_2}'A_1'+2A_1''-\frac{4{A_2}'}{r}+\frac{4A_1'}{r}-\frac{4e^{{A_2}}}{r^2}
+\frac{4}{r^2}\bigg),\\\nonumber
\mathcal{R}_{00}&=\frac{1}{4e^{{A_2}-A_1}}\bigg(A_1'^2-{A_2}'A_1'+2A_1''+\frac{4A_1'}{r}\bigg),\\\nonumber
\mathcal{R}_{11}&=\frac{1}{4}\bigg({A_2}'A_1'-A_1'^2-2A_1''+\frac{4{A_2}'}{r}\bigg),
\quad
\mathcal{R}_{22}=\frac{1}{2e^{{A_2}}}\bigg({A_2}'r-A_1'r+2e^{A_2}-2\bigg).
\end{align}

\section*{Appendix B}

The anisotropy corresponding to the model I is expressed by
\begin{align}\nonumber
\Pi&=\bigg\{\big(2 \beta r A_1''-\beta  {A_2} ' \big(r
A_1'+4\big)+\beta  r A_1'^2+4 \beta A_1'+32 \pi r e^{{A_2}}\big)
\big(2\big(8 \beta -\beta r^2 {A_2} '' \\\nonumber &+9\beta r^2
A_1''+64 \pi  r^2 e^{{A_2}}-8 \beta e^{{A_2}}\big)-\beta r^2 {A_2}
'^2+r^23 \beta  A_1 '^2-10 \beta  r {A_2} ' \big(r A_1'+4\big)
\\\nonumber &+16
\beta r A_1'\big)\bigg\}^{-1}\bigg\{3\beta
r^3(6\beta\mathrm{B_c}+1)A_1'^4+4r\big(\beta
{A_2}''\big(8\mathrm{B_c}\big(\beta+\big(8\pi r^2-\beta
\big)e^{{A_2}}\big)+r^2 \\\nonumber &\times (4 \beta \mathrm{B_c}-1)
A_1''\big)+8 A_1'' \big(\beta (1+\beta \mathrm{B_c} )+e^{{A_2}}
\big(8 \pi  r^2 (3\beta \mathrm{B_c} +1)-\beta (\beta \mathrm{B_c}
+1)\big)\big)\\\nonumber &+128 \pi \mathrm{B_c} e^{{A_2}}
\big(\big(8 \pi r^2-\beta \big)e^{{A_2}}+\beta\big) +\beta r^2 (10
\beta \mathrm{B_c} +9) A_1 ''^2\big)+2 r A_1'^2 \big(4 \big(2 \big(2
\beta
\\\nonumber &\times (3 \beta \mathrm{B_c} +1)+(2\mathrm{B_c} \beta +1)
\big(8 \pi r^2-\beta \big) e^{{A_2}}\big)+\beta r^2 (7 \beta
\mathrm{B_c} +3) A_1''\big)-\beta  r^2 {A_2}''
\\\nonumber &\times (6 \beta \mathrm{B_c}
+1) \big)+4A_1' \big(8 \big(\beta (2 \beta \mathrm{B_c}
+1)+e^{{A_2}} \big(4 \pi r^2 (8 \beta \mathrm{B_c} +1)-\beta  (2
\beta \mathrm{B_c}+1)\big)\big)
\\\nonumber &+ r^2 (24 \beta\mathrm{B_c} +19)\beta A_1''-3 \beta r^2 {A_2}
''\big)+\beta r^2 {A_2} '^3 \big(88 \beta \mathrm{B_c} +(22 \beta
\mathrm{B_c} r+r) A_1'+14\big)\\\nonumber &+\beta r {A_2} '^2
\big(-2 \big(8 \big(\mathrm{B_c} \big(\beta +24 \pi
r^2\big)e^{{A_2}}-9 \beta \mathrm{B_c} -4\big)+ (22 \beta
\mathrm{B_c} +1) A_1 ''(r)r^2\big)+9 r^2\\\nonumber & \times A_1'^2
(2 \beta \mathrm{B_c}+1) +6 r (8 \beta \mathrm{B_c} +11)
A_1'\big)+2\beta r^2 (16 \beta \mathrm{B_c} +19) A_1'^3-{A_2} '
\big(\beta r^3 A_1'^3
\\\nonumber &\times (34 \beta \mathrm{B_c} +13)+4 \big(\beta  r^2 (12 \beta \mathrm{B_c} +1) {A_2}
''+8 \big(\beta (2 \beta \mathrm{B_c} +1)-e^{{A_2}} \big(\beta  (1+2
\beta\mathrm{B_c})\\\nonumber &-12 \pi r^2 (4 \beta \mathrm{B_c}
+1)\big)\big)+\beta  r^2 (44 \beta \mathrm{B_c} +27) A_1''\big)+2 r
A_1' \big(\beta r^2 (6 \beta \mathrm{B_c} -1) {A_2} ''+8\\\nonumber
&\times  \big(\beta (13 \beta \mathrm{B_c} +10)+e^{{A_2}} \big(8 \pi
r^2 (3 \beta \mathrm{B_c} +1)-\beta (\beta \mathrm{B_c}
+1)\big)\big)+\beta  (22 \beta \mathrm{B_c} +19) \\\nonumber &\times
r^2A_1''\big)+2 \beta r^2 (92 \beta \mathrm{B_c}+35)
A_1'^2\big)\bigg\},
\end{align}
whereas it becomes for model II as
\begin{align}\nonumber
\Pi&=\bigg\{2 e^{{A_2}} r^2 \beta \mathcal{R}^2+4 \beta  \big(r
{A_2}'+2 e^{{A_2}}-r A_1'-2\big) \mathcal{R}+r \big(\beta \big(r
\mathcal{R}'-2 r \tau_6-4 \tau_3\big) A_1'\\\nonumber &-r \beta
\tau_3 A_1'^2+32 e^{{A_2}} \pi  r-4 r \beta \tau_9-4 \beta
\tau_6+\beta {A_2}' \big(r A_1' \tau_3+4 \tau_3+2 r \tau_6-r
\mathcal{R}'\big)+2 r\\\nonumber &\times \beta \mathcal{R}''-2 r
\beta \tau_3 A_1''\big)\bigg\}^{-1}\bigg\{r \big(r A_1'^2+2
A_1'-{A_2}' \big(r A_1'+2\big)+2 r A_1''\big)\bigg\}+\bigg\{8
e^{{A_2}} r \beta \mathcal{R}^2\\\nonumber &+\beta \big({A_2}'
\big(7 r A_1'-4\big)-r A_1'^2-20 A_1'-8 r A_1''\big) \mathcal{R}+2
\big(2 \beta \tau_8 r-3 \beta \tau_4 A_1' r-\beta \tau_5
A_1'r\\\nonumber &+2 \beta \mathcal{R}'' r+64 e^{{A_2}} \pi r-\beta
(6 \tau_1+2 \tau_2+3 r \tau_4+r \tau_5) {A_2}'-\beta \mathcal{R}'
\big(r {A_2}'+r A_1'-4\big)\\\nonumber &-6 \beta \tau_1 A_1'-2 \beta
\tau_2 A_1'\big)\bigg\}^{-1}\bigg\{4 \big(2 e^{{A_2}} r \beta
\mathrm{B_c} \mathcal{R}^2+\beta \mathrm{B_c} \big(2 {A_2}' \big(r
A_1'-2\big)-r A_1'^2-8 A_1'\\\nonumber &-2 r A_1''\big)
\mathcal{R}+32 e^{{A_2}} \pi r \mathrm{B_c} +4 r \beta \mathrm{B_c}
\tau_7-4 \beta \mathrm{B_c} \tau_1 {A_2}'-2 r \beta
\mathrm{B_c}\tau_4 {A_2}'-{A_2}'-4 \beta \mathrm{B_c}
\tau_1\\\nonumber &\times A_1'-2 r \beta \mathrm{B_c} \tau_4
A_1'-A_1'-\beta \mathrm{B_c} R' \big(r {A_2}'+r A_1'-4\big)+2 r
\beta \mathrm{B_c} \mathcal{R}''\big)\bigg\}+\bigg\{\big(8 e^{{A_2}}
r \beta \mathcal{R}^2\\\nonumber &+\beta \big(-r A_1'^2-20
A_1'+{A_2}' \big(7 r A_1'-4\big)-8 r A_1''\big) \mathcal{R}+2 \big(2
\beta \tau_8 r-3 \beta  \tau_4 A_1' r-\beta \tau_5 A_1' r\\\nonumber
&+2 \beta \mathcal{R}'' r+64 e^{{A_2}} \pi r-\beta  (6
\tau_1+2\tau_2+3 r \tau_4+r \tau_5) {A_2}'-6 \beta \tau_1 A_1'-2
\beta \tau_2 A_1'-\beta \mathcal{R}'\\\nonumber &\times \big(r
{A_2}'+r A_1'-4\big)\big)\big) \big(2 e^{{A_2}} r^2 \beta
\mathcal{R}^2+4 \beta \big(r {A_2}'+2 e^{{A_2}}-r A_1'-2\big)
\mathcal{R}-r \big(r \beta \tau_3 A_1'^2\\\nonumber &-\beta \big(-4
\tau_3-2 r \tau_6+r \mathcal{R}'\big) A_1'-32 e^{{A_2}} \pi r+4 r
\beta \tau_9+4 \beta\tau_6-\beta {A_2}' \big(r A_1' \tau_3+2 r
\tau_6\\\nonumber &+4 \tau_3-r \mathcal{R}'\big)-2 r \beta
\mathcal{R}''+2 r \beta \tau_3 A_1''\big)\big)\bigg\}^{-1}\bigg\{2 r
\beta \big(\big(-r (2 \tau_1+\mathcal{R}) A_1'^2-\big(4 \tau_1+r4
\tau_4\\\nonumber &-2 r \mathcal{R}'\big) A_1'-8 r \tau_7-8
\tau_4+{A_2}' \big(2 r A_1' \tau_1+4 \tau_1+4 r \tau_4+\mathcal{R}
\big(r A_1'+4\big)\big)-r4 \tau_1 A_1''\\\nonumber &-2 r \mathcal{R}
A_1''\big) \big(2 e^{{A_2}} r \beta \mathrm{B_c} \mathcal{R}^2+\beta
\mathrm{B_c} \big(2 r A_1'^2+4 A_1'+{A_2}' \big(r A_1'+8\big)-2 r
A_1''\big) \mathcal{R}-3 {A_2}'\\\nonumber &-32 e^{{A_2}} \pi r
\mathrm{B_c} +4 r \beta \mathrm{B_c}\tau_8-4 \beta \mathrm{B_c}
\tau_2 {A_2}'-2 r \beta \mathrm{B_c} \tau_5 {A_2}'-4 \beta
\mathrm{B_c} \tau_2 A_1'-2 r \beta \tau_5\mathrm{B_c}
A_1'\\\nonumber &-3 A_1'+\beta \mathrm{B_c} \mathcal{R}' \big(r
{A_2}'+r A_1'-4\big)-2 r \beta \mathrm{B_c}
\mathcal{R}''\big)+\big(2 r \tau_2 A_1'^2-r \mathcal{R}
A_1'^2+4\tau_2 A_1'+8 \tau_5\\\nonumber &+4 r \tau_5 A_1'-4
\mathcal{R} A_1'+8 r \tau_8+2 \mathcal{R}' \big(r {A_2}'-2 r
A_1'-4\big)+ \big( (\mathcal{R}-2 \tau_2)rA_1'-4 (\tau_2+r
\tau_5)\big)\\\nonumber &\times {A_2}' -4 r \mathcal{R}''+4 r \tau_2
A_1''-2 r \mathcal{R} A_1''\big) \big(-2 e^{{A_2}} r \beta
\mathrm{B_c} \mathcal{R}^2+\beta \mathrm{B_c} \big( A_1'^2r+8 A_1'+2
r A_1''\\\nonumber &+{A_2}' \big(4-2 r A_1'\big)\big) \mathcal{R}-32
e^{{A_2}} \pi  r \mathrm{B_c} -4 r \beta \mathrm{B_c} \tau_7+4 \beta
\mathrm{B_c} \tau_1 {A_2}'+2r \beta\mathrm{B_c} \tau_4
{A_2}'+{A_2}'\\\nonumber & +4 \beta \mathrm{B_c} \tau_1 A_1'+2 r
\beta \mathrm{B_c} \tau_4 A_1'+A_1'+\beta \mathrm{B_c} \mathcal{R}'
\big(r {A_2}'+r A_1'-4\big)-2 r \beta \mathrm{B_c}
\mathcal{R}''\big)\big)\bigg\}.
\end{align}\\
\textbf{Data Availability Statement:} This manuscript has no
associated data.

\end{document}